\RequirePackage[l2tabu, orthodox]{nag}
\documentclass[12pt,english,fleqn]{article}

\RequirePackage{mathtools}
\usepackage{pdfpages}
\usepackage{amsmath}
\usepackage{amssymb}
\RequirePackage{graphicx}
\RequirePackage[shortlabels]{enumitem}
\usepackage[T1]{fontenc}
\usepackage{float}
\RequirePackage{tabularray}
\usepackage{relsize}
\usepackage{setspace}
\usepackage{amsfonts, amstext, mathtools}
\usepackage{bbm}

\usepackage{dutchcal} 
\usepackage{enumitem}
\usepackage{color}
\usepackage{eurosym}
\usepackage{afterpage}
\usepackage{multirow}
\usepackage{amsfonts}
\usepackage{amssymb}
\usepackage{amstext}
\usepackage{amsmath,amsthm,amssymb}
\usepackage{yhmath}

\usepackage[english]{babel}
\usepackage{graphicx}
\usepackage[errorshow]{longtable}
\usepackage{array}
\usepackage{geometry}
\usepackage{authblk}
\usepackage{setspace}
\usepackage{rotating}
\usepackage{lscape}
\usepackage{verbatim}
\usepackage{hyperref}
\usepackage{booktabs}
\usepackage[T1]{fontenc}
\usepackage[utf8]{inputenc}
\usepackage{dsfont}
\usepackage[flushleft]{threeparttable}
\usepackage{float}
\usepackage{mathpazo}

\usepackage{tikz}
\usetikzlibrary{arrows.meta, positioning, shapes, calc}

\usepackage{bbm}

\usepackage{xcolor}
\usepackage{soul}  % for \st strikethrough
\hypersetup{
    colorlinks,
    linkcolor={blue!80!black},
    citecolor={blue!80!black},
    urlcolor={blue!80!black}
}
\hypersetup{
    colorlinks,
    linkcolor={blue!80!black},
    citecolor={blue!80!black},
    urlcolor={blue!80!black}
}

\usepackage[lining]{ebgaramond}
\iftutex
    \RequirePackage[Scale=MatchUppercase]{euler-math}
    \unimathsetup{warnings-off={mathtools-colon,mathtools-overbracket},mathbf=sym}
\else
    \RequirePackage[euler-digits,small]{eulervm} \RequirePackage{amssymb}
    \DeclareMathAlphabet{\mathbf}{U}{zeur}{b}{n}
\fi
\RequirePackage[largesc,nohelv]{newpxtext}
 \iftutex
     \RequirePackage{euler-math}
     \unimathsetup{warnings-off={mathtools-colon,mathtools-overbracket},mathbf=sym}
\else
     \RequirePackage[euler-digits]{eulervm} \RequirePackage{amssymb}
     \DeclareMathAlphabet{\mathbf}{U}{zeur}{b}{n}
 \fi

\usepackage[backend=biber, style=apa, sorting=nyt, natbib]{biblatex}

\title{\fontsize{17.5}{18.5}\selectfont{\textbf
{{\\Trade Relationships During and After a Crisis}}}\thanks{\fontsize{9.5}{9.5}\selectfont I am grateful to Isabelle Mejean, Dennis Novy, Carlo Perroni, and Christine Braun for their support. I am also grateful for comments by Isabela Manelici, Marta Santamaria, Wiji Arulampalam, James Fenske, Luis Candelaria,  as well as seminar participants at the Royal Economic Society Conference 2023, GEP-CEPR Postgraduate Conference, European Trade Study Group 2023, the 2023 Oxford Development Economics Workshop,2023 CEP-Warwick Junior Trade Workshop, the II Workshop on International Trends in Economic Research, and 2023 Junior workshop day at KCL. The author is affiliated with the University of Nottingham School of Economics and the Centre of Inclusive Trade Policy (CITP). Email address: \texttt{Alejandra.Martinez@nottingham.ac.uk}.}\vspace*{0.12in}}

\author{\vspace*{0.5cm}Alejandra Martinez \\ \vspace{0.5cm}} 

\date{\vspace*{0.1cm}{{\fontsize{13.5}{12.5}\selectfont{
May 2026} \\ \vspace{0.5cm}}}}

\begin{document}

\vspace*{-0.8in}

\renewcommand{\baselinestretch}{1.2}
\small \normalsize

{\let\newpage\relax\maketitle}

\begin{abstract}
 \noindent 
This paper provides causal evidence that temporary supply disruptions reshape firms' relationship portfolios in international trade. Using exogenous road disruptions during Colombia's 2010--11 La Niña episode, I identify exposure at the buyer-–seller relationship level, exploiting variation within importers' supplier portfolios. Exposed relationships are less likely to terminate when importers have alternative non-exposed suppliers. However, firms with broader portfolio exposure gradually reduce their relationship networks and may eventually exit the market. A framework linking relationship surplus to portfolio composition explains these contrasting responses and shows how the same shock generates opposite effects at relationship and firm levels.

\noindent 

\medskip
\medskip

{\normalsize \smallskip \noindent
\textbf{Keywords}: La Niña, Supply Chains, International Contracts, Weather Shocks%,\\ 
%\hspace*{0.82in} flower exporters \\
}

\bigskip\ 
\bigskip\ 
\bigskip\ 
\bigskip\ 
\bigskip\ 
\bigskip\ 
\end{abstract}

%\pagebreak
\thispagestyle{empty}

\setcounter{page}{0} 

{\let\newpage\relax\maketitle}

\renewcommand{\baselinestretch}{1.1}

\thispagestyle{empty}\pagebreak

\smallskip
\smallskip

\medskip
\medskip

\renewcommand{\baselinestretch}{1.15}
\small \normalsize

\section{Introduction}
\label{sec:intro}
 
In product markets with incomplete contracts, firms rely on buyer–seller relational contracts.\footnote{Relational contracts are defined by \citet*{baker2002} as ``informal agreements sustained by the value of future relationships.''} These relationships are  vulnerable to temporary disruptions that interfere with production or delivery. As extreme weather events become more frequent and severe, supply chains are increasingly exposed to such disruptions, including road closures, transport bottlenecks, and infrastructure outages. Even when shocks are local and transitory, they can substantially disrupt established production networks and generate aggregate effects.\footnote{\citet*{Elliott2022} show how idiosyncratic shocks propagate through production networks. \citet*{DiGiovanni2014} and \citet*{Magerman2016} document how firm-level shocks contribute to aggregate volatility.} Understanding how supply chain disruptions affect buyer--seller relationships is particularly important in international settings, where contracts are imperfectly enforceable and firms depend heavily on repeated interactions.
 
Existing work documents output losses among firms linked to regions affected by disruptions \citep*{Barrot2016, carvalho2021}. Much less is known about how firms adjust their relationship portfolios in response. These adjustments matter because buyer--seller relationships form the backbone of trade networks and shape the structure of world trade. The accumulation of relationships drives firm growth \citep*{Oberfield2018} and affects productivity through positive sorting \citep*{BernardMox2018b}. This paper exploits shocks at the level of individual buyer--seller relationships and shows how disruptions to specific relationships can propagate within firms to affect their overall relationship portfolio.
 
I focus on Colombian flower exporters and their U.S.\ importers during the extreme 2010--11 La Ni\~na episode, which disrupted road access to cargo terminals used by exporters. The Colombian flower industry exemplifies relational trade under limited contract enforceability: exporters access international markets almost exclusively through established direct relationships, and forming new connections is costly. The setting therefore provides a well-suited environment in which to study relationship portfolio adjustments. Importantly, flower production itself was largely unaffected by flooding at production sites, and the industry has minimal reliance on upstream inputs; allowing me to isolate disruptions operating through transport access rather than input shortages. This enables a clean comparison between buyer--seller relationships exposed to disrupted routes and those that were not.
 
I construct a novel dataset linking road disruptions to exporters' routes to cargo terminals, and combine it with transaction-level data from Colombian customs records covering 2007--2019. Exploiting variation in roads disrupted by the La Ni\~na floods, I define relationship \textit{exposure to the shock} based on whether a given buyer--seller relationship relied on disrupted transport routes in the 18 months preceding the shock. For importers, exposure varies within the firm because they typically source from multiple suppliers located across different regions. This design allows me to trace how disruptions to specific relationships transmit within firms to affect their broader relationship portfolios.\footnote{\citet*{Anindya2025}, \citet*{Balboni2023}, \citet*{Korovkin2024}, and \citet*{Fujiy2022} study the reorganization of firm linkages using firm-level shocks, which precludes analysis of how disruptions to individual relationships propagate within firms.}
 
I estimate the effect of road disruptions --- which reduced the likelihood of product delivery --- on the probability of a relationship ending using an event-study design. I estimate linear probability models throughout and interpret coefficients as percentage-point changes in the probability of a relationship ending. Because relationship endings are infrequent at the monthly level, outcomes are aggregated to six-month periods when analyzing relationship portfolio adjustments. I also examine adjustments along other margins using monthly data at the 6-digit HS product level, including transaction incidence, prices and traded quantities.
 
The results reveal a sharp contrast between relationship-level and  firm-level responses. Within an importer's portfolio, exposure reduces the probability that a relationship ends during the disruption by 8 percentage points --- a 48\% decline relative to a pre-shock baseline. The effect is driven by importers with partial exposure, who can still receive shipments through unaffected suppliers and therefore have an incentive to wait for disrupted ones to resume deliveries rather than terminate them. Consistent with this mechanism, exposed 
relationships are more likely to record a transaction once roads 
reopen, and exporters lower their prices per kilogram, compensating buyers for earlier non-deliveries.

At the firm level, the response reverses and is concentrated among high in-degree importers --- those with many pre-crisis relationships. During the crisis, moving from the 25$^{th}$ to the 75$^{th}$ percentile of exposure increases the probability of a relationship ending by 6.7 percentage points and the probability of firm exit by 6.5 percentage points. Low in-degree importers, by contrast, preserve existing links and instead form new ones during the shock, consistent with higher search and replacement costs making relationship preservation the dominant strategy. The crisis thus produces a persistent divergence: high in-degree survivors do not rebuild their portfolios, while low in-degree importers temporarily expand theirs, possibly as a hedge against future shocks. Aggregating these effects with pre-shock totals, the shock generates roughly 287 additional relationship terminations among high in-degree 
importers (an 56\% increase over baseline), the exit of about 4.3\% of high in-degree importers, and foregone exporter revenue of approximately \$55-77M US dollars per year, sustained over the two post-shock years in which firms adjust ($\approx$5--7\% of the sector's 2010 U.S.\ sales). 

The final section develops a simple framework to organize these findings. The central mechanism is that the surplus from any given relationship depends on the composition of the firm's entire relationship portfolio, not just on the performance of that relationship in isolation. I distinguish between an \emph{indirect} effect --- disruptions to other relationships raise the marginal value of any link that continues to deliver --- and a \emph{direct} effect --- disruptions within a relationship erode its own surplus. Which effect dominates depends on a single condition: whether the net cost of replacing the relationship exceeds the profit loss from the disruption. When exposure is limited to a subset of relationships, the \textit{indirect} effect dominates and firms preserve existing matches. When exposure is widespread, direct profit losses accumulate across the relationship portfolio and the condition reverses, leading to relationship termination and, in extreme cases, market exit. The framework thus provides a unified rationale for the opposing relationship-level and firm-level responses to the same shock.\\
 
\noindent \textbf{Contribution to the literature.}\
This paper contributes to the literature on the propagation of idiosyncratic shocks in production networks.\footnote{Recent work has examined shock propagation during the COVID-19 pandemic; see, among others \citet*{Anindya2025}, \citet*{Goldberg2023}, \citet*{Khanna2022}, and \citet*{Lebastard2023}.} Existing studies exploiting natural disasters and other exogenous shocks document substantial effects on firm outcomes, primarily along the intensive margin and over short horizons \citep*{Boehm2019, Barrot2016, carvalho2021, Todo2015, volpe2013, Kashiwagi2021, Gigout2021, Balboni2023}. Although disruptions may have persistent consequences, relatively little evidence speaks to how firms adjust their relationship portfolios over longer horizons, in part because such analysis requires detailed link-level data observed over time.\footnote{A small number of studies examine longer-run firm responses \citep[e.g.,][]{Gigout2021, Balboni2019, Todo2015}, but focus on transaction outcomes rather than relationship formation or termination.} Unlike \citet{Balboni2023}, who study reallocation across markets following firm-level shocks, I exploit variation at the level of individual buyer--seller relationships within a firm's relationship portfolio, allowing me to trace how disruptions to specific links transmit to the broader portfolio.
 
The paper also contributes to the literature on relational contracting in environments with limited contract enforceability.\footnote{See \citet*{Macchiavelo2022} and \citet*{Macchiavelo2023} for recent surveys.} While \citet*{Macciavello2015} studies the dynamics of contractual relationships in the Kenyan rose sector, I focus on a distinct margin: in \citet*{Macciavello2015}, firms face a choice between spot and relational contracts, whereas in my setting exporters access markets almost exclusively through established direct relationships and cannot switch to spot trade. This means the value of any given relationship is determined not only by its own performance but also by the composition of the entire relationship portfolio---a margin I identify using relationship-level variation in shock exposure.
 
Finally, this paper contributes to the literature on firm networks in international trade. A growing body of work documents that larger firms maintain more buyer--seller relationships but transact less intensively with each partner \citep*{BernardMoxnes2019, BernardDhyne2022, BernardDhin2018}. I complement this literature by showing that idiosyncratic shocks at the relationship level can propagate within firms, generating relationship portfolio-level adjustment and exit.\footnote{Related work examines how firms reorganize their networks following firm-level shocks \citep*{Korovkin2024, Huneus2020, Demir2024}. In contrast, I study temporary shocks that originate at the relationship level, allowing me to trace how disruptions to specific links transmit within firms.} My findings are also related to \citet*{Antras2014}, who emphasize that firms' sourcing decisions are interdependent across networks; I provide empirical evidence of such interdependence using relationship-level shocks.
 
To my knowledge, this is the first paper to identify the causal effect of supply disruptions at the level of individual buyer--seller relationships and to trace how such disruptions propagate within firms to affect relationship portfolio composition and exit. Existing work either uses firm-level shocks, which precludes relationship-level analysis, or focuses on short-run intensive-margin responses without tracking the longer-run reorganization of trading networks.
 
\bigskip
 
The remainder of the paper is organized as follows. Section~\ref{sec:context} describes the 2010--11 La Ni\~na event and the Colombian flower export sector. Section~\ref{sec:data_c1} presents the data, describes the construction of the main variables, and provides summary statistics. Section~\ref{sec:empirical} outlines the empirical strategy, while Section~\ref{sec:results} presents the main results. Section~\ref{sec:theory} presents the interpretive framework. The final section concludes.

\section{Context: 2010--11 La Ni\~na and Colombian flower exporters}
\label{sec:context}

\subsection{The 2010--11 La Ni\~na event}

Climate change is expected to intensify extreme weather events, causing significant infrastructure damage in developing countries with potential repercussions for global markets.\footnote{See \citet*{Geng2022} on infrastructure damages. For  evidence that flooding raises world prices, see \citet*{Fatica2022}, \citet*{Forslid2023}, \citet*{Chatzopoulos2020}, \citet*{BednarFriedl2022}, \citet*{Ghadge2019}, \citet*{Haile2017}.  \citet{Balboni2019} show that in Vietnam forward-looking infrastructure investments to avoid flood-prone regions could yield large welfare gains.} La Ni\~na, part of the El Ni\~no--Southern Oscillation (ENSO), is a major climate phenomenon, and the 2010--11 episode was among the strongest on record worldwide, associated with intense rainfall and flooding in Colombia.\footnote{This episode was also linked to extreme weather in other countries. In Australia, average temperatures reached their second- and third-highest levels since 1900, and the Western United States and Midwest recorded snowfall 180\% above average, with the exception of the Southern Rockies and Western Mountains\citep*{NCEI2011}.} 

\begin{figure}[t]
    \centering
   \includegraphics[width=\textwidth]{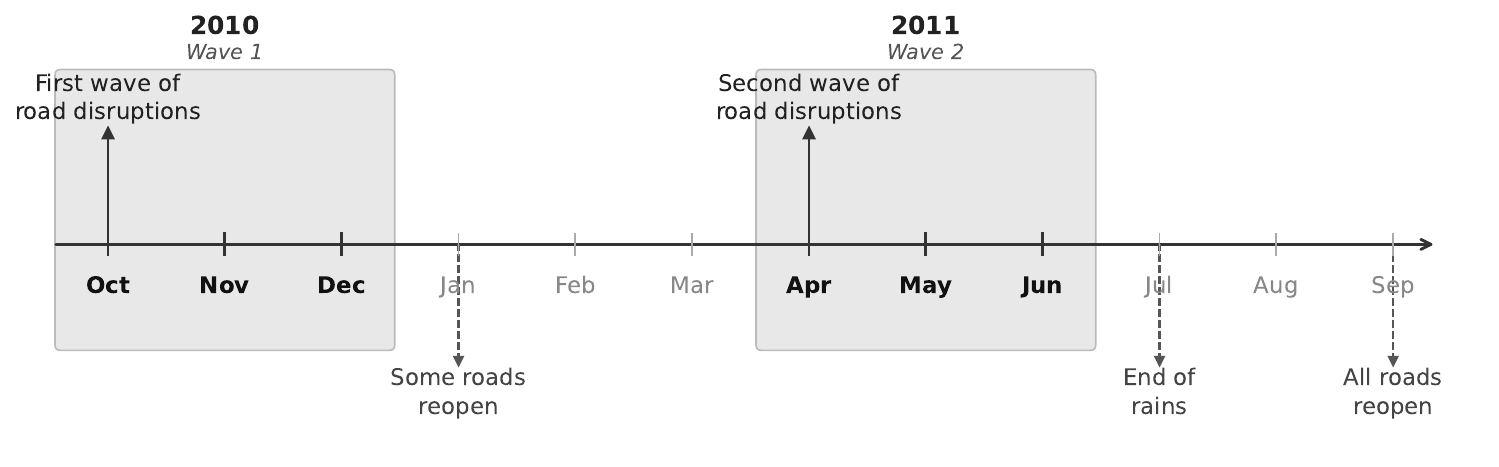}
     \vspace*{-0.35cm}
     
         \caption{Timeline of La Ni\~na 2010--11 in Colombia}
 \vspace*{0.1cm}   
     \scalebox{0.90}{
\begin{minipage}{1.1\textwidth}
\advance\leftskip 0.5cm
	{\footnotesize {\textit{\textbf{Notes:}}
    The figure shows the timeline of the 2010--11 La Ni\~na event in Colombia and the resulting road disruptions. The rains started in October 2010 and ended in June 2011. Flooding occurred in two waves: October to December 2010 and April to June 2011. During both waves, multiple road disruptions occurred across the country, lasting days, weeks, or even months. By September 2011, all roads disrupted by the flooding had reopened.
} \par }
\end{minipage}
} 
    \label{fig:timeline}
\end{figure}

Figure \ref{fig:timeline} shows the timeline of the La Ni\~na episode in Colombia during 2010--11 and the road disruptions it caused. Road disruptions occurred in two main waves coinciding with peak rainfall: (i) October--December 2010, (ii) April--June 2011. Some roads reopened within a few months of the initial wave, while some disruptions lasted until September 2011,  by which point all previously disrupted major roads had reopened. 

\begin{figure}[t]
    \centering
   \includegraphics[height=12cm]{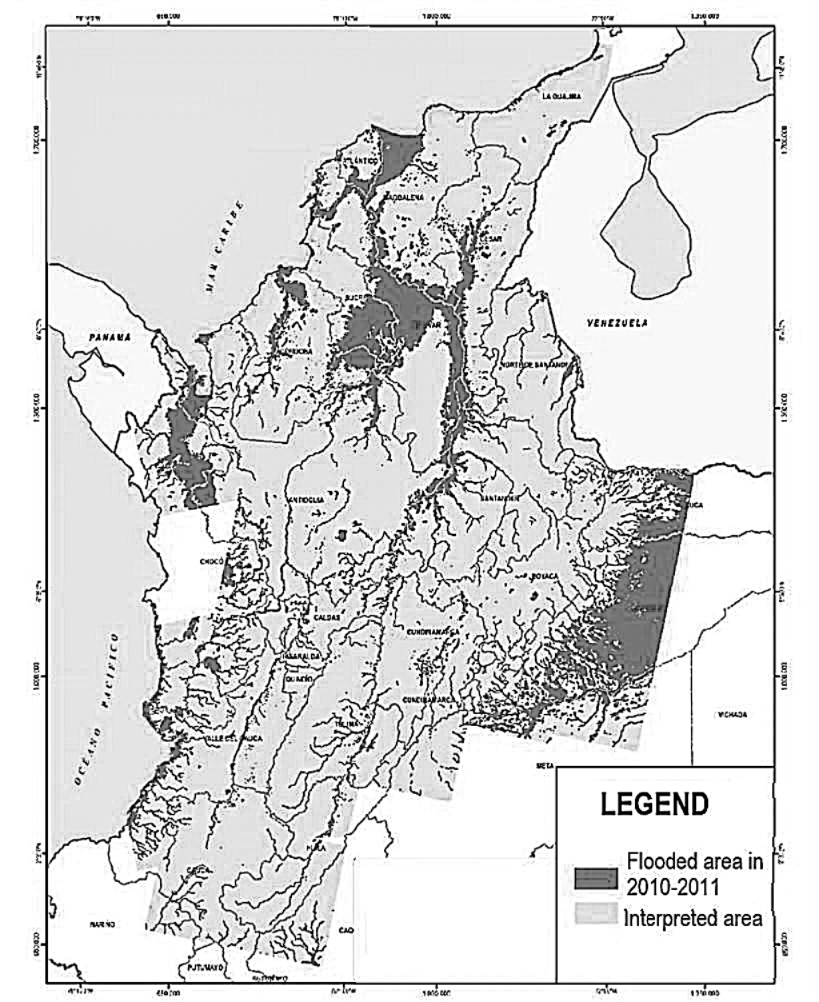}
    \vspace*{0.1cm}
\caption{Flooded area during La Ni\~na 2010--11}    
\vspace*{0.1cm}
   \scalebox{0.90}{
\begin{minipage}{1.1\textwidth}
\advance\leftskip 0.5cm
	{\footnotesize {\textit{\textbf{Source:}} Ideam 2010. \\
 \textit{\textbf{Notes:}} The map displays the accumulated flooded areas as of June 6, 2011, during the La Niña event. The satellite flooding data only includes information from the area highlighted in light grey, which excludes uninhabited regions like the Amazon. The dark grey shaded regions within this interpreted area represent the flooded zones during the event. 
 } \par }
\end{minipage}
} 
    \label{fig:flooding}
\end{figure}

Figure \ref{fig:flooding} illustrates the extent of the flooded area during the event. Approximately 3.5 million hectares were affected, representing roughly 3.3\% of Colombia's land area. Most flooding occurred in the northern and central regions, particularly along the Magdalena and Cauca rivers and the northeastern border with Venezuela. Damage to road infrastructure was estimated at \$6.5 million, modest relative to total reconstruction costs of \$1.5 billion.\footnote{Based on \citet{Cepal2012}; National Budget reports, December 2015. Truck volume estimates are from the 2011 Ministry of Transport report \citep{MinTransp2011}.} The affected road network accounted for roughly 36\% of total roads, including 9\% of primary roads. The flooding and road disruptions represented a major impediment to trade in Colombia: as of 2009, trucks carried approximately 73\% of freight by volume. According to transport sector reports, the sector suffered losses of roughly \$222 million; in Norte de Santander, transport costs increased by \$1 million \citep{Cepal2012}.

In response, the Colombian government implemented measures to improve climate resilience, including the ``Plan for Climate-Resilient Roads'' in 2012. The report indicate that without road upgrades, a 1°C increase by 2040 would leave 5.9\% of the road network unavailable on average, equivalent to 21 days of disruption per year due to higher precipitation.

\subsection{The Colombian flower sector}
\label{sec:flower_prod}

Colombia is the second-largest exporter of cut flowers worldwide, after the Netherlands. As of 2010, Colombia accounted for approximately 68\% of U.S.\ flower imports and held a 16.8\% share of the global export market. That year, the United States was the largest importer of flowers, with a market share of about 20\%.\footnote{Export data are from The Observatory of Economic Complexity (OEC).} 

Beyond its economic importance for both countries, the sector provides a particularly well-suited setting for studying buyer--seller relationships.  Trade in this sector relies on long-standing direct relationships between exporters and importers, many of which date back to the 1970s. Geographic proximity also facilitates repeated interaction and relationship formation between Colombian producers and U.S.\ buyers.

Most Colombian flower exporters are concentrated in two regions: the Savannah de Bogot\'a, near the capital, and the Antioquia region, near Medell\'{i}n. In 2009, roughly 70\% of export-oriented production was in the Savannah region, and 18\% in Antioquia. High-altitude areas are preferred, as flowers grow optimally at temperatures between 15°C and 25°C. Colombia produces approximately 1,600 flower varieties for export. The main species produced are roses (31\%), hydrangeas (15\%), carnations (13\%), and chrysanthemums (11\%).\footnote{Figure \ref{fig:production_flower} shows that Antioquia producers specialize in hydrangeas, which are often customized for events such as weddings and funerals, while Savannah producers predominantly grow more standardized varieties like roses and chrysanthemums.}

During the 2010--11 La Ni\~na episode, flooding at production sites was mitigated through protective measures including sandbags and drainage systems. Estimated production losses accumulated by June 2011 ranged from 5\% to 15\% nationally, primarily due to increased humidity affecting rose producers \citep{Cepal2012}. Flower producers in flood-prone areas received compensation from the Ministry of Agriculture and Rural Development (MADR).\footnote{According to the Colombian Ministry of Agriculture and Rural Development, 2010--11 aid and financial support totaled about \$1.6 million US dollars, of which 80\% was in credit-related programs. The industry's U.S.\ sales in 2010 were about 1 billion US dollars, making the subsidy small relative to the market.} 

\begin{figure}[ht]
\centering

    \includegraphics[height=2.8in]{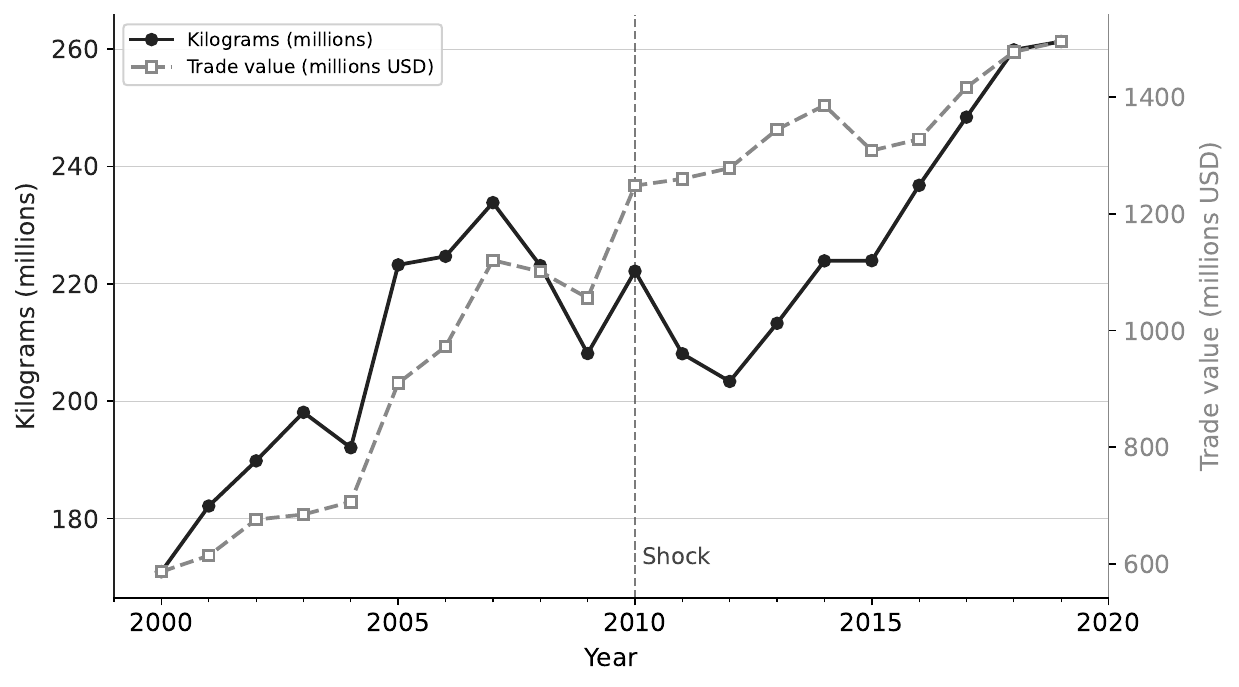}
            \vspace*{-0.2cm} 
\caption{Cut flower exports: quantity and trade value}
    \label{fig:export flowers}
        \vspace*{0.1cm}  
         \scalebox{0.90}{
\begin{minipage}{1.1\textwidth}
\advance\leftskip 0.5cm
	{\footnotesize {\textit{\textbf{Source:}} Customs data DIAN. \\
    \textit{\textbf{Notes:}} The figure plots total quantities of exported flowers in kilograms 
(millions) and total trade value in US dollars (millions), reported as FOB values to all destinations. The period covered is 2000--2019. 
Quantities are based on net export data (excluding packaging). Trade values are in nominal terms..} \par }
\end{minipage}
} 
 \end{figure}

Figure~\ref{fig:export flowers} shows total exported quantities (kilograms) and trade values (US dollars) for all export destinations. The figure reveals two notable disruptions to export volumes: the first, in 2009, coincides with the global financial crisis; the second, in 2011--12, with the  La~Ni\~na episode. Export volumes declined by approximately 8.5\% between 2010 and 2012, while nominal traded value rose by 12\% over the same period,  in line with the long-term trend. Rising prices alongside falling volumes are the classic signature of a supply-side shock.

Figure~\ref{fig:production} in the Appendix shows national production disaggregated by end use: exports account for almost all production, with only 5\% consumed domestically. Drawing on an independent data source (Colombian Association of Flower Exporters), the figure confirms the two-episode pattern visible in the customs data: declines in production in 2009 and 2011.

\section{Data and variables} 
\label{sec:data_c1}

\subsection{Data}
\label{sec:data_sources}
 
\noindent \textbf{Firm-to-firm relationships.} Using Colombian 
customs data over 2007--2019, I track all monthly transactions between Colombian flower exporters (identified by tax IDs) and U.S.\ importers (identified by firm names, since U.S.\ importers do not have Colombian tax IDs). The data include trade values, quantities, and customs offices used for each shipment. I construct an importer identifier through name standardization and generate a monthly buyer--seller panel. This panel tracks relationships from January 2007 through their last observed transaction, with December 2019 as the right-censoring date.

The dataset contains 23,347 buyer--seller relationships between U.S.\ importers and Colombian exporters, classified under HS 2-digit code 06 (live trees, plants, bulbs, roots, cut flowers, and ornamental foliage). Ninety-three percent of monthly transactions correspond to cut flowers (HS~0603), with the remainder involving plants and roots. The full dataset includes 1,113 exporters and 1,833 importers. The pre-shock analytical sample is smaller, as described in Section~\ref{sec:summary-stats}.\\

\noindent \textbf{Road disruptions and routes to cargo terminals.} 
Data on road disruptions come from multiple governmental sources. The Unidad Nacional para la Gestión del Riesgo de Desastres (UNGRD) produced a report in May 2011 documenting road disruptions, their locations, durations, and type (e.g., scheduled repair, landslide, or accident). To fill gaps in reopening dates, I reviewed Ministry of Transport reports, official news releases, and media coverage.

I focus on 37 road disruptions that affected exporters' routes to cargo terminals, shown in Figure~\ref{fig:road_disrupted}. Disruptions lasting less than a week or allowing one-way traffic are excluded, since flowers have a limited shelf life and must reach international markets promptly.

Transportation routes are constructed using the Ministry of Transport's SICE-TAC  (Sistema de Información de Costos Eficientes para el Transporte Automotor de Carga) system and customs data on cargo terminal usage. SICE-TAC provides routes from municipal capitals, aggregated across all transportation providers. In addition to the official SICE-TAC route, I compute the shortest route to the terminal using Google Maps and identify alternative routes when they exist. Alternative routes are retained only if they fall within roughly within similar time and distance of the SICE-TAC route, implying comparable travel times. For each buyer--seller--route combination (comprising one or more segments from farm to terminal), I identify disrupted roads separately for each wave of flooding: Wave~1 (October 2010--December 2011) and Wave~2 
(April--June 2011).\\

\noindent \textbf{Location, firm activity, and flooded areas.} Using IDEAM reports on La~Ni\~na-related flooding, I identify 
producers in flooded municipalities. Flooding relevant to flower production was concentrated in the Bogot\'a--Savannah region. The Antioquia region had more flooded municipalities, but the 
affected areas did not coincide with flower-producing locations. In the Savannah region, municipalities are classified as flooded 
if at least 0.7\% of their area was inundated. This cutoff retains 11 of the 16 flooded municipalities in the region, excluding 5 with inundated shares between 0.4\% and 0.6\%. In Antioquia, where flooding shares are higher (between 1.5\% and 9.5\%), the threshold does 
not affect the main flower-producing municipalities. In total, four municipalities with flower producers were flooded, affecting 
46 producers.\footnote{Statistics on municipal flooded areas are from joint reports by IGAC, IDEAM, and DANE (February 2011).}

Firm activity is collected from the Chamber of Commerce, covering 94\% of exporters selling to the U.S.\footnote{ Activity data are not collected for firms that first 
entered in 2018 or 2019, as these fall outside the estimation 
sample; they represent only 67 exporters out of 1,113.} Among these exporters, 20\% are intermediaries (non-producers), 68\% are producers, and 12\% have no registered activity. Producer farm locations are obtained from the Colombian Agricultural Institute (ICA) database. Missing information is supplemented using firm websites, Google Maps, and online directories. In total, farm municipalities are identified for 691 of the 762 producer-exporters, including 72 with multiple farm locations. Farm locations remain unknown for the remaining 71 producers. These firms are dropped from analyses that require location to construct exposure to road disruptions, but are retained otherwise.

\subsection{Construction of variables}
\label{sec:main_variables}

\noindent \textbf{Pre-shock period.}\
The pre-shock period for the main regressions consists of the 
18 months preceding the first road disruption in October 2010.\\

\noindent \textbf{Relationship status.}\ I construct a balanced 
panel of buyer--seller transactions. For each relationship, months 
without transactions are coded as active if the relationship 
transacts again at any future point, and as ended once no further 
transactions are observed. Formally, the monthly \textit{relationship 
status} indicator is defined as:
\begin{equation}
y_{ij,m} =
\begin{cases}
1 & \text{if } \sum_{s=m}^{M} T_{ij,s} > 0, \\
0 & \text{otherwise,}
\end{cases} \nonumber
\end{equation}
where $T_{ij,s}$ equals one if exporter $i$ and importer $j$ transact 
in month $s$ and zero otherwise, and $M$ is the last month of the 
data. $y_{ij,m}=1$ indicates that the relationship is active in 
month $m$, i.e., has not permanently ended. By construction, gaps 
between transactions within an active relationship are coded as 
active, not as endings.

To reduce noise at the monthly frequency, I aggregate $y_{ij,m}$ 
to six-month periods $t$. A relationship is coded as ongoing in 
period $t$ only if it is active in every month of the period:
\begin{equation}
y_{ij,t} =
\begin{cases}
1 & \text{if } \sum_{s=m}^{m+l-1} y_{ij,s} = l, \\
0 & \text{otherwise,}
\end{cases} \nonumber
\end{equation}
where $m$ is the first month of period $t$ and $l=6$ is the period 
length. To avoid classifying late-sample spells as endings due to 
right-censoring, I truncate the observation window at December 2017: 
relationships still active after this date are coded as ongoing 
rather than ended.\footnote{Long within-spell gaps are rare: 92\% of relationships 
transact at least once every year they are active. Across all relationship-year observations, only 3.5\% involve a gap of more 
than one year, and 2\% a gap of more than two years. This pattern justifies the December 2017 truncation: a two-year window is sufficient to identify the vast majority of relationship endings before the sample ends in December 2019.}\\

\noindent \textbf{Relationship exposure to the shock}.\
Customs data record actual shipments but do not indicate the terms 
of any underlying contracts or the reasons for missed shipments. 
To address this, I use pre-shock cargo terminal usage to infer 
which relationships could have been affected. For each relationship, 
I identify all routes connecting the exporter's farm to each cargo 
terminal used in the pre-shock period, including alternative routes. An origin--terminal pair's exposure is constructed by aggregating 
across three levels. First, a road segment is considered disrupted 
when blocked by flooding. Second, a route (made up of road segments) 
is considered affected when any of its segments is disrupted. 
Finally, an origin--terminal pair (the municipality of the farm 
paired with a cargo terminal, which may be connected by one or 
multiple routes) is considered exposed when all routes connecting 
the farm to the terminal are affected. The underlying assumption 
is that as long as at least one route remains open, shipments can 
be rerouted without significant delay. Exposure is assessed 
separately for each wave of flooding, since road segments may open 
or close over the course of the crisis.

I drop origin--terminal combinations accounting for less than 10\% 
of a relationship's pre-shock trade value. The remaining 
origin--terminal pairs are used to construct the binary 
relationship-level exposure measure: a relationship is 
\textit{exposed to the shock} if at least one of its origin--terminal 
pairs is \textit{exposed}. For exporters with multiple farms, I 
restrict attention to those whose farms are located in neighboring 
municipalities, so that the farm location used for classification 
does not affect the exposure assignment.\footnote{Fifteen of the 
72 exporters with multiple farms do not meet this criterion and 
are excluded.} Relationships involving intermediaries (non-producer 
exporters) or relationships terminating before the pre-shock period 
are excluded from the relationship-level exposure analysis. They 
are nevertheless retained when constructing importers' relationship 
portfolios, since they remain part of the firm's sourcing strategy.\\

\noindent \textbf{Importer exposure}.\
I aggregate the relationship-level exposure to construct an 
importer-level measure. Let $\text{IE}_j$ denote the share of 
importer $j$'s relationships that are \textit{exposed to the shock}:
\begin{equation}
 \text{IE}_j = \frac{\sum_{i}\mathbb{1}\{E_{ij,0}=1\}}
 {\sum_{i}\mathbb{1}\{E_{ij,0}=1\} + \sum_{i}\mathbb{1}\{E_{ij,0}=0\}}, \nonumber
\end{equation}
where the subscript $0$ denotes the pre-shock period and 
$E_{ij,0}=1$ indicates that the relationship is \textit{exposed 
to the shock}. The denominator sums over importer $j$'s 
classifiable relationships in the pre-shock period; relationships 
for which exposure cannot be determined ($E_{ij,0}=\text{NA}$) 
are excluded. The rationale is conceptual: $\text{IE}_j$ is meant 
to capture the share of the importer's portfolio whose disruption 
status is observable, since treating unclassified relationships as 
non-exposed would mechanically attribute resilience to firms whose 
exposure is simply unmeasured. In practice, NA relationships 
account for a modest share of trade value (mean 24\%, median 15\%, 
SD 0.22), so the denominator choice has limited effect on the 
economic interpretation of the measure.\footnote{As a robustness 
check, I re-estimate the main specifications with NAs included in 
the denominator (mean exposure 0.18 versus 0.36). 
Results are reported in the Supplemental Appendix.}

\subsection{Summary statistics}
\label{sec:summary-stats}

The estimation sample includes all relationships that trade at 
least twice over the full sample period and do not enter after 
September 2010, comprising 525 exporters and 674 importers. 
Within this sample, the exposure sample---a subset of relationships 
for which a relationship-level \textit{exposure to the shock} 
measure can be constructed---comprises 408 exporters and 624 
importers, accounting for 69\% of relationships and 81\% of 
pre-shock trade value. The exposure sample is used in the 
relationship-level regressions and in the construction of $IE_j$ 
(which excludes NA relationships, as discussed in 
Section~\ref{sec:main_variables}). The full estimation sample is 
used in the matching analysis and in the portfolio-adjustment 
regressions, which also allow relationships first observed after 
September 2010 to capture new relationship formation.

Table~\ref{tab:stat} reports summary statistics for the estimation 
sample, averaged across three consecutive six-month periods from 
April 2009 to September 2010.\footnote{Six-month periods are 
defined backward from the shock: $t=-1$ covers April--September 
2010, $t=-2$ covers October 2009--March 2010, and $t=-3$ covers 
April--September 2009.} Panel A reports network-level statistics. 
On average, 16.7\% of active relationships end in each pre-shock 
period. The share of new relationships is low because the 
denominator includes all potential buyer--seller pairs; in absolute 
terms, approximately 509 new matches form in each period. The 
share of active relationships---also known as network density---
reflects the sparsity of the data: fewer than 1.6\% of all 
potential buyer--seller pairs are active in each period. Together, 
exposed and non-exposed relationships account for approximately 
84\% of all active relationships.

Panel B reports relationship-level statistics, conditional on the 
relationship being active. For each relationship, I compute each 
statistic within a given six-month period and then average across 
the three periods. Transaction frequency is the average number of 
transactions per six-month period; the remaining statistics are 
monthly averages over the period. To ensure comparability across 
quantity measures, kilograms and units are computed for cut 
flowers only (HS 060311, 060312, 060314, 060319), since foliage 
and branches are reported in kilograms but not in comparable unit 
counts. Value per transaction, quantities, unit value, and value 
per kilogram all display substantial heterogeneity, reflecting the 
coexistence of small and large trading relationships alongside a 
wide variety of flower products and shipment units (stems and 
boxes).

Panel C reports firm-level statistics, computed per firm per 
six-month period and averaged across the three periods. The median 
number of connections is 3 for exporters and 2 for importers. I 
split importers into high in-degree and low in-degree groups at 
the median in-degree, where high in-degree importers are those 
above the median.\footnote{The exporter out-degree distribution 
is shown in the Supplemental Appendix since the heterogeneity analysis in the paper is conducted 
at the importer level.} The mean importer-level exposure share is 
0.36, indicating that on average importers have 36\% of their 
classifiable relationships \textit{exposed to the shock}. By 
construction, $IE_j$ is time-invariant, as it is computed using 
all active relationships in the pre-shock period.

\begin{table}[htbp]
\centering
\caption{Summary statistics, pre-shock period (April 2009--September 2010)}
\label{tab:stat}
\vspace*{0.2in}
\scalebox{1}{
\begin{tabular}{lcccc}
\toprule
\toprule
Statistic & Mean & Median & Min & Max  \\
\midrule
\addlinespace[3pt]
\textit{Panel A. Network-level statistics} & \\
\addlinespace[5pt]
Share of relationships ending            & 0.167  & 0.164   & - & - \\
\addlinespace[3pt]
Share of new relationships               & 0.0026 & 0.0027  & - & - \\
\addlinespace[3pt]
Share of active relationships            & 0.0166 & 0.0164  & - & -   \\
\addlinespace[3pt]
Total exposed relationships              & 1,555  & 1,506   & - & -   \\
\addlinespace[3pt]
Total non-exposed relationships          & 1,546  & 1,506   & - & -  \\
\addlinespace[3pt]
Total active relationships               & 3,676  & 3,596   & - & -   \\
\addlinespace[5pt]
\textit{Panel B. Relationship-level statistics} & \\
\addlinespace[5pt]
Transaction frequency                                & 0.39 & 0.33 & 0     & 1     \\
\addlinespace[3pt]
Monthly trade value (US\$ thousands)                 & 32.5 & 5.12 & 0.14  & 7,902 \\
\addlinespace[3pt]
Monthly quantity, kilograms (thousands)              & 11   & 0.89 & 0.006 & 31,700 \\
\addlinespace[3pt]
Monthly quantity, units (thousands)                  & 87   & 13.8 & 0.038 & 16,000 \\
\addlinespace[3pt]
Value per kilogram (US\$)                            & 6.9  & 6.07 & 0.78  & 34    \\
\addlinespace[3pt]
Unit value (US\$)                                    & 0.77 & 0.53 & 0.002 & 8.33  \\
\addlinespace[5pt]
\textit{Panel C. Firm-level statistics} & \\
\addlinespace[5pt]
Seller connections (out-degree)                      & 8.1  & 3    & 1 & 97 \\
\addlinespace[3pt]
Buyer connections (in-degree)                        & 6.4  & 2    & 1 & 86 \\
\addlinespace[3pt]
Importer exposure ($IE_j$)                           & 0.36 & 0.38 & 0 & 1  \\
\addlinespace[3pt]
\bottomrule
\end{tabular}
}
\vspace*{0.1in}

\scalebox{0.90}{
\begin{minipage}{1.1\textwidth}
\advance\leftskip 0.5cm
{\footnotesize{\textit{\textbf{Notes:}} All statistics are computed 
for the pre-shock period (April 2009--September 2010). \textit{Panel A} 
reports network-level statistics computed for each six-month period 
and averaged across the three periods; Mean and Median are reported 
across the three periods. The share of relationships ending is the 
fraction of active relationships that became inactive in a given 
period; the share of new relationships is the fraction of previously 
inactive buyer--seller pairs that became active, computed over all 
potential pairs; the share of active relationships (network density) 
is computed over all potential importer--exporter pairs. Exposed and 
non-exposed relationship counts refer to relationships with a 
constructed \textit{exposure to the shock} measure equal to one and 
zero respectively; the remaining active relationships have no 
exposure measure. \textit{Panel B} reports relationship-level 
statistics conditional on the relationship being active, computed 
per relationship per six-month period and averaged across the three 
periods. Transaction frequency is the average number of transactions 
per six-month period. Monthly trade value, monthly quantities, value 
per kilogram, and unit value are computed as monthly averages within 
each six-month period and then averaged across periods. Quantities 
(kilograms and units) are computed for cut flowers only 
(HS 060311, 060312, 060314, 060319) to ensure comparability, since 
foliage and branches are reported in kilograms but not in comparable 
unit counts. Units refer to quantities reported in customs records 
(boxes or stems). \textit{Panel C} reports firm-level statistics 
computed per firm per six-month period and averaged across the three 
periods. Seller and buyer connections are the average number of 
active relationships per exporter and importer respectively. 
Importer exposure ($IE_j$) is defined in 
Section~\ref{sec:main_variables}; by construction it is time-invariant, 
as it uses all active relationships in the pre-shock period. The 
estimation sample comprises 525 exporters and 674 importers; the 
exposure sample (used to construct $IE_j$ and the relationship-level 
exposure indicator) comprises 408 exporters and 624 importers.}\par}
\end{minipage}
}
\end{table}

\begin{figure}[ht]
    \centering
    \includegraphics[height=2.8in]{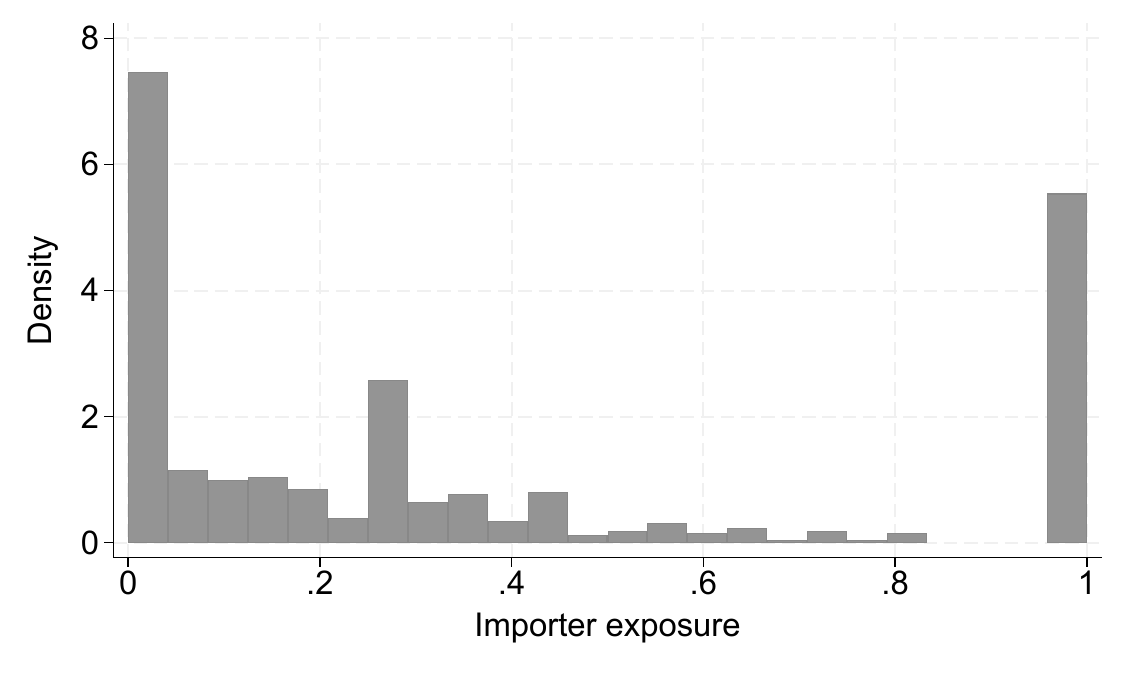}
    \vspace*{-0.2cm} 
    \caption{Distribution of importer exposure ($IE_j$)}
    \vspace*{0.1cm}
    \scalebox{0.90}{
    \begin{minipage}{1.1\textwidth}
    \advance\leftskip 0.5cm
    {\footnotesize{\textit{\textbf{Notes:}} The figure shows the 
    distribution of importer-level \textit{exposure to the shock} 
    ($IE_j$) in the pre-shock period (April 2009--September 2010). 
    Importer exposure is the share of an importer's classifiable 
    relationships that are exposed to road disruptions, as defined 
    in Section~\ref{sec:main_variables}. The sample consists of the 
    624 importers in the exposure sample (importers with at least 
    one classifiable relationship in the pre-shock period). 
    Importers with no classifiable relationships are excluded.}\par}
    \end{minipage}
    }
    
    \label{fig:distr_ie_all}
\end{figure}

Figure~\ref{fig:distr_ie_all} shows the distribution of importer 
exposure ($IE_j$), which is near-bimodal, with mass concentrated 
at zero and near one. Figure~\ref{fig:distr_ie_indegree} shows 
the distribution separately for high and low in-degree importers, 
where the split is based on the average number of active 
relationships in the twelve months preceding the pre-shock period 
(April 2008--March 2009; see Section~\ref{sec:empirical}). Low 
in-degree importers tend to source either entirely from exposed 
or entirely from non-exposed suppliers, while high in-degree 
importers are more likely to hold mixed relationship portfolios. 
This pattern is partly mechanical --- low in-degree importers have 
only a handful of relationships, leaving little room for mixed 
exposure --- but it is also descriptively relevant: the heterogeneity 
in portfolio composition shapes the firm-level responses documented 
in Section~\ref{sec:results}.

\section{Empirical strategy}
\label{sec:empirical}

\subsection{Identification and setup}
\label{sec:exogenous-variation}

I exploit relationship-level exposure to road disruptions between 
October 2010 and June 2011. These disruptions resulted from 
unprecedented rainfall during the 2010--11 La Ni\~na episode 
(see Section~\ref{sec:context}), which triggered landslides along 
multiple stretches of the road network and generated plausibly 
exogenous variation in which routes were affected. Because the 
episode was a tail event well beyond historical expectations, it 
is implausible that flower producers' farm-location decisions had 
internalized the risk of these specific disruptions. The identifying 
assumption is that, conditional on the fixed effects included in 
each specification, the timing and spatial incidence of road 
disruptions are uncorrelated with pre-existing trends in relationship 
survival. The event-study design allows this assumption to be 
assessed empirically through the pre-trend test reported in 
Section~\ref{sec:results}. The relationship-level \textit{exposure 
to the shock} measure compares exposed relationships to non-exposed 
ones within the same importer's portfolio. The importer-level 
measure ($IE_j$) captures the share of a firm's portfolio that is 
exposed, enabling the analysis of firm-level responses.

Three potential threats to this identification strategy deserve 
consideration. First, if importers had anticipated the shock and 
pre-adjusted their relationship portfolios, the pre-period 
event-study coefficients would not be flat --- a pattern not observed 
in the data. Second, exposed exporters could differ from non-exposed 
ones in ways that independently affect relationship survival; 
relationship fixed effects absorb all time-invariant 
relationship-specific heterogeneity, and pre-trend tests at both 
the relationship and firm levels further support comparability. 
Differences across importers themselves are addressed by the 
in-degree split, which allows results to be interpreted within 
more homogeneous subgroups rather than pooled across heterogeneous 
firms. Third, general equilibrium effects could contaminate the 
control group if non-exposed exporters gained market share during 
the crisis. This is unlikely in the short run, as flower production 
requires extended growing cycles and producers cannot quickly scale 
up output to replace disrupted competitors. Moreover, production-site 
flooding and route disruptions affected different regions: flooding 
of production sites was concentrated in the Savannah region (see 
Section~\ref{sec:flower_prod}), while road network disruptions 
primarily affected route access for Antioquia producers, who 
specialize in hydrangeas. Less-affected Savannah producers 
predominantly grow more standardized varieties such as roses. These 
products are imperfect substitutes, limiting the scope for 
within-sample reallocation.\footnote{A fourth concern is measurement 
error in route classification, which may arise from imperfect 
mapping of routes via Google Maps or from the choice of distance 
threshold for alternative routes. The 
estimated coefficients can be interpreted as lower bounds (in absolute value) on the true effects.}

\begin{figure}[t]
    \centering
    \includegraphics[width = 0.95\textwidth]{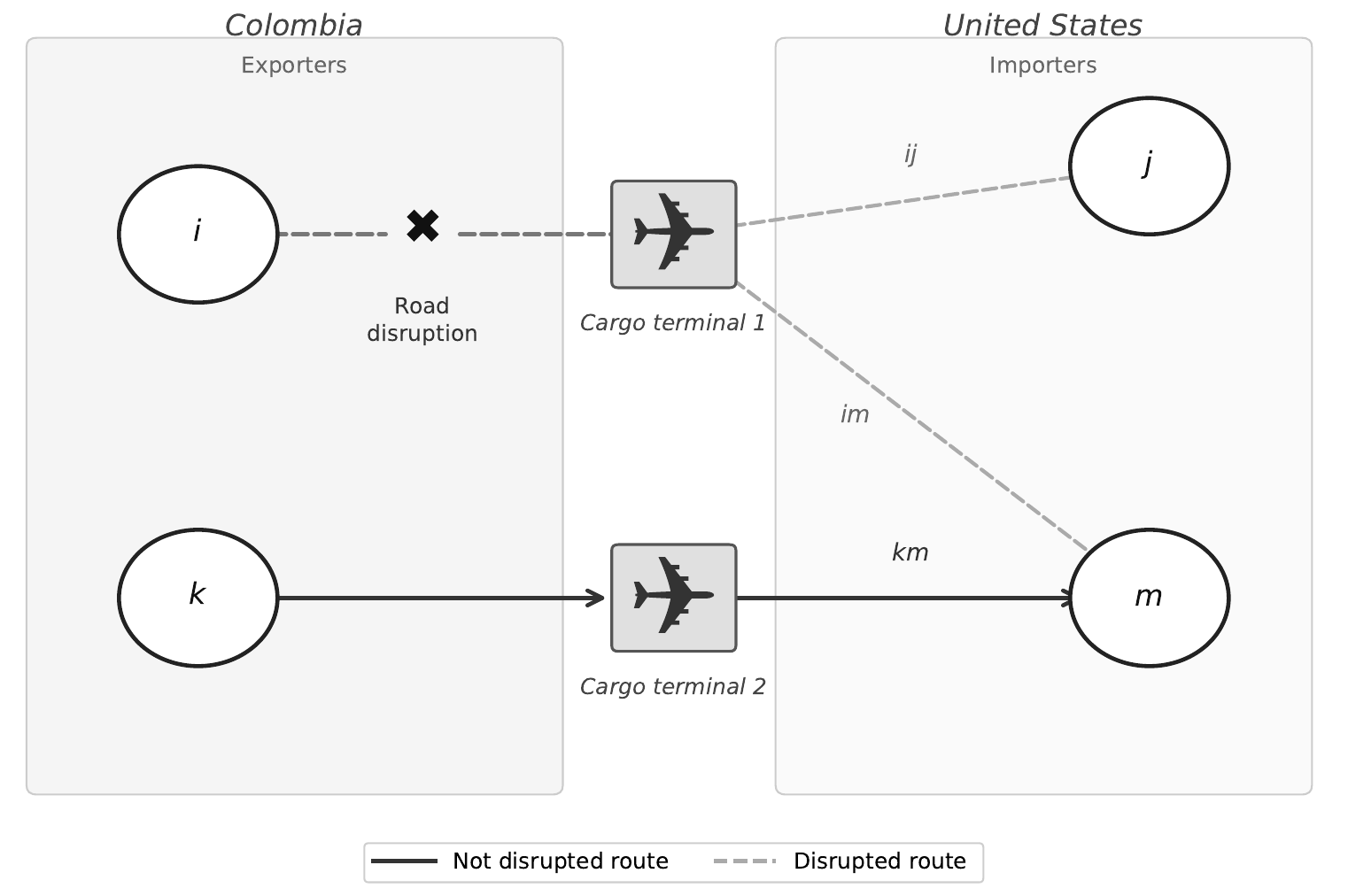}
    \vspace*{-0.2cm} 
    \caption{Illustration of relationship-level and importer-level exposure to road disruptions}
    \label{fig:setting}
    \vspace*{0.1cm}
    \scalebox{0.90}{
    \begin{minipage}{1.1\textwidth}
        \advance\leftskip 0.5cm
        {\footnotesize {\textit{\textbf{Notes:}} The figure 
        illustrates the variation exploited to identify the effect 
        of road disruptions. For simplicity, each exporter is shown 
        with a single route to the cargo terminal; in the data, a 
        relationship is exposed only if all routes connecting an 
        exporter to a given cargo terminal are affected. Dashed 
        lines denote disrupted routes; solid lines denote unaffected 
        routes. Relationships $ij$ and $im$ are exposed because the 
        route from exporter $i$ to the cargo terminal is disrupted, 
        while relationship $km$ is unaffected. Importer-level 
        exposure ($IE_j$) is the share of an importer's relationships 
        that are exposed: importer $m$ has 50\% exposure (one of two 
        relationships exposed) and importer $j$ has 100\% exposure 
        (one of one).} \par }
    \end{minipage}
    }
\end{figure}

Figure~\ref{fig:setting} illustrates the setup. For simplicity, 
the figure shows a single route from each exporter to the cargo 
terminal; in the data, a relationship is considered exposed when 
all routes connecting the exporter to at least one of its cargo 
terminals are affected. Relationships $ij$ and $im$ are exposed 
because exporter $i$'s route to the cargo terminal is disrupted 
(shown by the dashed line), while relationship $km$ is unaffected. 
Within-importer comparisons exploit variation across relationships 
connected to the same importer: relationships $im$ and $km$ both 
connect to importer $m$, allowing a tighter comparison that holds 
the importer fixed. Importer-level exposure reflects the share of 
an importer's relationships that are \textit{exposed to the shock}: 
importer $m$, with two relationships of which one is exposed, has 
50\% firm-level exposure, while importer $j$, whose single 
relationship is exposed, has 100\% firm-level exposure.

\subsection{Estimating equations}
\label{sec:main-equation}

\noindent \textbf{Relationship ending.} I estimate the effect of 
road disruptions on the probability of a relationship ending using 
an event-study framework. To simplify notation, I define a 
relationship-ending indicator on the subsample of relationships 
active in the previous period:
\begin{equation}
e_{ij,t} \equiv \mathbb{1}\{y_{ij,t}=0\} \quad \text{for } 
(i,j,t) \text{ with } y_{ij,t-1}=1, \nonumber
\end{equation}
which equals one if the relationship ends in period $t$ and zero 
if it remains active. The conditioning ensures that $\beta_l$ is 
interpreted as the change in the ending hazard rather than as the 
change in the stock of inactive relationships. The estimating 
equation is:
\begin{equation}
e_{ij,t} = \sum_{l \neq -1} \beta_l \, E_{ij,0} \cdot 
\mathbb{1}\{t=l\} + \theta_{ij} + \zeta_{c,t} + \varepsilon_{ij,t},
\label{Y1}
\end{equation}
where $E_{ij,0}$ is the relationship-level \textit{exposure to the 
shock}, equal to one if at least one origin--terminal pair used by 
the relationship had all its routes affected by the disruptions, 
and zero otherwise. 
The coefficients of interest, $\beta_l$, measure the effect of 
exposure at each lead and lag relative to the shock, with $t=0$ 
corresponding to the six-month period containing the first wave 
of disruptions. Because periods are defined at the six-month 
frequency, each event period fully encompasses a three-month wave 
of disruptions; conveniently, the two waves are themselves spaced 
six months apart, so each maps cleanly to one event period. Period 
$t=-1$ is omitted as the reference; periods $t<-1$ are pre-shock 
periods used to assess parallel trends, and periods $t \geq 0$ 
capture the dynamic post-shock response.

Fixed effects control for potential confounders: $\theta_{ij}$ are 
relationship fixed effects, which subsume importer and exporter 
fixed effects and absorb all time-invariant pair-specific 
heterogeneity. Identification therefore comes from within-relationship 
variation in survival outcomes over time. $\zeta_{c,t}$ denotes 
relationship-age-by-time fixed effects, where $c$ indexes 
relationship age --- defined as the number of months between the 
first observed transaction and the shock in October 2010 --- grouped 
into eight bins.\footnote{Even 
with relationship fixed effects, relationships from different 
cohorts may follow different baseline survival paths. The 
age-by-time fixed effects flexibly absorb these differential 
trends across cohorts.  Results are not driven by the construction of the cohort groups, using different groupings of cohorts yield similar results.The age groups used in the main specifications are shown in the Supplemental Appendix.} The $\beta_l$ coefficients are interpreted 
as the percentage-point change in the probability of a relationship 
ending for exposed relationships relative to non-exposed ones.

I report two specifications. The baseline uses equation~\eqref{Y1} 
and identifies $\beta_l$ from variation across all exposed and 
non-exposed relationships in the sample. The second specification 
adds importer-by-time fixed effects $\alpha_{j,t}$, which absorb 
all time-varying importer-specific shocks. Identification then comes 
from comparing exposed and non-exposed relationships within the 
same importer's portfolio, providing a within-importer test of how 
shocks propagate across relationships.\\

\noindent \textbf{Relationship extensive and intensive margins.} 
Much of the existing literature focuses on short-term adjustment 
--- changes in prices and quantities, total sales (intensive margin), 
or transaction patterns (extensive margin) --- typically over a few 
months after a shock \citep{Anindya2025, Khanna2022, Balboni2023}. 
These short-term responses can act as temporary buffers, giving 
firms time to adjust their relationship portfolios. The granularity 
of the data allows me to study both extensive- and intensive-margin 
responses during and after the shock.

First, I consider the extensive margin: whether relationships 
\textit{exposed to the shock} are more or less likely to record 
shipments. A month without trade does not necessarily imply that 
the relationship has ended; it may instead reflect a temporary 
disruption. Using monthly transaction data, I estimate:
\begin{equation}
T_{ij,m} = \sum_{l \neq -1} \beta_l \, E_{ij,0} \cdot 
\mathbb{1}\{m=l\} + \theta_{ij} + \zeta_{c,m} + \varepsilon_{ij,m},
\label{Y3}
\end{equation}
where $T_{ij,m}$ equals one if a transaction between exporter $i$ 
and importer $j$ is observed in month $m$, and zero otherwise. 
September 2010 is omitted as the reference month. As before, I 
control for relationship fixed effects $\theta_{ij}$ and 
cohort-by-month fixed effects $\zeta_{c,m}$. In a tighter 
specification, I add importer-by-month fixed effects $\alpha_{j,m}$ 
to isolate within-importer-month variation.\footnote{Results are 
robust across both specifications and are reported in the Supplemental Appendix.}

Second, I exploit 6-digit HS data to examine intensive-margin 
responses. For each outcome, I estimate:
\begin{equation}
X_{ijp,m} = \sum_{l \neq -1} \beta_l \, E_{ij,0} \cdot 
\mathbb{1}\{m=l\} + \boldsymbol{\Gamma} + \zeta_{c,m} + 
\varepsilon_{ijp,m},
\label{Y4}
\end{equation}
where $X_{ijp,m}$ is measured at the relationship--product--month 
level, $p$ indexes 6-digit HS products, and $\boldsymbol{\Gamma}$ 
denotes a set of fixed effects whose composition varies by outcome. 
September 2010 is again omitted as the reference month. The choice 
of fixed-effects structure reflects the source of residual 
variation each outcome exhibits within a 6-digit HS code.

Trade value and total kilograms measure the \textit{scale} of a 
pair's trade in a given product. Within an HS code, the level of 
these outcomes varies systematically across pair--product 
combinations: a given buyer--seller pair may consistently ship 
one sub-variety while another ships a different one under the 
same classification, with persistent differences in shipment size. 
Absorbing only relationship and product--month fixed effects 
leaves residual variation correlated with each pair's product 
specialization, generating apparent pre-trends. I therefore 
estimate scale outcomes with relationship--product fixed effects, 
$\boldsymbol{\Gamma} = \delta_{ijp}$, which restrict identification 
to within-relationship--product variation over time.\footnote{Appendix 
Figure~\ref{fig:rob_kilogram} reports scale outcomes under the 
less restrictive structures $\theta_{ij} + \delta_{p,m}$ and 
$\delta_{p,m}$ alone (used for composition outcomes below). 
Pre-trends emerge in both specifications, while post-shock 
coefficients remain statistically indistinguishable from zero, 
consistent with the within-pair-across-product heterogeneity 
described above.}

Price per kilogram, unit value, units and kilograms per unit are 
rate-based outcomes sensitive to \textit{composition}: shifts in 
variety, quality, or shipment size within an HS category. Because 
these outcomes divide out the scale of trade, they are more 
comparable across pairs within an HS code, and the dominant 
non-relationship source of variation is market-wide --- input-cost 
changes, seasonal demand, or variety substitution at the industry 
level. For these outcomes, $\boldsymbol{\Gamma} = \theta_{ij} + 
\delta_{p,m}$, where relationship fixed effects absorb 
time-invariant pair-specific levels and product--month fixed 
effects absorb common dynamics within an HS code. Kilograms per 
unit captures shifts in shipment composition (e.g., heavier vs. 
lighter boxes or stems).

In a tighter specification, I additionally include importer-by-month 
fixed effects $\alpha_{j,m}$, which restrict identification to 
within-importer-month variation across exporters serving the same 
importer. To preserve sample comparability across specifications, 
I restrict the estimation sample to relationship--product cells 
with at least two monthly observations.

Together, equations~\eqref{Y3} and~\eqref{Y4} track whether firms 
temporarily stop shipments or adjust shipment volumes and 
composition in response to road disruptions.\\
 
\noindent \textbf{Importer exposure.} The shock is defined at the 
relationship level. However, it remains unclear whether the effect 
operates at the relationship level or whether an importer's exposure 
across its relationship portfolio shapes its overall response. To 
examine this channel, I estimate the following at the six-month 
frequency ($t$):
\begin{equation}
e_{ij,t} = \sum_{l \neq -1} \beta_l \, IE_{j,0} \cdot 
\mathbb{1}\{t=l\} + \theta_{ij} + \zeta_{c,t} + \varepsilon_{ij,t},
\label{Y2}
\end{equation}
where $e_{ij,t}$ is the relationship-ending indicator defined in 
equation~\eqref{Y1} (conditional on the relationship being active 
in $t-1$, so that the estimand is a hazard) and $IE_{j,0}$ is the 
importer exposure as defined in Section~\ref{sec:main_variables}. 
Period $t=-1$ is omitted as the reference.

To examine heterogeneity by pre-shock portfolio size, I split the 
sample into high in-degree and low in-degree importers based on 
the average number of active relationships in the twelve months 
before the pre-shock period (April 2008--March 2009). I use this 
earlier window --- rather than the pre-shock period itself --- to avoid 
classifying importers that entered during the pre-shock period and 
had not yet accumulated their portfolios. The median in-degree in 
this window is 3, somewhat above the within-pre-shock median of 2 
reported in Panel C of Table~\ref{tab:stat}.\\

\noindent \textbf{Buyer's relationship portfolio.} Adjusting a 
relationship portfolio involves substantial search costs and time 
to establish trusted connections \citep*{Mejean2023, Cajal-Grossi2023, 
RauchTrinidade2002}. Ending existing relationships and 
forming new ones is therefore a gradual process. Building on the 
previous analysis, I examine whether relationship-level shocks 
aggregated at the firm level affect future portfolio composition. 
This captures both the ending of existing relationships and the 
formation of new ones after the shock. As in the relationship-level 
analysis, I use the six-month frequency because relationship endings 
and new formations are rare events at the monthly level.

I construct two relationship-level indicators for each importer:

\noindent (i) \textit{New relationship formation} ($n_{ij,t}$): 
equals one if the pair forms for the first time in period $t$, 
zero if the pair has never been active by $t$ (a potential entrant 
that has not yet formed), and is excluded for pairs that were 
already active in any period $\tau<t$:
\begin{equation}
n_{ij,t} =
\begin{cases}
1 & \text{if } y_{ij,t}=1 \text{ and } y_{ij,\tau}=0 \;\forall \tau<t,\\
0 & \text{if } y_{ij,t}=0 \text{ and } y_{ij,\tau}=0 \;\forall \tau<t,\\
\text{NA} & \text{otherwise.}
\end{cases}
\label{y_new}
\end{equation}
This restriction confines the universe of pairs to first-time 
formations, excluding continuations of already-active relationships 
as well as re-matches of previously ended ones, which would 
otherwise generate an accumulated-stock effect rather than a 
formation hazard.

\noindent (ii) \textit{Relationship ending} ($e_{ij,t}$): defined 
as in equation~\eqref{Y1}, conditional on the relationship being 
active in $t-1$ so that the estimand is an ending hazard.

For new-formation outcomes, relationship fixed effects cannot be 
identified, since each pair is observed only from the moment of 
formation. I therefore use importer fixed effects to control for 
firms' time-invariant propensity to form new relationships. The 
two specifications are:

\begin{align}
n_{ij,t} &= \sum_{l \neq -1} \beta_l \, IE_{j,0} \cdot 
\mathbb{1}\{t=l\} + \theta_j +\theta_i + \delta_t + \varepsilon_{ij,t},
\label{Y5n} \\
e_{ij,t} &= \sum_{l \neq -1} \beta_l \, IE_{j,0} \cdot 
\mathbb{1}\{t=l\} + \theta_{ij} + \zeta_{c,t} + \delta_t + \varepsilon_{ij,t}, \label{Y5e}
\end{align}

where $\theta_{ij}$ are relationship fixed effects, $\theta_j$  and $\theta_i$
are importer and exporter fixed effects, $\zeta_{c,t}$ are 
cohort-by-time fixed effects, and $\delta_t$ are time fixed 
effects. In both cases, the outcome equals one when the event 
(ending or formation) occurs in period $t$.\\

\noindent \textbf{Firm-level extensive margin.} The final analysis 
examines firm adjustment at the extensive margin: whether temporary 
shocks affect importers' market participation. An importer is 
classified as having exited the market in period $t$ if none of its 
relationships are active in $t$:
\begin{equation}
\text{Exit}_{j,t} =
\begin{cases}
1 & \text{if } e_{ij,t}=1 \;\; \forall\, i,\\
0 & \text{otherwise,}
\end{cases}
\label{y_exit}
\end{equation}
where $e_{ij,t}$ is the relationship-ending indicator from 
equation~\eqref{Y1}. Because relationship endings are permanent by 
construction (Section~\ref{sec:main_variables}), $\text{Exit}_{j,t}=1$ 
implies that the importer is observed with no further transactions 
through the end of the sample. I estimate the effect of importer-level 
exposure on firm exit using a linear probability model:
\begin{equation}
\text{Exit}_{j,t} = \sum_{l \neq -1} \beta_l \, IE_{j,0} \cdot 
\mathbb{1}\{t=l\} + \alpha_j + \delta_t + \varepsilon_{j,t},
\label{Y6}
\end{equation}
where $\alpha_j$ are importer fixed effects and $\delta_t$ are time 
fixed effects. The $\beta_l$ coefficients are interpreted as the 
percentage-point change in the probability of firm exit.

\section{Results} 
\label{sec:results}
In this section, I show that buyer--seller relationships adjust 
in response to disruptions. I distinguish between relationship 
endings and short-run adjustments that allow firms to preserve 
existing relationships despite temporary delivery shortfalls. I 
then examine how relationship-level shocks aggregate to affect 
importers' relationship portfolios, assessing whether more exposed 
importers experience greater relationship turnover. Finally, I 
analyze whether these disruptions translate into firm exit.\\

\begin{figure}[h]
\begin{center}
\textbf{\footnotesize{\textsc{All relationships}}}
\hspace{1.8in}
\textbf{\footnotesize{\textsc{Within importers}}}\\
\hspace*{-0.35in}
\includegraphics[height=1.95in]{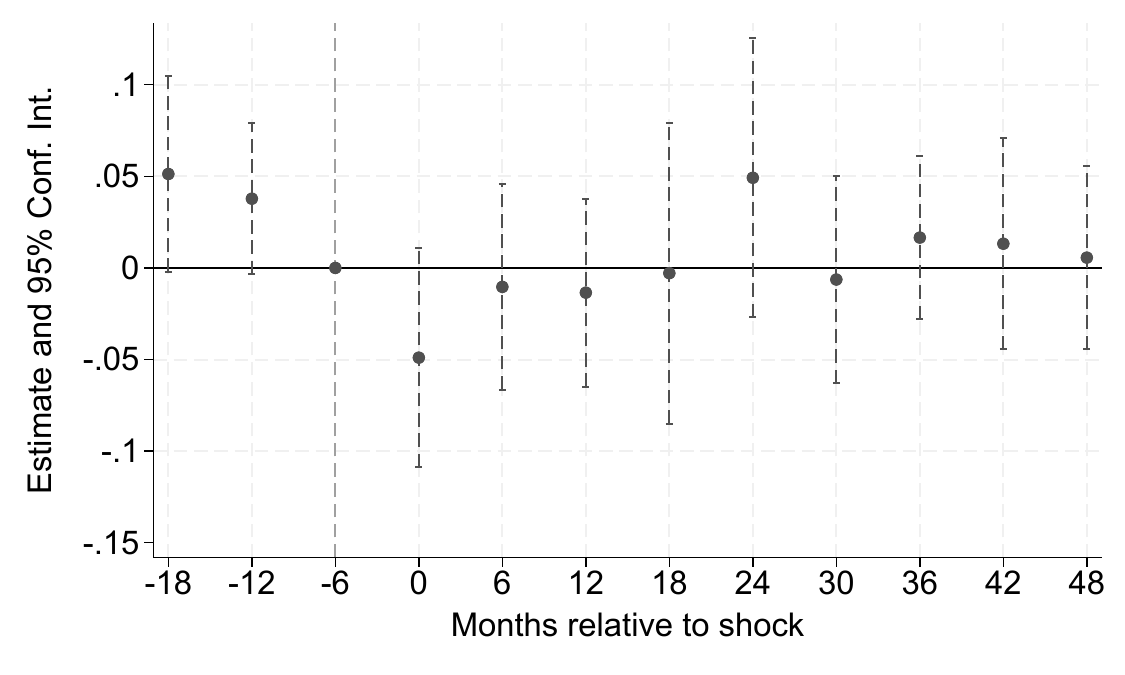}
\includegraphics[height=1.95in]{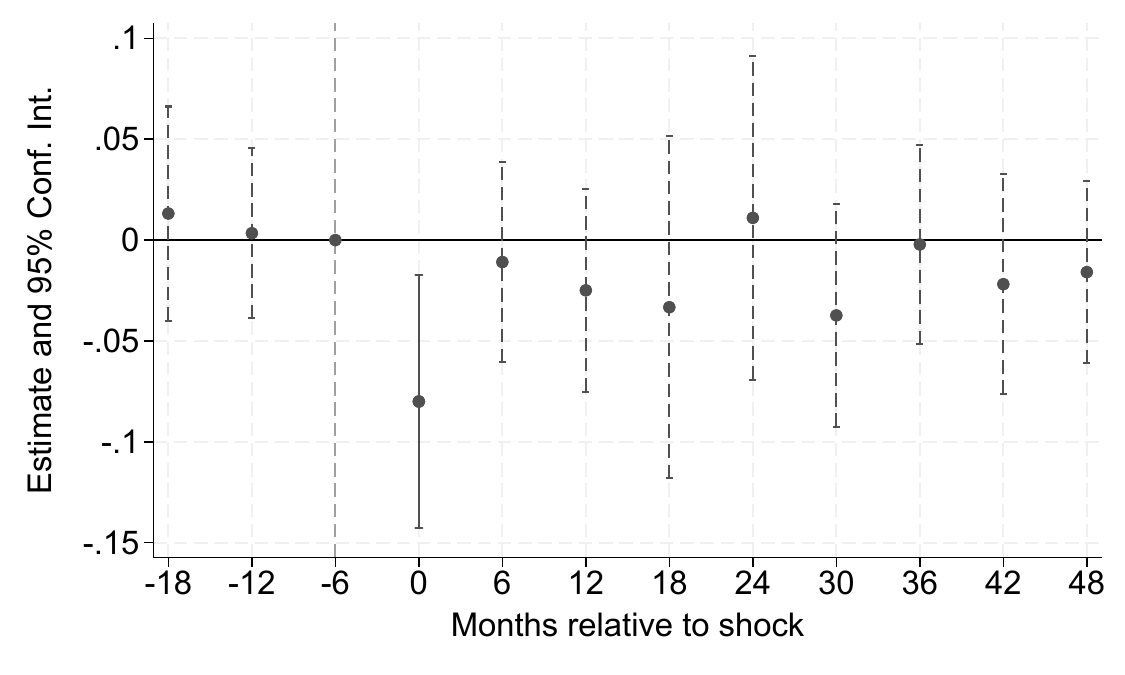}
\end{center}
\vspace*{-0.2cm}
\caption{Effect of road disruptions on the probability of 
a relationship ending}
\vspace*{0.1cm}
\scalebox{0.90}{
\begin{minipage}{1.1\textwidth}
\advance\leftskip 0.5cm
{\footnotesize{
\textit{\textbf{Notes:}} The figure plots the $\beta_l$ 
coefficients from estimating equation~\eqref{Y1}, where the 
outcome is the relationship-ending hazard (equal to one if a 
relationship active in $t-1$ ends in $t$). The sample is the 
exposure sample defined in Section~\ref{sec:summary-stats}. All 
specifications include relationship and cohort-by-time fixed effects. The 
right panel additionally includes importer-by-period fixed effects 
$\alpha_{j,t}$, restricting identification to variation across 
relationships linked to the same importer in the same period. 
Sample sizes are $N=20{,}003$ relationship--period observations 
in the left panel (324 exporter clusters) and $N=18{,}322$ in the 
right panel (314 exporter clusters). Coefficients are reported 
with 95\% confidence intervals based on standard errors clustered 
at the exporter level.}\par}
\end{minipage}
}
\label{fig:main}
\end{figure}

\noindent \textbf{Effect of road disruptions on relationship 
ending.} Figure~\ref{fig:main} displays estimates of the 
differential effect of road disruptions on the probability of a 
relationship ending, comparing relationships \textit{exposed to 
the shock} with those that are not. The left panel compares all 
exposed and non-exposed relationships, while the right panel 
restricts to variation within an importer's relationship portfolio.

When comparing all relationships, the effect at the shock ($t=0$) 
is small and statistically indistinguishable from zero. However, 
focusing within an importer's portfolio, the probability of a 
relationship ending decreases by approximately 8 percentage 
points---a 48\% decline relative to the pre-shock baseline of 
0.167. The 
within-importer specification holds the importer fixed, so the 
estimate reflects variation in exposure across relationships 
linked to the same buyer. This is consistent with importers with 
partial portfolio exposure delaying the baseline turnover process: 
because they can still receive shipments through non-exposed 
relationships, they have an incentive to wait for disrupted 
suppliers to resume deliveries rather than terminate them 
immediately.

The results are robust to (i) an alternative two-year pre-shock 
window, (ii) including exporters with farms in flooded production 
areas (excluded from the baseline because their disruptions stem 
from production rather than transport, and may trigger different 
responses such as \textit{force majeure} clauses or government compensation), 
and (iii) excluding the earliest relationship cohort.\footnote{The results are reported in the Supplemental Appendix}\\

\begin{figure}[h]
\begin{center}
\textbf{\footnotesize{\textsc{All relationships}}}
\hspace{1.8in}
\textbf{\footnotesize{\textsc{Within importer}}}\\
\hspace*{-0.35in}
\includegraphics[height=1.95in]{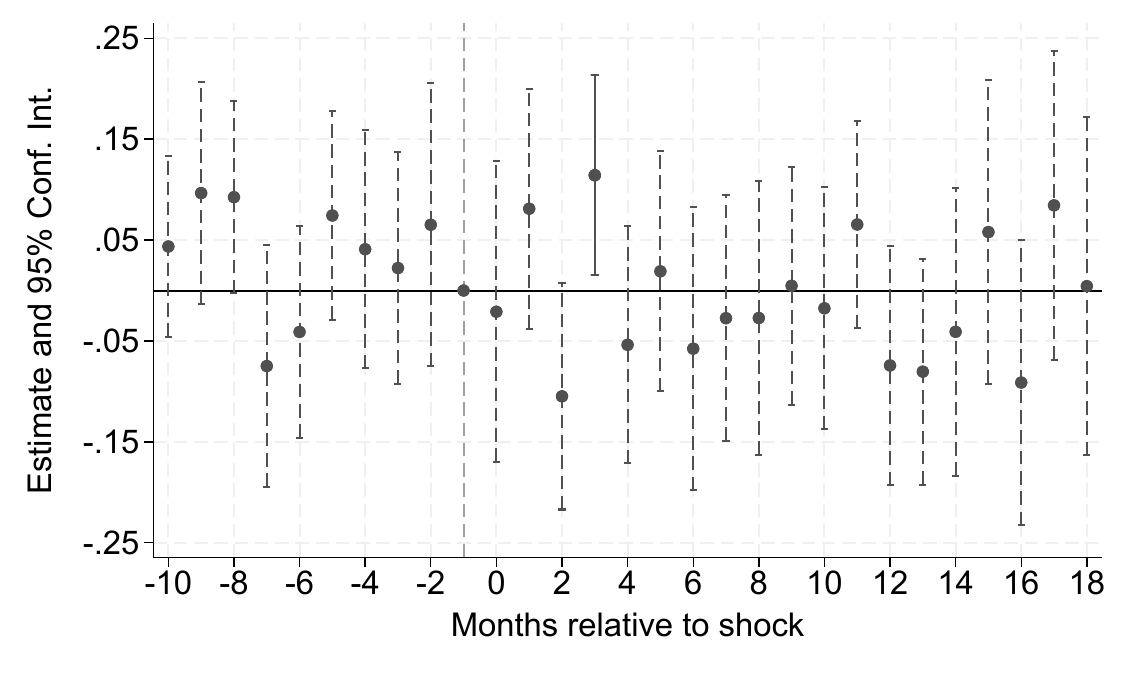}
\includegraphics[height=1.95in]{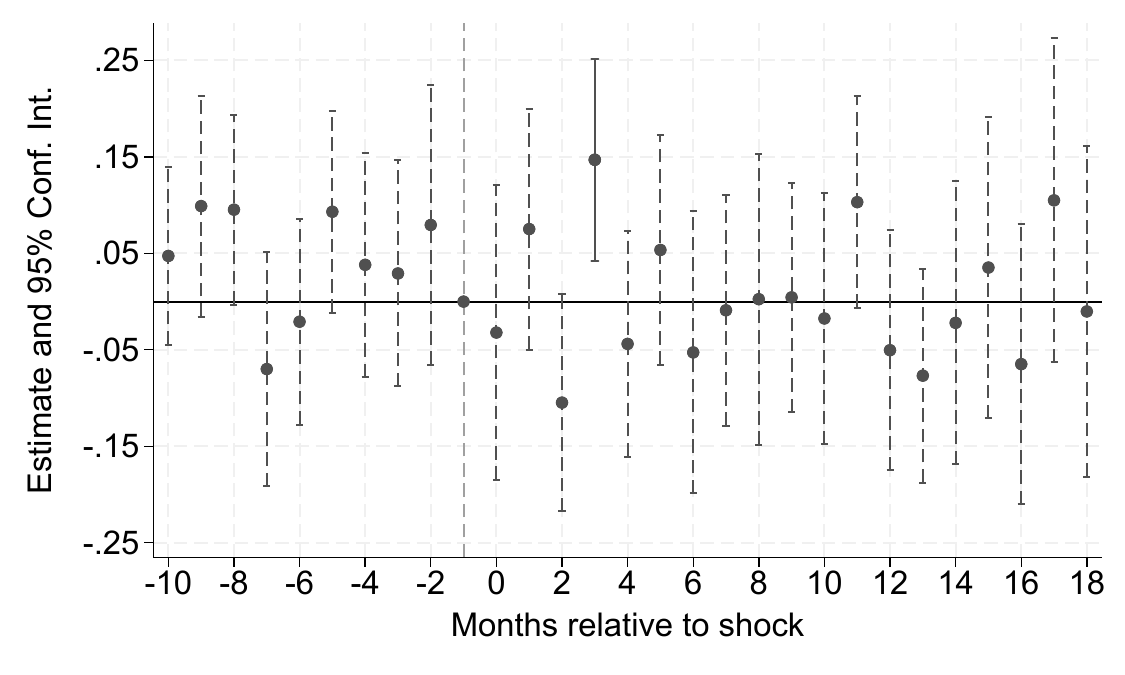}
\end{center}
\vspace*{-0.2cm}
\caption{Effect of road disruptions on the probability of transacting}
\vspace*{0.1cm}
\scalebox{0.90}{
\begin{minipage}{1.1\textwidth}
\advance\leftskip 0.5cm
{\footnotesize{
\textit{\textbf{Notes:}} The figure plots the $\beta_l$ coefficients 
from equation~\eqref{Y3}, where the outcome equals one if a 
transaction is observed in month $m$ and zero otherwise. The 
estimation sample is the panel of relationships active at some 
point in the sample, excluding relationships first observed after 
September 2010. All specifications include relationship and 
cohort-by-month fixed effects. The right panel additionally includes 
importer-by-month fixed effects $\alpha_{j,m}$, restricting 
identification to within-importer-month variation. Sample sizes 
are $N=65{,}684$ relationship--month observations in the left panel 
(323 exporter clusters) and $N=61{,}260$ in the right panel (314 
clusters). Coefficients are reported with 95\% confidence intervals 
based on standard errors clustered at the exporter level. }\par}
\end{minipage}
}
\label{fig:transaction}
\end{figure}

\noindent \textbf{Effects on extensive and intensive margins.}\
Firms can delay portfolio adjustments because more flexible 
margins --- such as temporary shipment delays --- absorb shocks without 
forcing immediate termination, particularly when exposure varies 
across an importer's links. Figure~\ref{fig:transaction} presents 
estimates from equation~\eqref{Y3} of the probability that a 
relationship transacts at the monthly frequency. The estimation 
window covers $t = -10$ to $t = +18$ months relative to the shock 
($t=0$ corresponds to October 2010). In both panels, the probability of 
transacting declines slightly in December 2010 (end of the first 
wave of rains): the coefficient is approximately a 10 percentage 
point decrease, significant only at the 10\% level.

Between the first and second waves of disruptions, when some roads 
reopen, relationships \textit{exposed to the shock} are more likely 
to transact. The effect at $t=3$ (January 2011) is an increase of 
11--14 percentage points, a 22--28\% increase over the baseline 
monthly transaction probability among active relationships in the 
pre-shock period (0.50).\footnote{The relevant baseline is the 
monthly probability that an active relationship records a transaction, 
since equation~\eqref{Y3} is estimated on the panel of active 
relationships. This differs from the six-month transaction frequency reported in Table~\ref{tab:stat}.} The results indicate 
that shipments are delayed at the end of the first wave of road 
disruptions (December 2010) but resume once rains temporarily 
decrease (January 2011). For the second wave of disruptions 
(April--June 2011), there are no substantial effects.

\begin{figure}[h]
\begin{center}
\textbf{\footnotesize{\textsc{Kilograms}}}
\hspace{2in}
\textbf{\footnotesize{\textsc{Trade value}}}\\
\hspace*{-0.35in}
\includegraphics[height=1.95in]{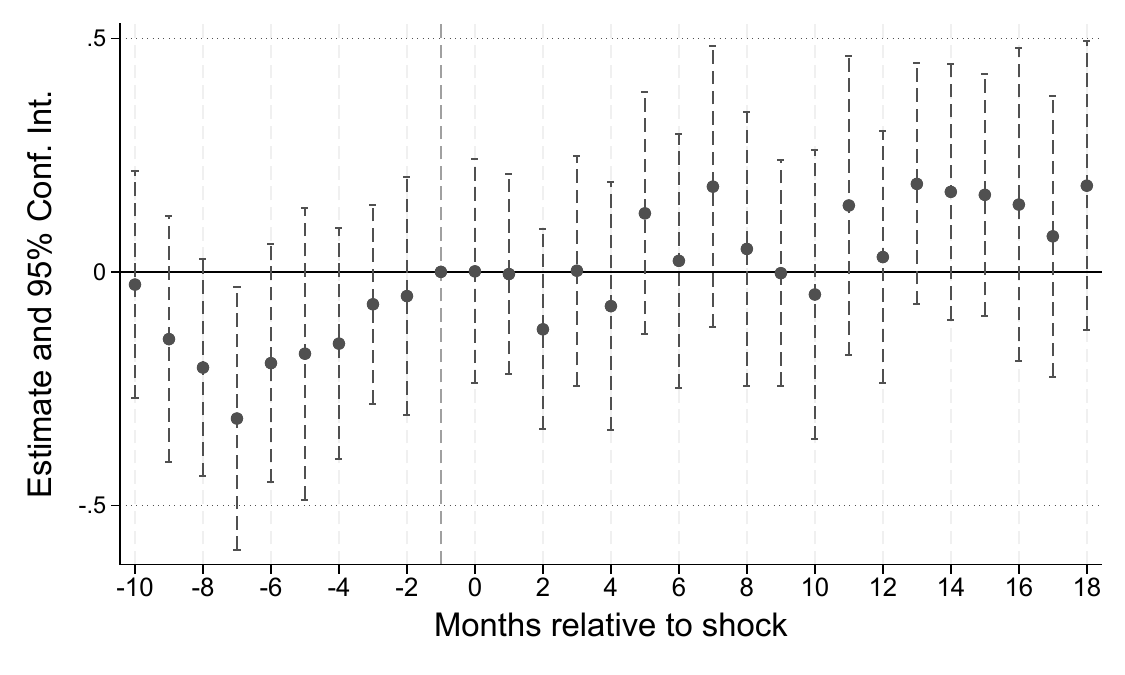}
\includegraphics[height=1.95in]{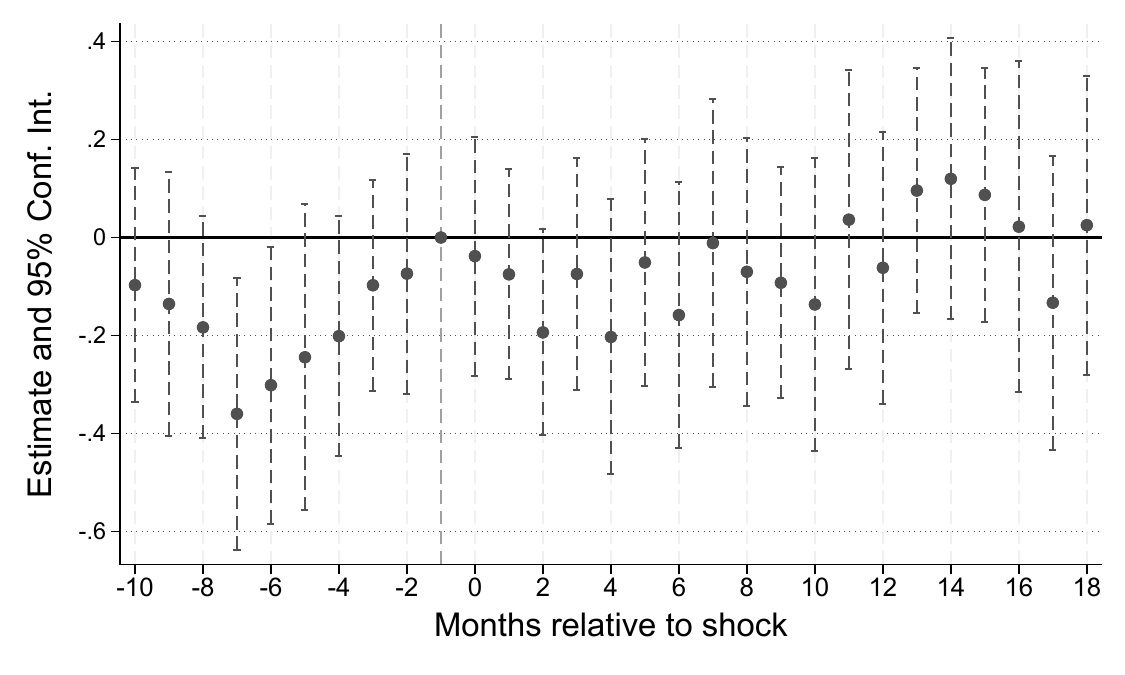}\\
\end{center}
\vspace*{-0.2cm}
\caption{Effect of road disruptions on kilograms and trade value}
\label{fig:kilograms}
\vspace*{0.1cm}
\scalebox{0.90}{
\begin{minipage}{1.1\textwidth}
\advance\leftskip 0.5cm
{\footnotesize{
\textit{\textbf{Notes:}} Both panels plot the $\beta_l$ coefficients 
from equation~\eqref{Y4} with 95\% confidence intervals, where the 
dependent variable is the inverse hyperbolic sine of kilograms 
(left panel) and trade value (right panel). The sample is restricted 
to cut-flower HS codes (same HS6 codes as in Panel B of 
Table~\ref{tab:stat}) and to relationship--product cells observed 
in at least two months. Regressions include relationship--product 
and cohort-by-month fixed effects. $N=40{,}904$ relationship--product--month 
observations in each panel (293 exporter clusters). Standard errors 
are clustered at the exporter level. }\par}
\end{minipage}
}
\end{figure}

Figure~\ref{fig:kilograms} reports estimates from equation~\eqref{Y4} 
for trade value and kilograms, on cut-flower products (same HS6 codes 
as in Panel B of Table~\ref{tab:stat}) and restricting the sample to 
relationship--product cells observed in at least two months. All 
outcomes throughout this section are transformed with the inverse 
hyperbolic sine, and coefficients are interpreted as approximate 
semi-elasticities (percent changes in the underlying variable). 
Kilograms show no statistically significant response throughout the 
post-shock period. Trade value shows a modest decline at $t=2$ 
(December 2010), significant only at the 10\% level. Results are 
qualitatively similar --- though with a slight pre-trend --- when 
relationship and product--month fixed effects are used in place of 
$\delta_{ijp}$, and when identification is further restricted to 
within-importer variation through the inclusion of importer--month 
fixed effects (Figure~\ref{fig:rob_kilogram} in the Appendix).

\begin{figure}[h]
\begin{center}
\hspace{0.25in}
\textbf{\footnotesize{\textsc{Unit value}}}
\hspace{2.5in}
\textbf{\footnotesize{\textsc{Price per kg}}}\\
\hspace*{-0.35in}
\includegraphics[height=1.95in]{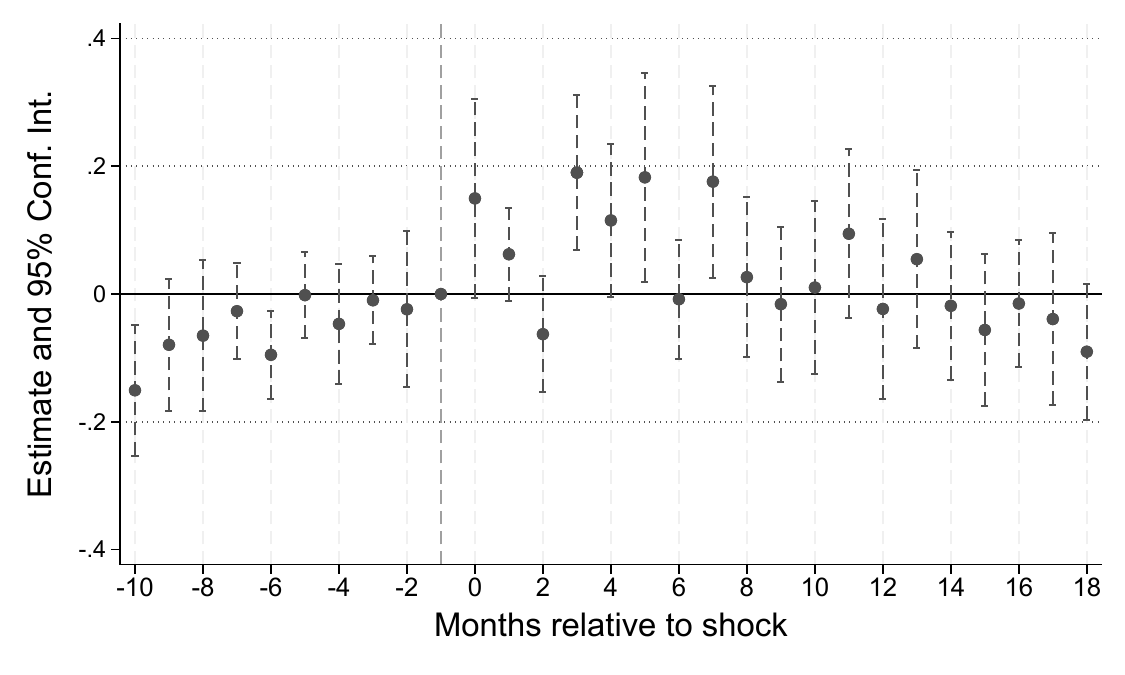}
\includegraphics[height=1.95in]{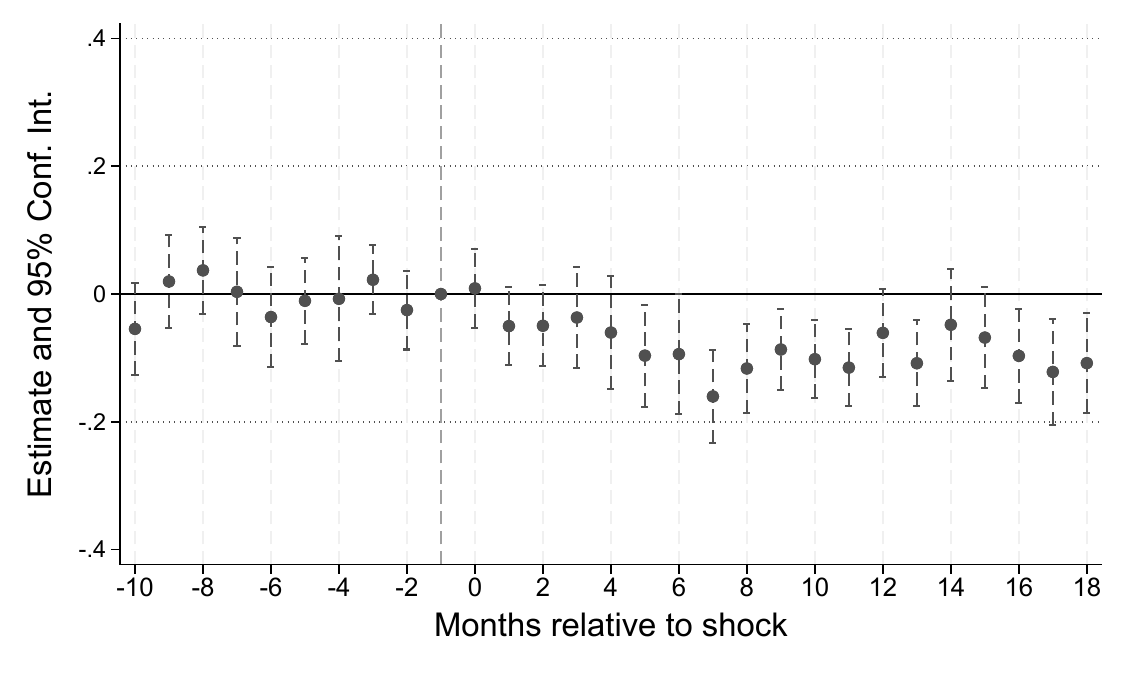}
\end{center}
\vspace*{-0.2cm}
\caption{Effect of road disruptions on prices}
\vspace*{0.1cm}
\scalebox{0.90}{
\begin{minipage}{1.1\textwidth}
\advance\leftskip 0.5cm
{\footnotesize{
\textit{\textbf{Notes:}} Both panels plot the $\beta_l$ coefficients 
from equation~\eqref{Y4} with 95\% confidence intervals, where the 
dependent variable is the inverse hyperbolic sine of unit value 
(left panel) and price per kilogram (right panel). The sample is 
restricted to cut-flower HS codes (same HS6 codes as in Panel B of 
Table~\ref{tab:stat}) and to relationship--product cells observed 
in at least two months. Regressions include relationship, 
cohort-by-month, and product-by-month fixed effects. $N=40{,}904$ 
relationship--product--month observations (300 exporter clusters). 
Standard errors are clustered at the exporter level. }\par}
\end{minipage}
}
\label{fig:units}
\end{figure}

Figure~\ref{fig:units} reports responses for unit value and price 
per kilogram on the same sample. For relationships \textit{exposed to the shock}, within the first six months of the 
shock, experience a unit value rises by approximately 11--19\%, with smaller 
within-importer estimates suggesting that part of this rise reflects 
across-importer composition. Over the medium run (approximately two 
years after the shock), price per kilogram falls by approximately 
6--16\%, consistent with exporters absorbing disruption costs to preserve 
their trading relationships \citep{Macciavello2015}.\footnote{This 
mechanism relates to two strands of the literature on exporter 
pricing. \citet*{Berman2012} show that high-performance exporters 
absorb a larger share of cost shocks in their markups rather than 
passing them through to prices in order to preserve market share. 
\citet*{Heise2018} documents in U.S.\ customs data that exchange-rate 
pass-through increases with relationship length, implying that 
exporters in more fragile relationships absorb more of any cost 
shock to retain buyers --- consistent with the price concessions 
observed here for exposed relationships.} The decline two years 
after the shock is economically meaningful: approximately \$0.4--1.10 US dollars 
per kilogram when evaluated at the pre-shock average of \$6.9 US dollars per 
kilogram (Table~\ref{tab:stat}).\footnote{The fall in prices per kilogram is not permanent or a feature of non-stationarity. Two-years after the shock the differential effect dissappears. Estimations using 36 months after the shock are in the Supplemental Appendix.}

The two price measures move in opposite directions, but this is not 
a contradiction. Unit value is the ratio of trade value to units; 
price per kilogram is the ratio of trade value to kilograms. As I 
show below, the apparent rise in unit value is largely a denominator 
effect from a sharp decline in units, while the decline in price per 
kilogram reflects exporter concessions on the actual price of the 
product shipped.\footnote{All outcomes in this section are transformed 
with the inverse hyperbolic sine to handle variables with substantial 
mass near zero. The log specification yields qualitatively similar 
results for trade value, kilograms, and price per kilogram (which 
have means well above zero), but generates pre-trends in unit value, 
kilograms per unit, where small values are more common. }

\begin{figure}[H]
\begin{center}
\hspace{0.30in}
\textbf{\footnotesize{\textsc{Units}}}
\hspace{2.5in}
\textbf{\footnotesize{\textsc{Kilograms per unit}}}\\
\hspace*{-0.35in}
\includegraphics[height=1.95in]{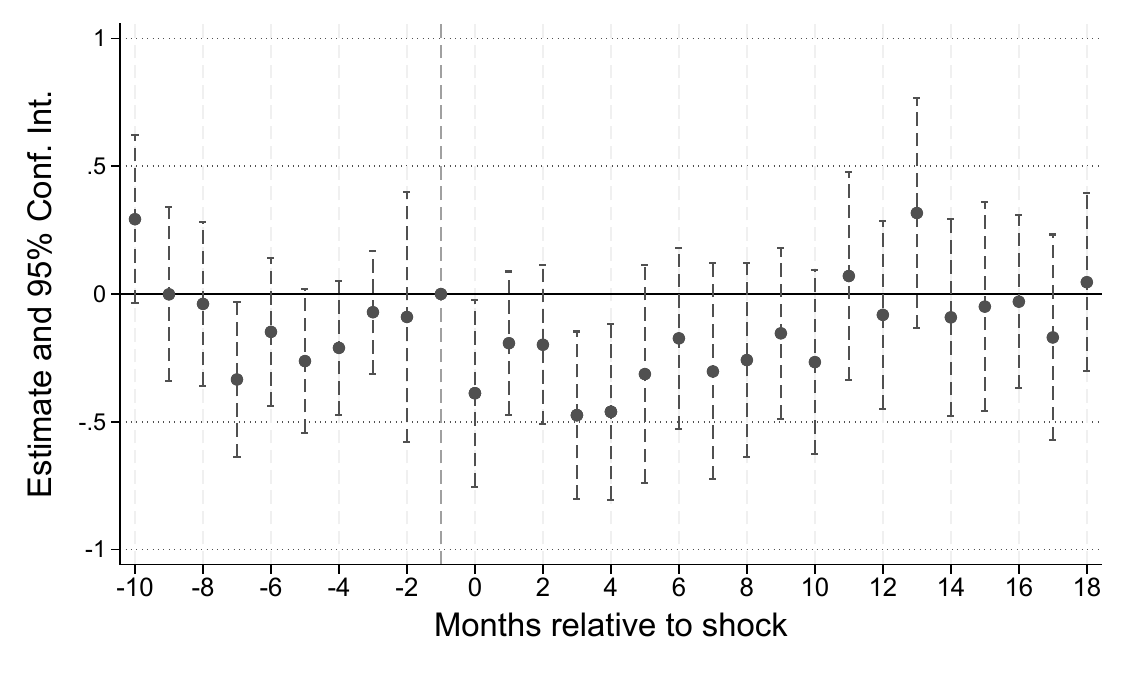}
\includegraphics[height=1.95in]{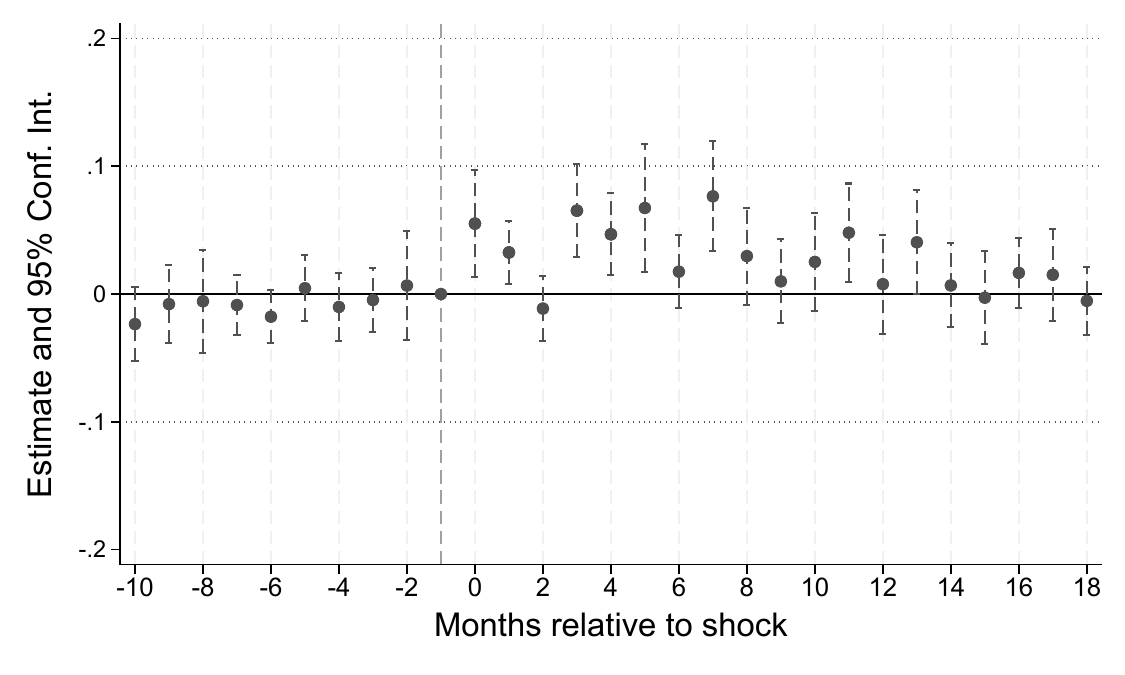}
\end{center}
\vspace*{-0.2cm}
\caption{Effect of road disruptions on units and kilograms per unit}
\label{fig:rob_kg_un}
\vspace*{0.1in}
\scalebox{0.90}{
\begin{minipage}{1.1\textwidth}
\advance\leftskip 0.5cm
{\footnotesize{
\textit{\textbf{Notes:}} Both panels plot the $\beta_l$ coefficients 
from equation~\eqref{Y4} with 95\% confidence intervals, where the 
dependent variable is the inverse hyperbolic sine of units (left 
panel) and kilograms per unit (right panel). The sample is restricted 
to cut-flower HS codes (same HS6 codes as in Panel B of 
Table~\ref{tab:stat}) and to relationship--product cells observed 
in at least two months. Regressions include relationship, 
cohort-by-month, and product-by-month fixed effects. $N=40{,}904$ 
relationship--product--month observations (300 exporter clusters). 
Standard errors are clustered at the exporter level. }\par}
\end{minipage}
}
\end{figure}

To distinguish the price-per-kilogram decline from a denominator 
effect, I separate units and kilograms per unit. Figure~\ref{fig:rob_kg_un} 
shows that within six months of the shock, units decline by 
approximately 20--58\% while kilograms per unit rise by approximately 
3--6\%. The decline in units, combined with stable kilograms, indicates 
that once exporters can resume shipments they consolidate contracted 
quantities into fewer, larger units. The rise in unit values observed in Figure~\ref{fig:rob_kg_un} therefore reflects a sharp decline in shipped units rather than a genuine increase in unit values: trade values either remain relatively stable or decline by less than shipped units. The 
medium-run decline in price per kilogram, by contrast, persists 
through the two-year window and reflects exporter concessions on 
the actual price per kilogram delivered.

In summary, temporary disruptions affect both the extensive and 
intensive margins of trade. Exposed relationships exhibit a 
temporary, short-lived reduction in transaction frequency. On the 
intensive margin, the short-run response is dominated by a sharp 
contraction in units shipped, accompanied by a modest consolidation 
into larger units. Over the medium run, exporters absorb part of 
the disruption costs through lower prices per kilogram, consistent 
with effort to preserve long-run relationships.\\

\begin{figure}[h]
\begin{center}
\hspace{0.25in}
\textbf{\footnotesize{\textsc{All importers}}}
\hspace{2.5in}
\textbf{\footnotesize{\textsc{By in-degree}}}\\
\hspace*{-0.35in}
\includegraphics[height=1.95in]{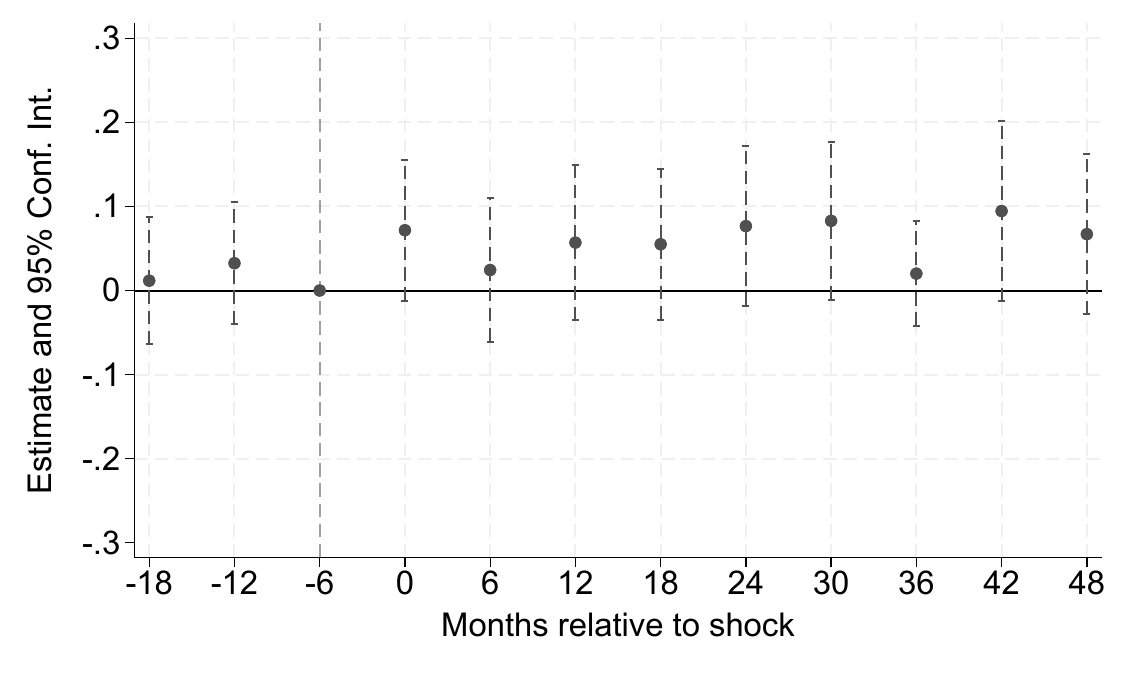}
\includegraphics[height=1.95in]{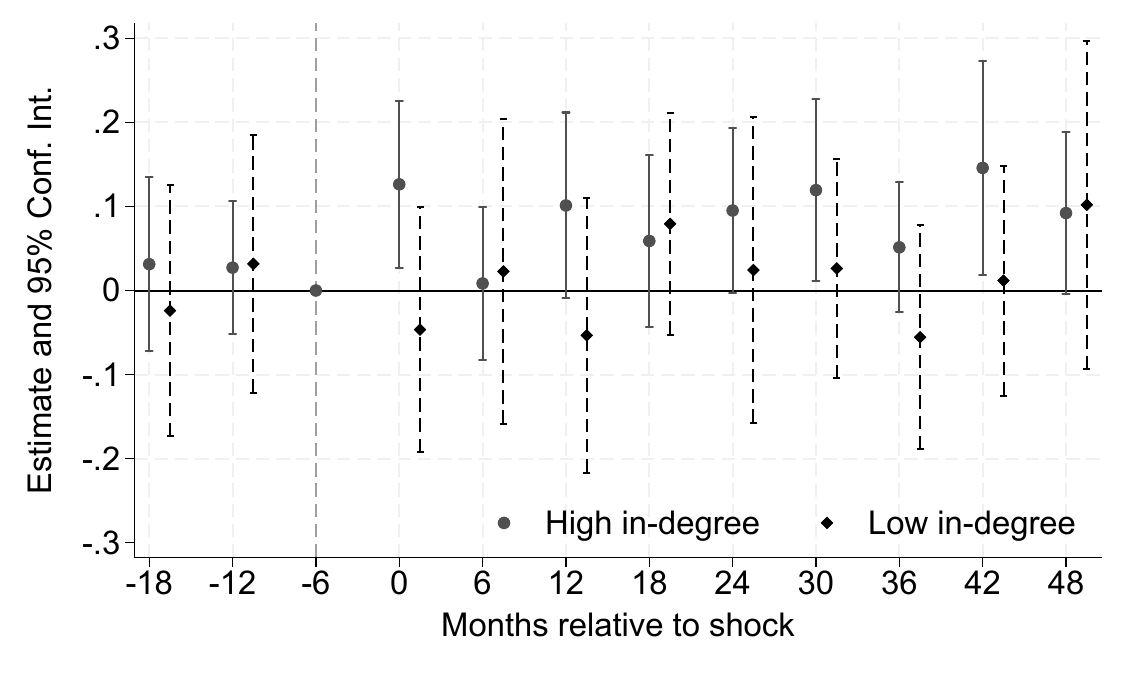}
\end{center}
\vspace*{-0.2cm}
\caption{Effect of importer-level exposure ($IE_j$) on the probability 
of a relationship ending}
\label{fig:firm_Exposure}
\vspace*{0.1cm}
\scalebox{0.90}{
\begin{minipage}{1.1\textwidth}
\advance\leftskip 0.5cm
{\footnotesize{
\textit{\textbf{Notes:}} The figure plots the $\beta_l$ coefficients 
from equation~\eqref{Y2}, where the outcome is the relationship-ending 
hazard (equal to one if a relationship active in $t-1$ ends in $t$). 
The left panel uses all relationship--period observations 
($N=20{,}003$, 324 exporter clusters); the right panel estimates 
effects separately for high in-degree importers ($N=17{,}760$, 317 
exporter clusters) and low in-degree importers ($N=2{,}243$, 170 
exporter clusters), with the in-degree split defined in 
Section~\ref{sec:empirical}. All regressions include relationship 
and cohort-by-time fixed effects. Coefficients are reported with 
95\% confidence intervals based on standard errors clustered at the 
exporter level. }\par}
\end{minipage}
}
\end{figure}

\noindent \textbf{Relationship shocks and importers' response.}\
Figure~\ref{fig:firm_Exposure} shows estimates from 
equation~\eqref{Y2}, which replaces relationship-level exposure with 
importer-level exposure ($IE_j$) to examine whether shocks propagate 
through a firm's entire relationship portfolio. The left panel shows 
results for all importers; the right panel splits the sample into 
high in-degree and low in-degree importers based on the in-degree 
window defined in Section~\ref{sec:empirical} (average number of 
active relationships in the twelve months preceding the pre-shock 
period).

For all importers, a higher share of exposed relationships increases 
the probability of a relationship ending. After splitting by 
in-degree, the effect is concentrated among high in-degree importers: 
at the shock period ($t=0$), a 10 percentage point increase in 
firm-level exposure raises the probability of a relationship ending 
by 1.26 percentage points. Equivalently, moving an importer from the 25$^{th}$ to the 75$^{th}$ percentile of the high in-degree exposure distribution (0.045 to 0.58), raises the probability of a relationship ending by 6.7 percentage points. This is roughly a 40\% increase relative to the pre-shock baseline of 0.167. The effect attenuates six months after 
the shock ($t=6$, coefficient near zero) but remains positive and 
statistically significant at $t=12, 24, 30, 42$, and $48$, indicating 
that portfolio-level adjustments occur immediately and continue over 
the following years. For low in-degree importers, I find no 
statistically significant effect at any horizon. Results using a 
binary indicator for whether importer-level exposure exceeds 50\% 
show a similar pattern: no effect for low in-degree importers, and 
a 7 percentage point increase in the probability of relationships 
ending for high in-degree importers, this corresponds to a a 42\% increase on the pre-shock baseline (Figure~\ref{fig:firm_Exposure_rob} 
in the Appendix).

The relationship-level and importer-level results appear to point 
in opposite directions but are in fact complementary. The 
relationship-level result (Figure~\ref{fig:main}) identifies the 
effect of exposure \textit{within} an importer's portfolio: 
conditional on the importer, exposed relationships are \textit{less} 
likely to end than non-exposed ones, consistent with importers 
actively preserving their disrupted suppliers. The importer-level 
result (Figure~\ref{fig:firm_Exposure}) identifies the effect of an 
importer's \textit{portfolio-level} exposure on the survival of any 
of its relationships (whether exposed or non-exposed): firms with a 
higher share of exposed relationships end more relationships overall. 
The two results therefore capture distinct margins of adjustment: 
the within-importer result reflects short-run relationship 
preservation, while the importer-level result reflects medium-run 
portfolio rebalancing by the most exposed firms.\\

\noindent \textbf{Portfolio reallocation.}\
Temporary disruptions to relationships may affect a firm's longer-run 
accumulation of supplier links. One potential response is reallocation: 
firms may shift activity toward suppliers in less-exposed regions, 
updating their beliefs about supply reliability \citep{Balboni2023}, 
or move toward less relationship-intensive sourcing channels that 
are more resilient to supply disruptions \citep{Cajal-Grossi2023}.

I estimate equations~\eqref{Y5e} and~\eqref{Y5n} for relationship 
endings ($e_{ij,t}$) and the formation of new relationships 
($n_{ij,t}$), allowing both margins to evolve over time. I hold the 
set of importers fixed at the 624 in the exposure sample, while 
allowing all 1,113 exporters (Section~\ref{sec:data_sources}) to 
potentially form new relationships, including both incumbents and 
entrants. 

\begin{figure}[h]
\begin{center}
\hspace{0.25in}
\textbf{\footnotesize{\textsc{All importers}}}
\hspace{2.5in}
\textbf{\footnotesize{\textsc{By in-degree}}}\\
\hspace*{-0.35in}
\includegraphics[height=1.95in]{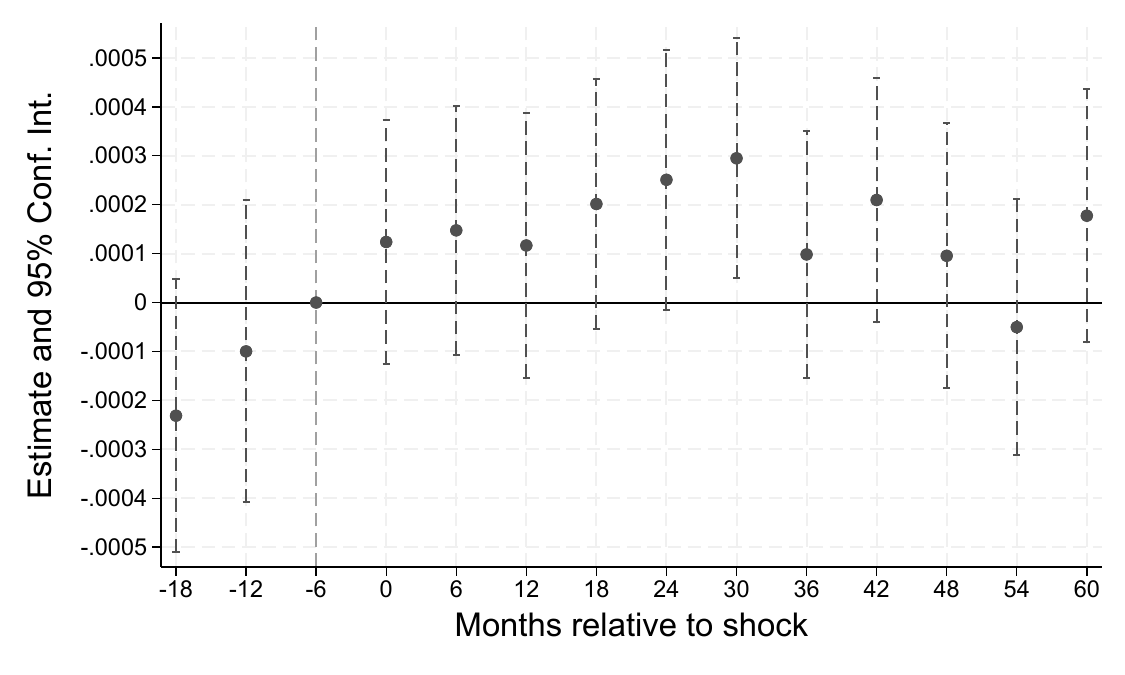}
\includegraphics[height=1.95in]{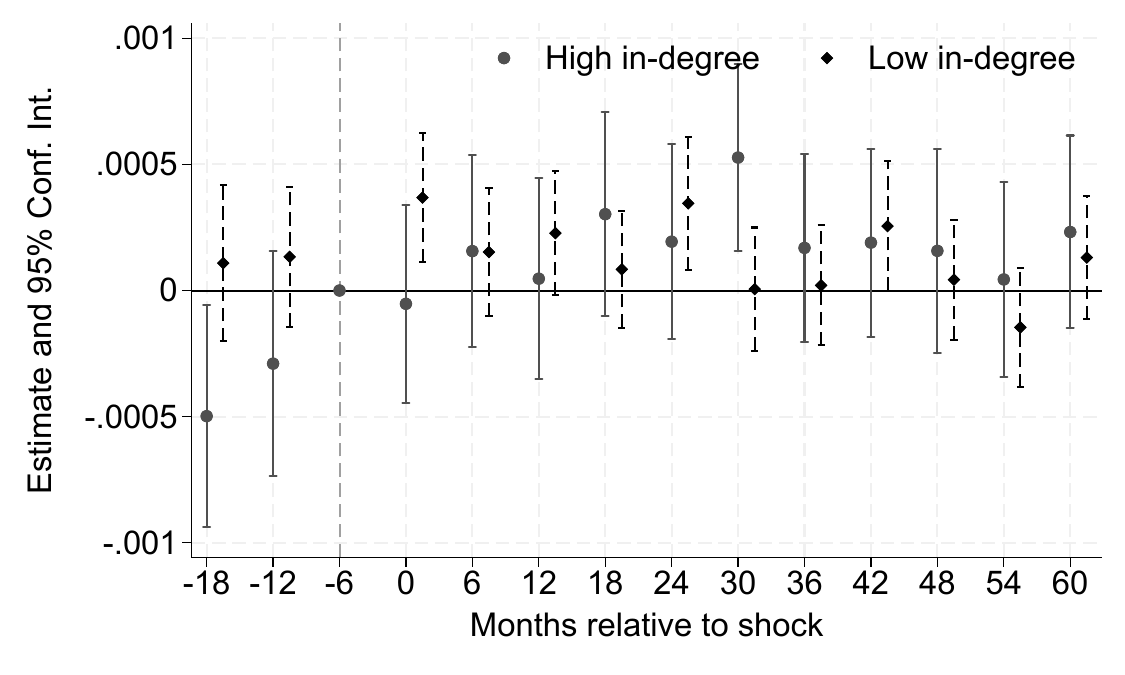}
\end{center}
\vspace*{-0.2cm}
\caption{Effect of road disruptions on new relationship formation}
\vspace*{0.1cm}
\scalebox{0.90}{
\begin{minipage}{1.1\textwidth}
\advance\leftskip 0.5cm
{\footnotesize{
\textit{\textbf{Notes:}} The figure plots the $\beta_l$ coefficients 
from equation~\eqref{Y5n}, where the outcome is the new-relationship 
formation indicator ($n_{ij,t}$), for all importers (left panel) 
and by in-degree (right panel), with 95\% confidence intervals. The 
estimation universe is the set of potential exporter--importer pairs, 
holding the 624 importers in the exposure sample fixed and allowing 
all 1,113 exporters (Section~\ref{sec:data_sources}) to form new 
relationships. Sample sizes are $N=9{,}578{,}505$ in the left panel, 
$N=6{,}045{,}987$ for high in-degree importers, and $N=3{,}532{,}518$ 
for low in-degree importers. All regressions include importer, exporter and 
time fixed effects. Standard errors are clustered at the exporter level with 1,113 clusters.}\par}
\end{minipage}
}
\label{fig:new_matches}
\end{figure}

Figure~\ref{fig:new_matches} presents results for the probability 
of forming a new relationship. In the left panel, controlling for 
importer and time fixed effects, the probability of new matching 
rises gradually starting roughly two years after the shock and then 
declines. This is consistent with relationship adjustment operating 
over the long run rather than as a short-term response. The right 
panel splits the sample by in-degree, using the split defined in 
Section~\ref{sec:empirical}. For low in-degree importers, the effect 
is a small but persistent increase in new matches at the shock and 
in subsequent periods: the coefficient at the shock ($t=0$) is 0.0003. Equivalently,   moving an importer from no exposure to full exposure, raises the probability of matching by 11.5\% increase relative to the pre-shock baseline matching 
probability of 0.0026 (Panel A of Table~\ref{tab:stat}).  The effect is still significant in the medium-run at $t=12$ and 24.
For high in-degree importers, pre-trends 
are present (complicating interpretation), and post-shock coefficients 
are statistically indistinguishable from zero except at $t=30$, 
which accounts for the spike visible in the left panel.\footnote{The estimations using a binary exposure above 50\% also show similar pattern, where exposed low in-degree importers are more likely to match in the first two years after the shock. High in-degree importers show pre-trends at $t=-18$ in both specifications. The results are shown in the Supplementary Appendix. }

These results suggest that low in-degree importers, unable to rely 
on alternative relationships within their portfolio are more likely to seek new suppliers to 
replace disrupted ones. High in-degree importers, by contrast, can 
draw on their non-exposed relationships during the disruption and 
therefore have less incentive to incur the search costs of forming 
new matches.

\begin{figure}[h]
\begin{center}
\hspace{0.25in}
\textbf{\footnotesize{\textsc{All importers}}}
\hspace{2.5in}
\textbf{\footnotesize{\textsc{By in-degree}}}\\
\hspace*{-0.35in}
\includegraphics[height=1.95in]{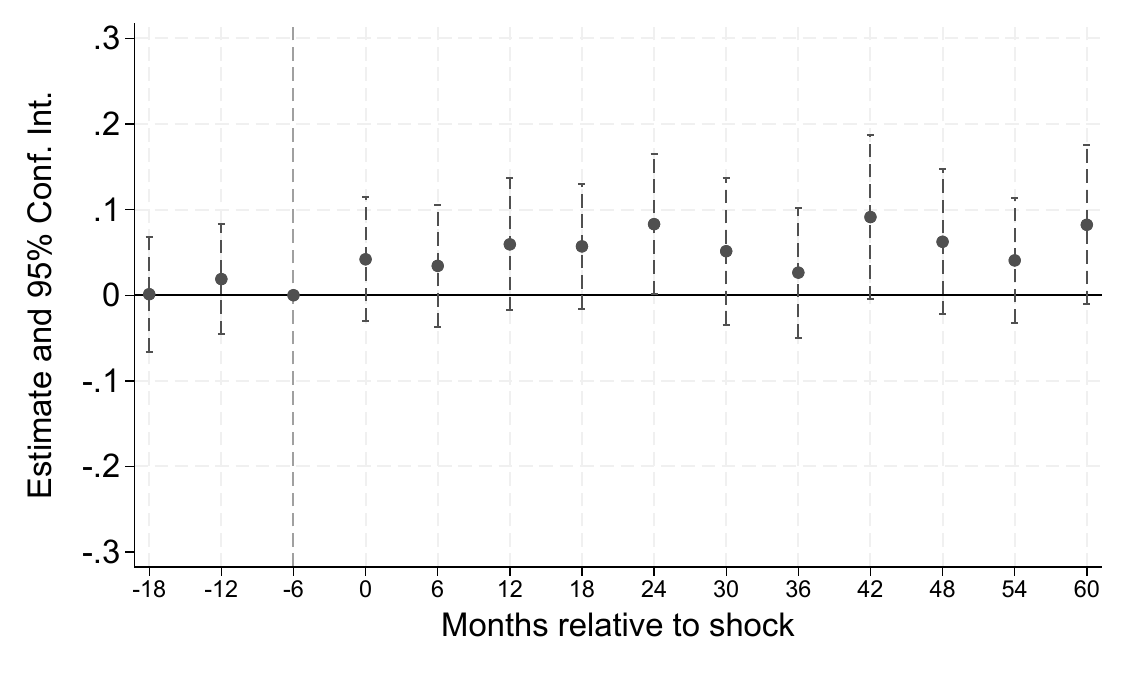}
\includegraphics[height=1.95in]{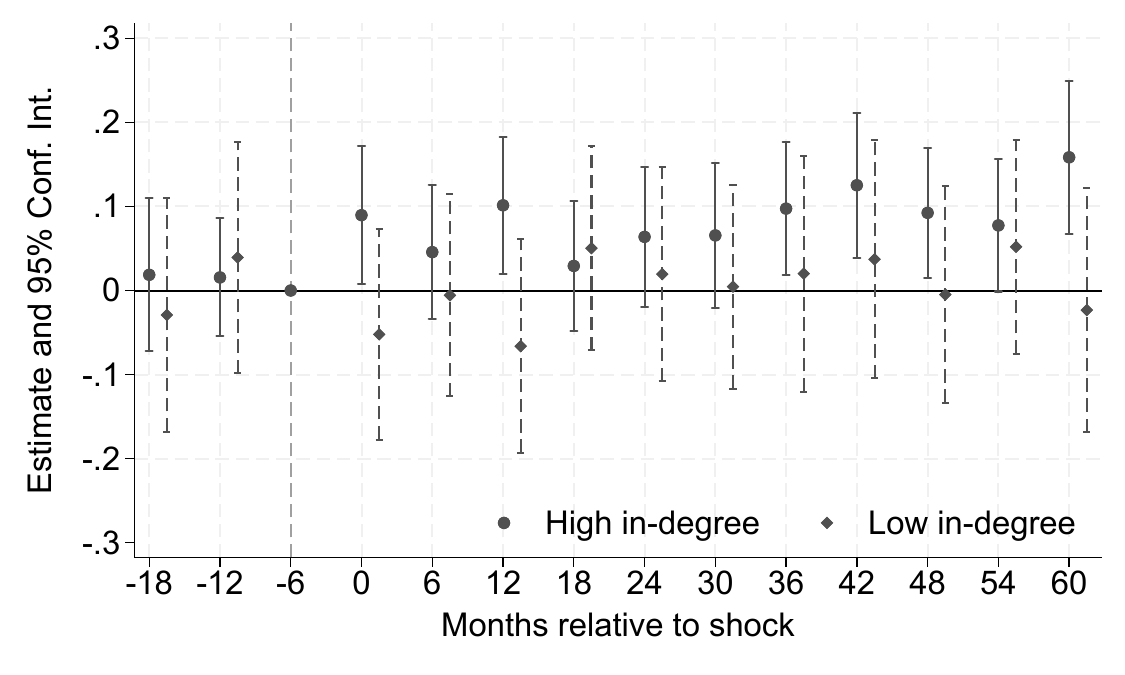} 
\end{center}
\vspace*{-0.2cm}
\caption{Effect of road disruptions on relationship ending}
\vspace*{0.1cm}
\scalebox{0.90}{
\begin{minipage}{1.1\textwidth}
\advance\leftskip 0.5cm
{\footnotesize {
\textit{\textbf{Notes:}} The figure plots the $\beta_l$ coefficients 
from equation~\eqref{Y5e} for the relationship-ending indicator 
($e_{ij,t}$), for all importers (left panel) and by in-degree 
(right panel), with 95\% confidence intervals. Unlike 
Figure~\ref{fig:firm_Exposure}, the sample includes relationships 
formed after the shock. Sample sizes are $N=42{,}391$ in the left 
panel (648 exporter clusters), $N=36{,}584$ for high in-degree 
importers (621 clusters), and $N=5{,}807$ for low in-degree 
importers (382 clusters). All regressions include relationship and 
cohort-by-time fixed effects. Standard errors are clustered at the 
exporter level. } \par }
\end{minipage}
} 
\label{fig:end_matches_all}
\end{figure}

Figure~\ref{fig:end_matches_all} shows results from 
equation~\eqref{Y5e} for the probability of relationship ending, 
$e_{ij,t}$. This analysis differs from equation~\eqref{Y2} in the 
sample: it includes relationships formed after the shock, in 
addition to those active before. The probability of relationships 
ending rises over time (left panel), driven by high in-degree 
importers and consistent with Figure~\ref{fig:firm_Exposure}. For 
high in-degree importers, the coefficient at the shock ($t=0$) shows that a 10 percentage point increase in firm-level exposure raises the probability of a relationship ending by 0.89 percentage points (2.8\% interquantile increase on the pre-shock baseline). The effect remains positive through the medium ($t = 12, 24$) and 
long run ($t=>42$).\\

\begin{figure}[h]
\begin{center}
\hspace{0.25in}
\textbf{\footnotesize{\textsc{All importers}}}
\hspace{2.5in}
\textbf{\footnotesize{\textsc{By in-degree}}}\\
\hspace*{-0.35in}
\includegraphics[height=1.95in]{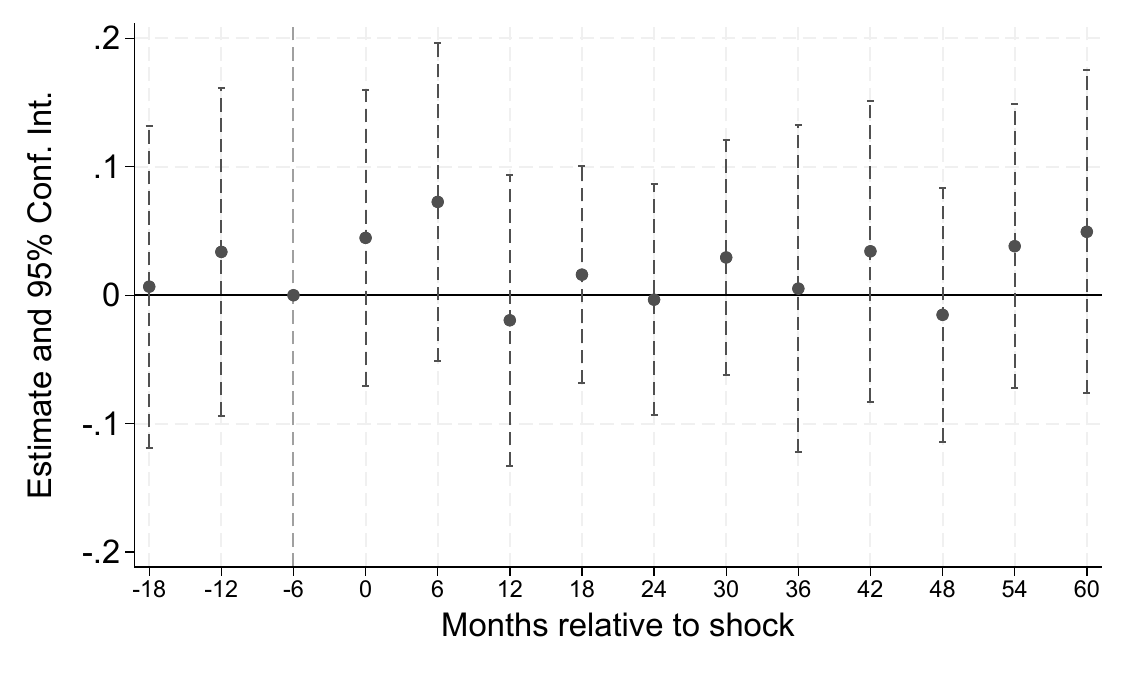}
\includegraphics[height=1.95in]{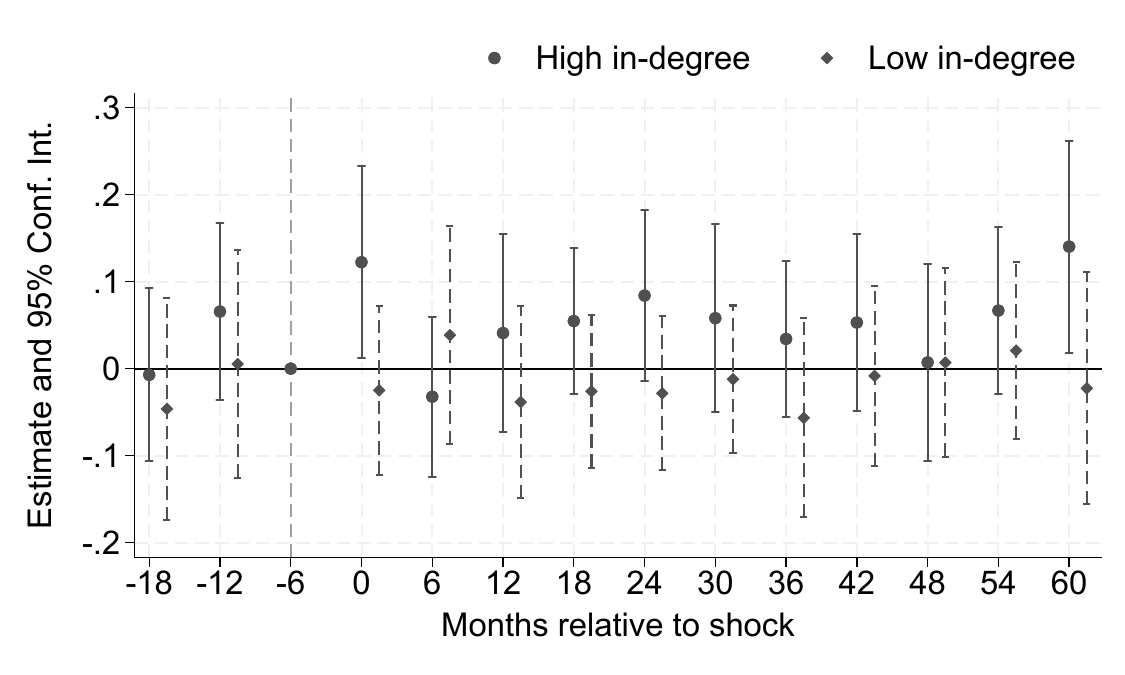} 
\end{center}
\vspace*{-0.2cm}
\caption{Effect of importer-level exposure on the probability of 
importer exit}
\vspace*{0.1cm}
\scalebox{0.90}{
\begin{minipage}{1.1\textwidth}
\advance\leftskip 0.5cm
{\footnotesize{
\textit{\textbf{Notes:}} The figure plots the $\beta_l$ coefficients 
from equation~\eqref{Y6} with 95\% confidence intervals, where the 
outcome is the importer exit indicator ($\text{Exit}_{j,t}$) defined 
in Section~\ref{sec:empirical}. The left panel shows results for 
all importers; the right panel splits the sample by in-degree, with 
the split defined in Section~\ref{sec:empirical}. All regressions 
include importer and time fixed effects. Sample sizes: $N=4{,}688$ 
importer--period observations in the left panel; $N=4{,}168$ for 
high in-degree importers and $N=2{,}155$ for low in-degree importers. High in-degree importers are 396 and low in-degree importers are 228.
 Standard errors are not clustered, as the unit of 
observation is already at the importer level.}\par}
\end{minipage}
}  
\label{fig:firm_Exit}
\end{figure}

\noindent \textbf{Firm exit.}\
Finally, I study whether portfolio contraction translates into 
importer exit. Figure~\ref{fig:firm_Exit} reports estimates from 
equation~\eqref{Y6}. High in-degree importers experience a sharp 
increase in exit probability at the time of the shock ($t=0$): a 
10 percentage point increase in $IE_j$ raises the probability of 
importer exit by 1.22 percentage points. This indicates that an importer moving from the 25$^{th}$ to the 75$^{th}$ percentile of exposure distribution (0.045 to 0.58), is 6.5 percentage points more likely to exit during the crisis. This is consistent with 
the observe patterns of relationship endings among high in-degree 
firms documented in Figure~\ref{fig:firm_Exposure}.\footnote{Estimations using the binary exposure are shown in the Supplemental Appendix.}

Notably, the magnitude of the firm-exit response among high 
in-degree importers coincides with the magnitude of the 
relationship-ending response in Figure~\ref{fig:firm_Exposure} 
(1.22 pp vs.\ 1.26 pp per 10 pp increase in $IE_j$). When exposure 
is widespread, ending a relationship and exiting the market become 
the same margin of adjustment for high in-degree firms --- a corner 
case of the framework I develop next. For low in-degree importers, 
firm-level exposure to the shock does not significantly increase exit probabilities. These 
firms appear more resilient: as documented earlier 
(Figure~\ref{fig:firm_Exposure}), they preserve existing 
relationships and even form new relationships during the shock (Figure~\ref{fig:new_matches}), which keeps them above the exit 
threshold. For high in-degree firms exposed to disruptions that do not exit immediately, 
continue to end relationships in subsequent periods without increasing their relationships. In the medium and long run they are more likely to exit.\\

\noindent \textbf{Aggregate implications.}\
To gauge the aggregate magnitude of the documented effects, I 
combine the estimated coefficients with pre-shock totals from 
Table~\ref{tab:stat}. Three margins matter, all concentrated among 
high in-degree importers. First, the extensive margin at the 
relationship level: applying the importer-exposure coefficient 
($\hat\beta \approx 0.126$) to the average firm exposure ($\bar{IE}_j$= 0.36) 
implies approximately 287 additional relationships terminated 
during the disruption period, a 56\% increase over the baseline 
of 506 expected terminations among high in-degree 
relationships.\footnote{Additional endings: 
$ \hat\beta \times \hat{IE}_j \times \bar{N}_j \times N_{\text{firms}} 
= 0.126 \times 0.36 \times 16 \times 396 = 287$, where $\bar{N}_j=16$ 
is the average number of relationships per high in-degree importer 
and $N_{\text{firms}}=396$ is the total number of high in-degree 
importers in the pre-shock exposure sample.  The 
baseline is $0.167 \times N_{\text{rel}}^{\text{high}} = 0.167 
\times 3{,}034 = 506$, where $N_{\text{rel}}^{\text{high}}=3{,}034$ 
is the total number of relationships held by high in-degree 
importers in the pre-shock period.}

Second, the firm-level extensive margin: the same calculation 
applied to firm exit implies approximately 17 high in-degree 
importers exit at the time of the shock, representing 4.3\% of 
this group.\footnote{Additional exits: 
$\hat\beta \times \bar{IE}_j \times N_{\text{firms}} = 0.122 
\times 0.36 \times 396 = 17$.}

Third, the intensive margin: the 6--16\% price decline among 
surviving exposed relationships, applied to their pre-shock trade 
value, corresponds to roughly \$11M--30M US dollars in foregone exporter revenue 
per year. Combined with the extensive-margin loss of approximately \$44M US dollars per year in exporter revenue, the total annual loss is roughly \$55M–74M US dollars. This loss is sustained over the two post-shock years during which the effects remain statistically significant, and is equivalent to 5–7\% of the sector's 2010 U.S.\ sales of approximately \$1B US dollars.\footnote{To estimate the annual value 
of a relationship I use Table~\ref{tab:stat} Panel B. Taking the 
average monthly trade value of \$32.5K and the transaction 
frequency of 0.39 (months out of 6 with a positive transaction), 
the annual relationship value is $0.39 \times 6 \times 2 \times 
32.5K \approx \$152K$, where the factor $0.39 \times 6 \times 2$ 
converts the per-transacting-month average to an annual value. 
For the intensive margin, I use the 1,555 exposed relationships 
from Panel A and the 6--16\% price decline from 
Figure~\ref{fig:units}, subtracting the 287 lost relationships 
to obtain the surviving exposed pool: $(1{,}555 - 287) \times 
(\hat{\beta}=0.06-0.16) \times \$152K \approx \$11.5M-30.8M$. For the extensive margin: 
$287 \times \$152K \approx \$43.6M$. This calculation approximates 
the lost-relationship pool with high in-degree terminations, which 
understates the total if low in-degree terminations.}

Together, these calculations indicate that the relationship- and 
firm-level adjustments translate into economically meaningful 
aggregate effects.

\section{Theoretical framework}
\label{sec:theory}

This section develops a deliberately stylized framework to organize the empirical findings. The goal is not to provide a full structural model, but to clarify why the same shock generates opposite responses at the relationship and firm levels --- a pattern that is not immediately intuitive. The key insight is that when a firm holds a portfolio of relationships, the surplus from any single relationship depends on what is happening across the rest of the portfolio. Road disruptions, by simultaneously affecting multiple links within a firm, activate two forces that operate in opposite directions. I label these the \emph{indirect} and \emph{direct} effects, and show that which force dominates depends on how broadly exposure is distributed across a firm's relationships.\\

\noindent\textbf{Setting and assumptions.}\ Throughout this section, the terms \textit{relationship} and \textit{match} are used interchangeably to refer to a buyer--seller pair; \textit{link} is used synonymously in the network analysis above. The market comprises two types of firms: domestic sellers (exporters) and foreign buyers (importers). I assume that markets clear in each period $t$, so final demand and prices in the US are taken as given. Buyers and sellers meet randomly in $t$ and, once matched, choose prices and quantities optimally and trade in both $t$ and $t+1$.\footnote{As in \citet*{Macciavello2015}, buyers and sellers form optimal contracts each period. As in \citet*{Eaton2022}, I assume efficient bargaining so that buyers and sellers split the surplus.}

The discount factor is unity, so revenues and costs across both periods are known and constant given that final demand for each buyer is the same. Revenue is a concave function of quantities, $R'(Q)>0$ and $R''(Q)<0$, and costs are convex, $C'(Q)>0$ and $C''(Q)>0$.

Relationship revenue net of production costs is always positive and exceeds the maintenance cost $\lambda_x$. Firms also pay a formation cost $\rho_x$ (where $x$ denotes importers or exporters) whenever they establish a new match.\footnote{Formation costs can be conceptualized as search costs arising from firms' incomplete information about potential partners, as in \citet*{aker2010}, \citet*{allen2014}, and \citet*{Goyal2010}. In \citet{Monarch2014}, \citet*{Alessandria2009}, \citet*{Drozd2012}, and \citet{Rauch2003}, the information friction is on the buyer side. The maintenance cost $\lambda$ reflects fixed costs such as account maintenance, technical support, and client-specific product adjustments, as in \citet*{Eaton2022}. \citet{BernardMox2018b} also model relationship-specific costs paid in labor units by the seller.} The key assumption is that maintaining an existing relationship is always cheaper than replacing it: $\rho_x > \lambda_x$.\\

\noindent\textbf{Firm decisions.}\ Firms make two decisions in period $t$: whether to keep or drop an existing relationship (denoted $e_{ij}=1$ for ending and $e_{ij}=0$ for keeping), and whether to add a new match to their portfolio for $t+1$ (denoted $n_{ij}=1$ for forming a new link and $n_{ij}=0$ otherwise). Keeping a relationship requires the maintenance cost $\lambda_x$; adding a new match requires the formation cost $\rho_x$, paid at $t$. In $t+1$, firms collect the payoffs from their period-$t$ choices without making further decisions.

I focus on the relationship surplus for the importers, since this is the active margin of adjustment and is also the margin targeted by the empirical estimations. The four possible combinations of decisions are: (i)~keep the existing relationship and do not add a new match ($e_{ij}=0,\, n_{ij}=0$); (ii)~drop the relationship and form a new match ($e_{ij}=1,\, n_{ij}=1$); (iii)~keep the relationship and add a new match ($e_{ij}=0,\, n_{ij}=1$); and (iv)~drop the relationship without forming a new one ($e_{ij}=1,\, n_{ij}=0$). The profits of importer $j$ in period $t$, given these decisions, are
\begin{equation}
\begin{split}
    \pi_{t}^j= &
    \, \mathlarger{\mathbb{1}}\{e_{ij}=0\}\, \mathlarger{\mathbb{1}}\{n_{ij}=0\} \,\big(R(S_{j,-i}+q_{ij})-C(S_{j,-i}+q_{ij}) - N \lambda_j\big) \;     \\
  &  +\mathlarger{\mathbb{1}}\{e_{ij}=1\}\,\mathlarger{\mathbb{1}}\{n_{ij}=1\} \,\big(R(S_{j,-i})-C(S_{j,-i}) -(N-1)\lambda_j -\rho_j\big) \;  \\
   &  +\mathlarger{\mathbb{1}}\{e_{ij}=0\}\,\mathlarger{\mathbb{1}}\{n_{ij}=1\} \,\big(R(S_{j,-i}+q_{ij})-C(S_{j,-i}+q_{ij}) - N\lambda_j -\rho_j\big)  \\
    &  +\mathlarger{\mathbb{1}}\{e_{ij}=1\}\,\mathlarger{\mathbb{1}}\{n_{ij}=0\} \,\big(R(S_{j,-i})-C(S_{j,-i}) -(N-1)\lambda_j\big), \nonumber  \\
    \end{split}
\end{equation}
where $\mathbb{1}\{e_{ij}=0\}$ indicates that the firm keeps the relationship and $\mathbb{1}\{n_{ij}=1\}$ indicates that it adds a new match.\footnote{Time subscripts are suppressed since final demand is identical across periods: $Q_{ij,t}=Q_{ij,t+1}$ and therefore $q_{ij,t}=q_{ij,t+1}$.} $S_{j,-i}$ is the total quantity importer $j$ buys from all relationships other than $i$. Importers bear a fraction of the maintenance cost $\lambda_j$, which is assumed to be the same across all relationships. $N$ is the total number of active relationships in the firm's portfolio.

Profits in $t+1$ following the decisions of $j$ in $t$ are
\begin{equation}
\begin{split}
    \pi_{t+1}^j= & \;
  \mathlarger{\mathbb{1}}\{e_{ij}=0\}\, \mathlarger{\mathbb{1}}\{n_{ij}=0\}\,\big(R(S_{j,-i}+q_{ij})-C(S_{j,-i}+q_{ij})\big) \;    \\
  & + \mathlarger{\mathbb{1}}\{e_{ij}=1\}\,\mathlarger{\mathbb{1}}\{n_{ij}=1\}\, \big(R(S_{j,-i}+q_{kj})-C(S_{j,-i}+q_{kj})\big) \;    \\
 & + \mathlarger{\mathbb{1}}\{e_{ij}=1\} \, \mathlarger{\mathbb{1}}\{n_{ij}=0\}\,\big(R(S_{j,-i})-C(S_{j,-i})\big)\; \\
 &  + \mathlarger{\mathbb{1}}\{e_{ij}=0\}\,\mathlarger{\mathbb{1}}\{n_{ij}=1\}\, \big(R(S_{j,-i}+\omega q_{ij}+ (1-\omega) q_{kj})  \\
   & \qquad -C(S_{j,-i}+\omega q_{ij}+ (1-\omega) q_{kj})\big), \nonumber    \\
    \end{split}
\end{equation}
where $q_{kj}$ denotes the quantity from the new match $k$. When a firm keeps its existing relationship and adds a new one simultaneously, it can supply only a fraction $\omega$ of contracted demand through the new partner, since total final demand is fixed and contractually determined at $t$.\\

\noindent\textbf{Relationship surplus.}\ I define relationship surplus at the firm level with respect to a particular $ij$ relationship. Comparing maintaining the current relationship without adding a new match ($e_{ij}=0,\, n_{ij}=0$) against dropping it and forming a new one ($e_{ij}=1,\, n_{ij}=1$) yields the surplus from maintaining a relationship:
\begin{equation}
\begin{split}
\Delta\pi^j(S_{j,-i};q_{ij}) = \big(R^j(S_{j,-i}+q_{ij}) - C^j(S_{j,-i}+q_{ij}) - R^j(S_{j,-i}) + C^j(S_{j,-i})\big) + \rho_j - \lambda_j > 0,
    \end{split}
    \label{YC}
\end{equation}
where the first four terms capture the net revenue gain from trading $q_{ij}$ units --- the additional profit $j$ earns from this relationship relative to not having it. The last term, $\rho_j - \lambda_j > 0$, is positive by assumption: maintaining the relationship is always less costly than replacing it. This is consistent with search being unattractive when there are no gains from additional links (under fixed final demand).\footnote{The surplus from the other combinations follows the same logic and is derived in Appendix~\ref{sec:math-appendix_surplus}. The remaining comparisons involve either $\lambda_j$ or $\rho_j$ alone, which makes them less informative.}\\

\noindent\textbf{Comparative statics.}\ Under road disruptions, some transactions are not delivered, reducing $q_{ij}$ for any $i$ in the firm's portfolio ($S_{j,-j}+q_{ij}$). The economically relevant comparison under a disruption is between ($e_{ij}=0,\, n_{ij}=0$) and ($e_{ij}=1,\, n_{ij}=1$): keeping the disrupted relationship versus terminating it and forming a new match, since the sign of \eqref{YC} can reverse with the magnitude of the shock. Given this sign reversal, two distinct effects can rationalize firms' decisions to keep or end a relationship.

The relationship surplus in \eqref{YC} can be decomposed into two channels. The \emph{indirect} effect operates through $S_{j,-i}$: disruptions to other relationships in the firm's portfolio reduce aggregate delivered quantities, changing the marginal value of the focal relationship. The \emph{direct} effect operates through $q_{ij}$: a disruption within the focal relationship itself reduces the quantities delivered and hence the net surplus it generates.\\

\noindent\textbf{Effect 1: Indirect channel.}\ \textit{A disruption that reduces deliveries in a firm's other relationships increases the surplus from any given relationship.}\\
When $S_{j,-i}$ falls, the marginal revenue of each unit delivered through the focal relationship rises, since $R''<0$. Simultaneously, the marginal cost of those units falls, since $C''>0$. Both forces raise the net surplus from maintaining the focal relationship. Formally, $\partial \Delta\pi^j(q_{ij}) / \partial S_{j,-i} < 0$: a deterioration in the rest of the portfolio (a decrease in $S_{j,-i}$) raises the value of keeping any relationship that continues to deliver.\footnote{The proof is in Appendix~\ref{sec:math-appendix-p1}.}

The intuition maps directly to the empirical setting. When some of an importer's suppliers are disrupted but others are not, the remaining non-disrupted relationships become relatively more valuable. Importers therefore have a stronger incentive to preserve those existing links and absorb temporary delivery shortfalls rather than churn their portfolio. Crucially, this force is stronger when the overall share of exposed relationships is low --- so that there are still many non-disrupted links through which portfolio re-balancing can occur.\\

\noindent\textbf{Effect 2: Direct channel.}\ \textit{A disruption that reduces deliveries within a relationship decreases its surplus. The relationship is maintained if and only if the net replacement cost exceeds the profit loss induced by the disruption.}\\
The relationship surplus remains positive after a reduction in $q_{ij}$ --- and the relationship is preserved --- if and only if
\begin{equation}
\underbrace{\rho_j - \lambda_j}_{\text{replacement cost}} > \; \underbrace{MR^j(S_{j,-i}+q_{ij}) - MC^j(S_{j,-i}+q_{ij})}_{\text{profit loss}}. \nonumber
\end{equation}
If the replacement cost for relationship $ij$ exceeds the relationship's profit loss, the firm prefers to maintain the disrupted relationship and wait for deliveries to resume. When the loss exceeds the cost of replacement, the inequality reverses if: the relationship surplus becomes negative and dominates the \emph{indirect} effect. Under these conditions the relationship is terminated.\footnote{The proof is in Appendix~\ref{sec:math-appendix-p2}.}

The critical determinant is therefore not the disruption per se, but the \emph{relative} magnitudes of replacement costs and profit losses. In settings like the flower sector, where relationship-specific knowledge and product customization make forming new matches costly and slow, replacement costs tend to be high. The condition is therefore likely to bind for most relationships --- unless the disruption is severe enough that delivered quantities collapse entirely.\\

\noindent\textbf{From results to empirical evidence.}\ The two results jointly rationalize the heterogeneous responses documented in the empirical analysis. Whether the \textit{indirect} or \textit{direct} effect dominates depends on the share of a firm's portfolio that is exposed.

When exposure is limited (a low share of disrupted relationships within the portfolio), the indirect effect dominates. Disruptions to other relationships raise the marginal value of any link that continues to function, and firms optimally preserve disrupted matches in anticipation of resumed deliveries. This is consistent with Figure~\ref{fig:main} (right panel): conditional on the importer, exposed relationships are less likely to end at the time of the shock.

When exposure is widespread (a high share of disrupted relationships), the \textit{direct} effect dominates, as profit losses accumulate across the portfolio. With many simultaneously disrupted links, the \textit{indirect} channel --- which depends on the existence of well-functioning alternatives --- weakens. For high in-degree importers, this manifests as both relationship termination (Figure~\ref{fig:firm_Exposure}) and firm exit (Figure~\ref{fig:firm_Exit}).

Low in-degree importers respond differently. Because their portfolios are smaller, ending a single relationship would force immediate exit; consequently, they preserve existing matches and instead pay the search cost $\rho_x$ to form new ones (Figure~\ref{fig:new_matches}). The framework rationalizes this through the relative magnitudes of the replacement cost and the disruption-induced loss: when the firm has few links, the option to absorb temporary delivery shortfalls and search elsewhere dominates termination.

The framework is deliberately stylized and abstracts from several margins observed in the data, most notably delayed shipments and shipment reorganization within existing relationships. These intensive-margin responses operate within the period of disruption and are not modeled here. The framework therefore captures firms' extensive-margin decisions over relationships, but not the full set of short-run adjustments used to absorb temporary supply shocks. A more complete model incorporating dynamic inventory management or endogenous search would be a natural extension, but lies beyond the scope of this paper.

\section{Conclusion}
\label{sec:conclusion_c1}

This paper studies how temporary, relationship-specific supply disruptions affect the organization of international trade. In environments where relational contracts underpin market access, shocks that impair individual buyer--seller matches can generate effects that extend well beyond the directly affected transactions. Understanding these dynamics is increasingly important as climate-related, political, and trade-policy disruptions become more frequent and severe. The degree to which firms can absorb such disruptions without unwinding long-standing relationships has direct implications for the resilience of global value chains and the design of policies aimed at supporting firms through crises.

The central finding is that the same shock generates heterogeneous responses depending on portfolio composition: temporary disruptions can simultaneously increase the survival of individual relationships, when portfolio exposure is limited, and accelerate firm-level exit, when exposure is widespread. The difference between these two results depends on pre-shock connections.

High in-degree importers --- those with many pre-shock relationships --- are the firms for whom widespread exposure binds. Their adjustment combines accelerated relationship termination, market exit at the time of the shock, and a persistent decline in new relationship formation. The shock leaves them with smaller portfolios in the medium run, with no recovery toward their pre-shock trajectory. Low in-degree importers behave on the opposite margin: they preserve existing relationships during the disruption and respond instead by forming new matches, consistent with replacement costs being large relative to a single relationship's disruption-induced loss for less-connected buyers.

The results highlight the role of portfolio composition in firm responses to temporary disruptions: resilience depends not only on the severity of the shock, but also on the structure of firms' trading networks. Policies that aim to support firms through climate-related or other transient disruptions would benefit from targeting firms whose portfolio composition leaves them most exposed to widespread shocks, rather than treating exposure uniformly.

\printbibliography

\pagebreak
\appendix
\newpage
\section{Figure appendix}
\label{sec:figure-appendix_c1}

\renewcommand{\thefigure}{B.\arabic{figure}}
\setcounter{figure}{0}
\renewcommand\theHfigure{Appendix.\thefigure}

\begin{figure}[H]
\centering
    \includegraphics[height=2.5in]{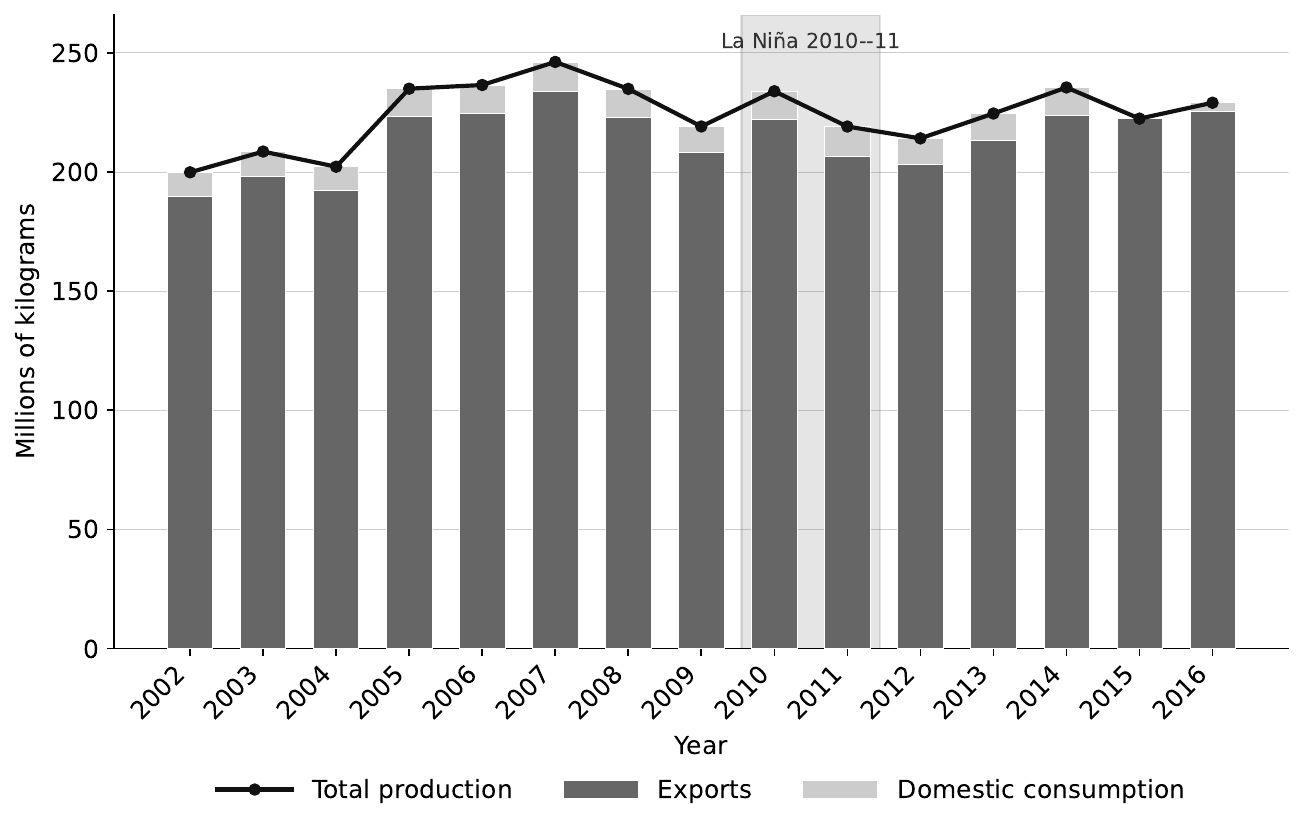}

 \vspace*{-0.2in}
    \caption{Production and exports of cut flowers}
      \vspace*{0.1in}
         \scalebox{0.90}{
\begin{minipage}{1.1\textwidth}
\advance\leftskip 0.5cm
	{\footnotesize {\textit{\textbf{Source}}: MADR - ICA (2016). \\
 \textit{\textbf{Notes:}} The figure plots the total 
kilograms of flowers produced in Colombia, disaggregated by 
end use: exports (dark grey bars) and domestic consumption 
(light grey bars). The black line plots total production.} \par }
\end{minipage}
} 
    \label{fig:production}
\end{figure}

\begin{figure}[H]
\centering

     \hspace*{0.4in}\includegraphics[height=3in]{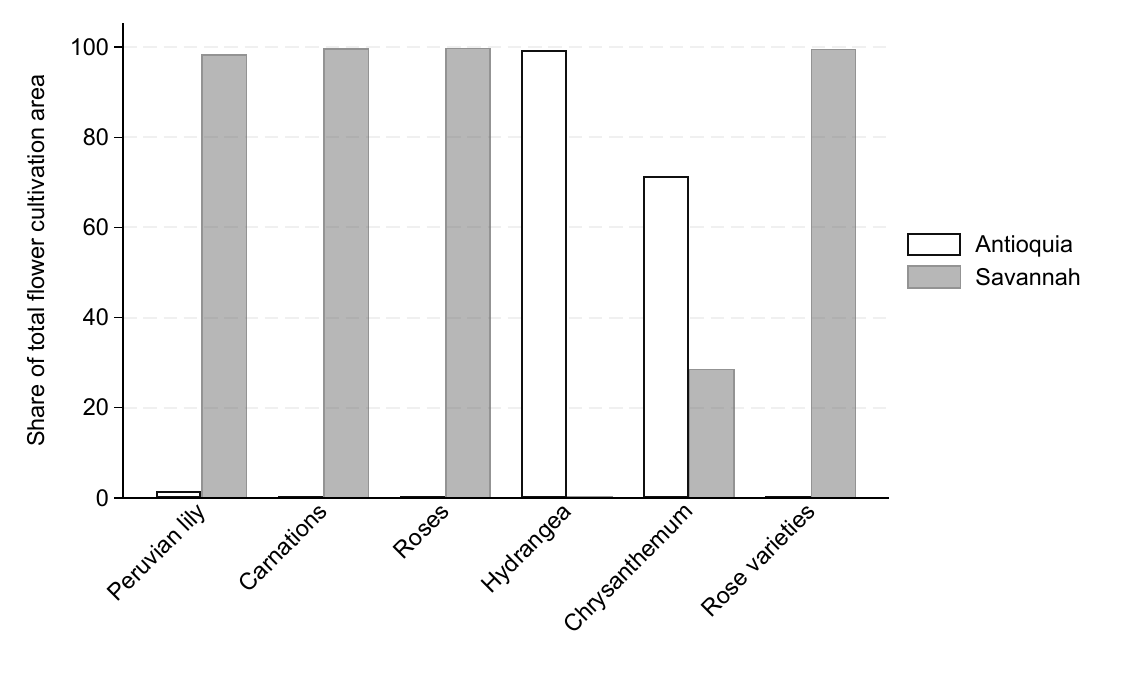}   
\vspace*{-0.2in}
\caption{Flower species region of cultivation}
\vspace*{0.1in}     
     \scalebox{0.90}{
\begin{minipage}{1.1\textwidth}
\advance\leftskip 0.5cm
	 {{\footnotesize{  \textit{\textbf{Source:}}  Colombian Association of Flower Exporters (Asocolflores). \\ \textit{\textbf{Notes:}} The figure plots the share of land used for production (in hectares) for each type of flower in the Savannah and Antioquia regions. The remaining percentages represent production from other regions. Production data from 2016. The flowers shown account for 85\% of the total land used for flower production.}\par } }
\end{minipage}
}
\label{fig:production_flower}
\end{figure}

\begin{figure}[H]
    \centering
 \vspace*{0.2in}   
   \includegraphics[width=0.8\textwidth]{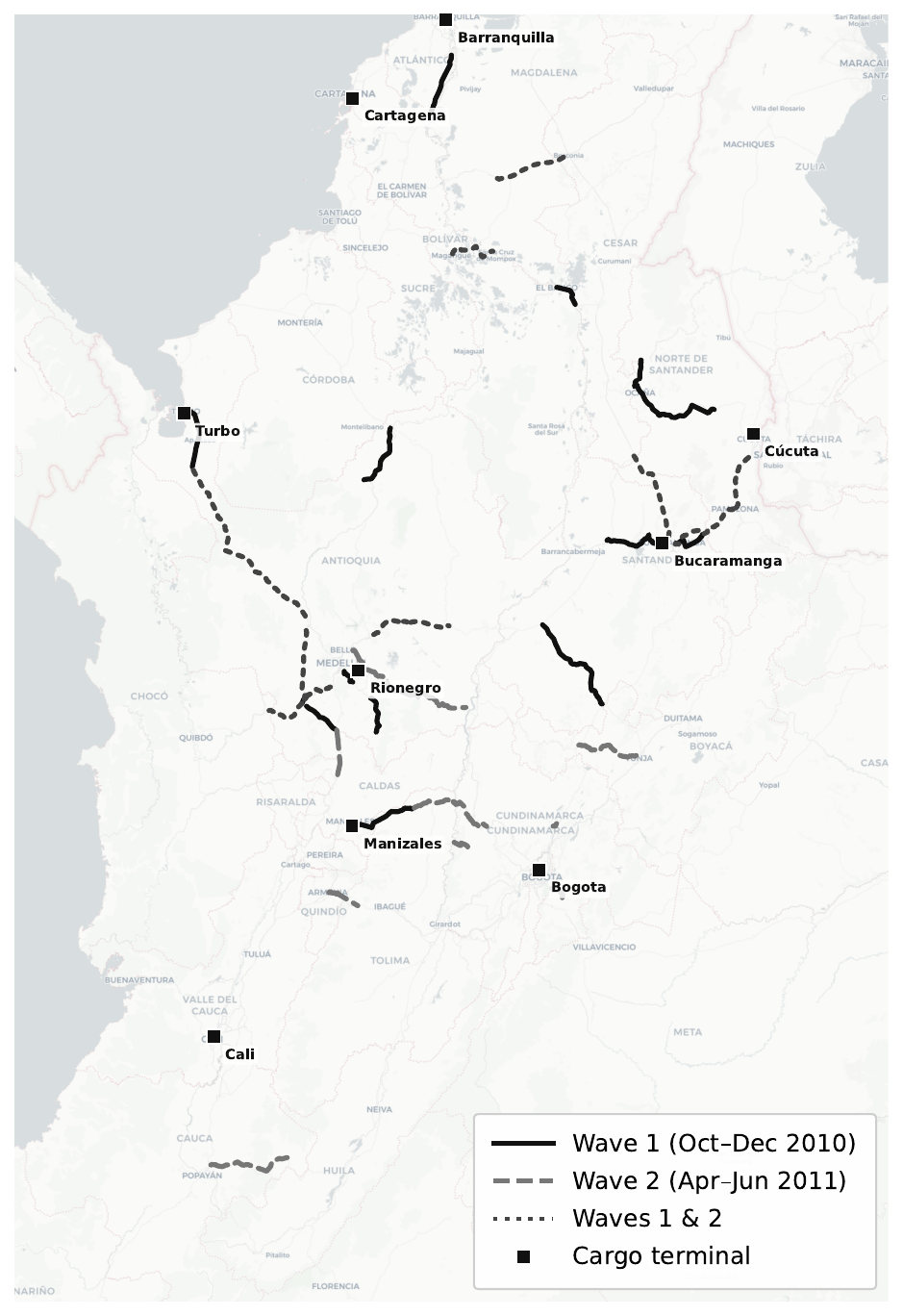}
         \vspace*{-0.2in}
        \caption{Road disruptions and cargo terminals affecting flower exporters 2010--11}

     \vspace*{0.1in}
     \scalebox{0.90}{
\begin{minipage}{1.1\textwidth}
\advance\leftskip 0.5cm
	{\footnotesize {\textit{\textbf{Source}}: CARTO (CartoDB Positron), © OpenStreetMap contributors. \\ \textit{\textbf{Notes}}: The map shows road disruptions affecting flower exporters during the 2010–-11 La Niña episode, by wave, along with cargo terminals used prior to the first wave. Wave 1 corresponds to October–-December 2010, and wave 2 to April-–June 2011. Roads disrupted in both waves are indicated accordingly.} \par }
\end{minipage}
} 
    \label{fig:road_disrupted}
\end{figure}

%%%%%%%%%%% ROBUSTNESS

\begin{figure}[H]
 \begin{center}
    \hspace{0.25in}
\textbf{\footnotesize{\textsc{A. Kilograms}}}
\hspace{2.5in}
\textbf{\footnotesize{\textsc{B. Trade value }}}\\
\hspace*{-0.35in}
\includegraphics[width=0.5\linewidth]{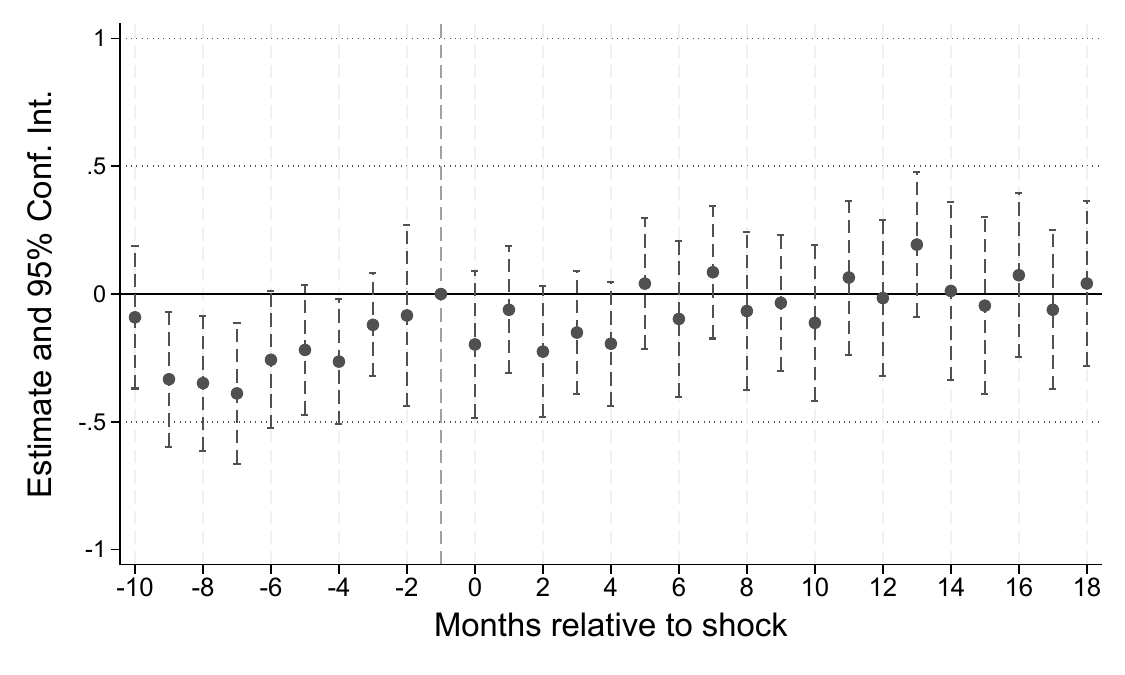}
\includegraphics[width=0.5\linewidth]{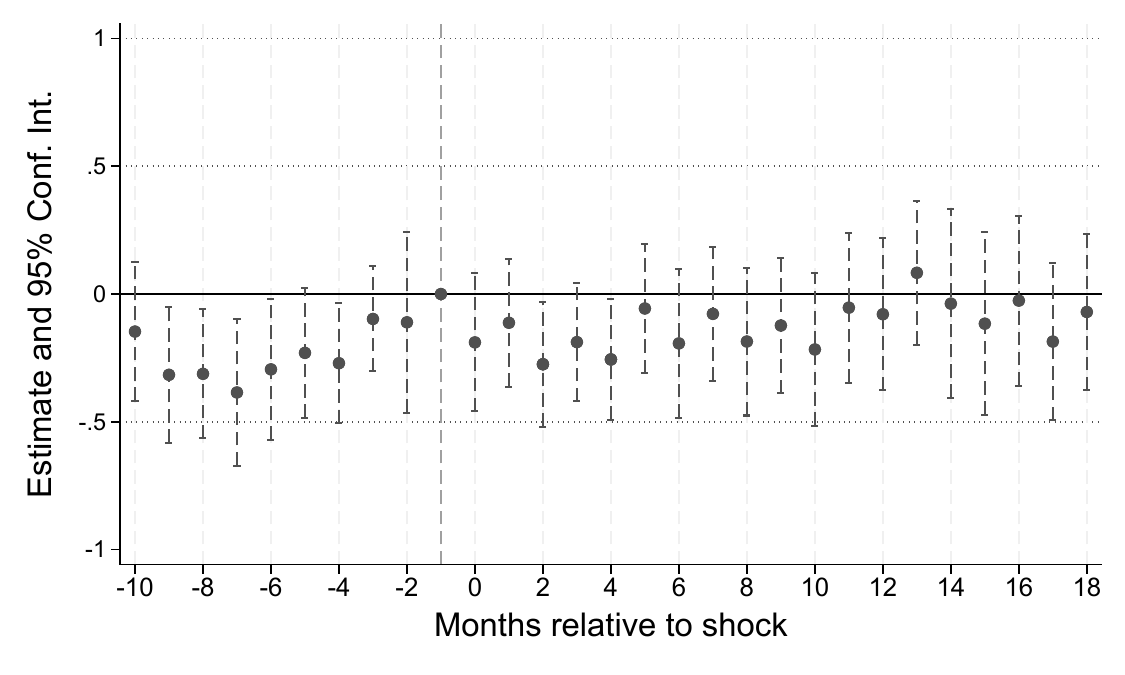}
 \end{center}
    \caption{Robustness: ($\Gamma = \theta_{ij} + \delta_{p,m})$}
      \label{fig:rob_kilogram}
        \vspace*{0.1in}
\scalebox{0.90}{
\begin{minipage}{1.1\textwidth}
\advance\leftskip 0.5cm
	{\footnotesize{
\textit{\textbf{Notes:}} All panels plot the 
$\beta_l$ coefficients from equation~(\ref{Y4}) 
with 95\% confidence intervals. All regressions
include relationship, product-month and cohort-month fixed effects. Standard errors are clustered at 
the exporter level. } \par}
\end{minipage}
}  
\end{figure}

\newpage
\section{Mathematical appendix}
\label{sec:math-appendix}

\subsection{Relationship surplus: additional cases}
\label{sec:math-appendix_surplus}

\noindent\textit{Keeping and not  matching ($e=0, n = 0$) compared to  not keeping and not matching ($e=1, n= 0$)}.\ This yields 
\begin{align}
 & \Delta(\pi^{j})=\Delta(e = 0,n = 0) - \Delta(e=1,n=0) = \nonumber \\
 & R^j(S_{j,-i}+q_{ij})-R^j(S_{j,-i}) -C^j(S_{j,-i}+q_{ij}) + C^j(S_{j,-i}) - \lambda_j >0 \nonumber
\end{align}
Since \[R^j(S_{j,-i}+q_{ij})>R^j(S_{j,-i}), \]
and \[C^j(S_{j,-i}+q_{ij})>C^j(S_{j,-i}). \]

 The inequality can only hold if cost of maintenance is lower than the net revenue of $\Delta \pi^j(q_{ij})-\lambda_j>0$. \\

\noindent\textit{Keeping and  not matching ($e=0, n = 0$) compared to keeping and  matching ($e=0, n= 1$)}.\ 
This yields 
\[
 \Delta(\pi^{j})= \Delta(e = 0,n = 0) - \Delta(e=0,n=1) = \lambda_j >0
\]

Which is always positive since cost of keeping a relationships is assumed positive.\\ 

\noindent\textit{Not keeping and  matching ($e=1, n = 1$) compared to  not keeping and not matching ($e=1, n= 0$)}.\ This yields 
\begin{align}
 &\Delta(\pi^{j})=\Delta(e = 1,n = 1) - \Delta(e=1,n=0) = \nonumber\\
 & R^j(S_{j,-i}+q_{ij})-R^j(S_{j,-i}) -C^j(S_{j,-i}+q_{ij}) + C^j(S_{j,-i})  - \rho_j >0 \nonumber
\end{align}
Since \[R^j(S_{j,-i}+q_{ij})>R^j(S_{j,-i}), \]
and \[C^j(S_{j,-i}+q_{ij})>C^j(S_{j,-i}). \]

 The inequality can only hold if cost of search is lower than the net revenue of $\Delta \pi^j(q_{ij})-\rho_j>0$. \\

\noindent\textit{Not keeping and  matching ($e=1, n = 1$) compared to   keeping and  matching ($e=0, n= 1$)}.\ This yields 
\begin{align}
 & \Delta(\pi^{j})=\Delta(e = 0,n = 1) - \Delta(e=1,n=1) = \nonumber \\
 & R^j(S_{j,-i}+q_{ij})-R^j(S_{j,-i}) -C^j(S_{j,-i}+q_{ij}) + C^j(S_{j,-i}) - \lambda_j >0, \nonumber
\end{align}
Since \[R^j(S_{j,-i}+q_{ij})>R^j(S_{j,-i}), \]
and \[C^j(S_{j,-i}+q_{ij})>C^j(S_{j,-i}). \]

 The inequality can only hold if cost of maintenance is lower than the net revenue of $\Delta \pi^j(q_{ij})-\lambda_j>0$. \\

\noindent\textit{Not keeping and  not matching ($e=0, n = 1$) compared to  keeping and  matching ($e=0, n= 1$)}.\ This yields 
\begin{align}
 &\Delta(\pi^{j})=\Delta(e = 0,n = 1) - \Delta(e=1,n=0) = \nonumber \\ & 2\bigg(R^j(S_{j,-i}+q_{ij})-R^j(S_{j,-i}) -C^j(S_{j,-i}+q_{ij}) + C^j(S_{j,-i})\Bigg)  - \rho_j - \lambda_j >0. \nonumber
\end{align}
Since \[R^j(S_{j,-i}+q_{ij})>R^j(S_{j,-i}), \]
and \[C^j(S_{j,-i}+q_{ij})>C^j(S_{j,-i}). \]

 The inequality can only hold if costs are lower than the net revenue of $\Delta \pi^j(q_{ij})-\rho_j -\lambda_i>0$. \\

\subsection{Result 1}
\label{sec:math-appendix-p1}
 \begin{proof}

Revenues are concave in $Q$, with $MR^j: \frac{\partial{R(.)}}{\partial{Q}}>0$ and $\frac{\partial^2{R(.)}}{\partial{Q}}<0$, and that costs are convex in $Q$, with $MC^j: \frac{\partial{C(.)}}{\partial{Q}}>0$ and $\frac{\partial^2{C(.)}}{\partial{Q}}>0$.
Given these convexity and concavity assumptions, the following holds:
\[\frac{\partial{R(S_{j,-i}+q_{ij})}}{\partial{S_{j,-i}}}<\frac{\partial{R(S_{j,-i})}}{\partial{S_{j,-i}}};
\]
as well as
\[
\frac{\partial{C(S_{j,-i}+q_{ij})}}{\partial{S_{j,-i}}}>\frac{\partial{C(S_{j,-i})}}{\partial{S_{j,-i}}}.
\]

\[\Big(R^j(S_{j,-i}+q_{ij}) - C^j(S_{j,-i}+q_{ij})\Big) < \Big( R^j(S_{j,-i}) - C^j(S_{j,-i})\Big).
\]

 \noindent A decrease in $S_{j,-i}$  and $\rho_j-\lambda_j>0$, follows that:
\begin{align}
 &\Delta(\pi^{j})=\Delta(e = 0,n = 0) - \Delta(e=1,n=1) = \nonumber\\
 & \underbrace{R^j(S_{j,-i}+q_{ij}) - R^j(S_{j,-i}) -C^j(S_{j,-i}+q_{ij}) + C^j(S_{j,-i})}_{(+)} \underbrace{- \lambda_j+\rho_j}_{(+)} >0 \nonumber
    \end{align}

\end{proof}

\subsection{Result 2}
\label{sec:math-appendix-p2}
  \begin{proof}
I assume revenues are concave in $Q$, with $\frac{\partial{R(.)}}{\partial{Q}}>0$ and $\frac{\partial^2{R(.)}}{\partial{Q}}<0$, and that costs are convex in $Q$, with $\frac{\partial{C(.)}}{\partial{Q}}>0$ and $\frac{\partial^2{C(.)}}{\partial{Q}}>0$. Given these convexity and concavity assumptions, the following holds:
\[\frac{\partial{R(S_{j,-i}+q_{ij})}}{\partial{q_{ij}}}>\frac{\partial{R(S_{j,-i})}}{\partial{q_{ij}}};\]
as well as
\[
\frac{\partial{C(S_{j,-i}+q_{ij})}}{\partial{q_{ij}}}>\frac{\partial{C(S_{j,-i})}}{\partial{q_{ij}}}.
\]
A decrease in $q_{ij}$ therefore results in
\[\Big(R^j(S_{j,-i}+q_{ij}) -  C^j(S_{j,-i}+q_{ij}) \Big) < R^j(S_{j,-i}) -  C^j(S_{j,-i}).
\] 
Assuming that $\rho_j-\lambda>0$ the effect on the surplus is then
\begin{itemize}
\item[(i)] negative when
\[\Big(R^j(S_{j,-i}+q_{ij}) -  C^j(S_{j,-i}+q_{ij}) - R^j(S_{j,-i}) +  C^j(S_{j,-i})\Big)> \rho_j-\lambda_j;
\] 
\item[(ii)] positive when 
\[\Big(R(S_{j,-i}+q_{ij}) -  C(S_{j,-i}+q_{ij}) - R(S_{j,-i}) +  C(S_{j,-i})\Big)< \rho_j-\lambda_j.
\] 
\end{itemize}
\end{proof}

\begin{comment}

\noindent\textit{Keeping and  not matching ($e=0, n = 0$) compared to not keeping and matching ($e=1, n = 1$) }.\ The relationship surplus is 
\begin{align}
 &\Delta(\pi^{j})=\Delta(e = 0,n = 0) - \Delta(e=1,n=1) = \nonumber\\
 & R^j(S_{j,-i}+q_{ij}) - R^j(S_{j,-i}) -C^j(S_{j,-i}+q_{ij}) + C^j(S_{j,-i}) - \lambda_j+\rho_j >0 \nonumber
    \end{align}

Since \[R^j(S_{j,-i}+q_{ij})>R^j(S_{j,-i}) \]
and \[C^j(S_{j,-i}+q_{ij})>C^j(S_{j,-i}) \].

And $\lambda_j<\rho_j$ by assumption.  \\
\end{comment}
\newpage
\section{Supplemental Appendix}
\subsection{Classification of road disruptions}
\setcounter{page}{1}
\pagenumbering{arabic}

This section describes the methodology used to classify road disruptions and construct routes to cargo terminals. Routes are estimated using the municipality centroid or main city of each farm as the origin. I rely on the SICE-TAC software provided by the Colombian Ministry of Transport, which reports routes supplied by transportation service companies operating inter-municipal freight services. As a baseline, I assume a two-axle truck with a container trailer and include one hour of waiting time for both loading and unloading.

The software provides route information between specific origin--destination pairs, primarily connecting farms to distribution centers and cargo terminals in major cities such as Bogotá and Medellín, as well as secondary cities. For each route, SICE-TAC reports travel time, distance, cost per kilometer, and the precise path taken, including tolls. When multiple feasible routes are available for a given origin--destination pair, all alternatives are retained.

For expositional simplicity, road disruptions are grouped into two waves: October 2010 to April 2011, and May 2011 to June 2011. For each disruption episode, I first estimate benchmark routes using the standard SICE-TAC configuration for inter-municipal freight transport. Routing is conducted for all possible origin--cargo terminal pairs and repeated whenever a disruption occurs.

When SICE-TAC does not report a feasible route, or when additional comparable routes are available, I supplement the analysis using Google Maps. These alternative routes are selected to be similar in distance and travel time and reflect feasible options exporters could plausibly use to reach cargo terminals. All such routes are treated as potential benchmark routes when classifying exposure to road disruptions.

\newpage

\begin{figure}[H]
\centering
    \caption{Example of a reported road closure}
    \vspace*{0.2in}   
    \includegraphics[height=2.5in]{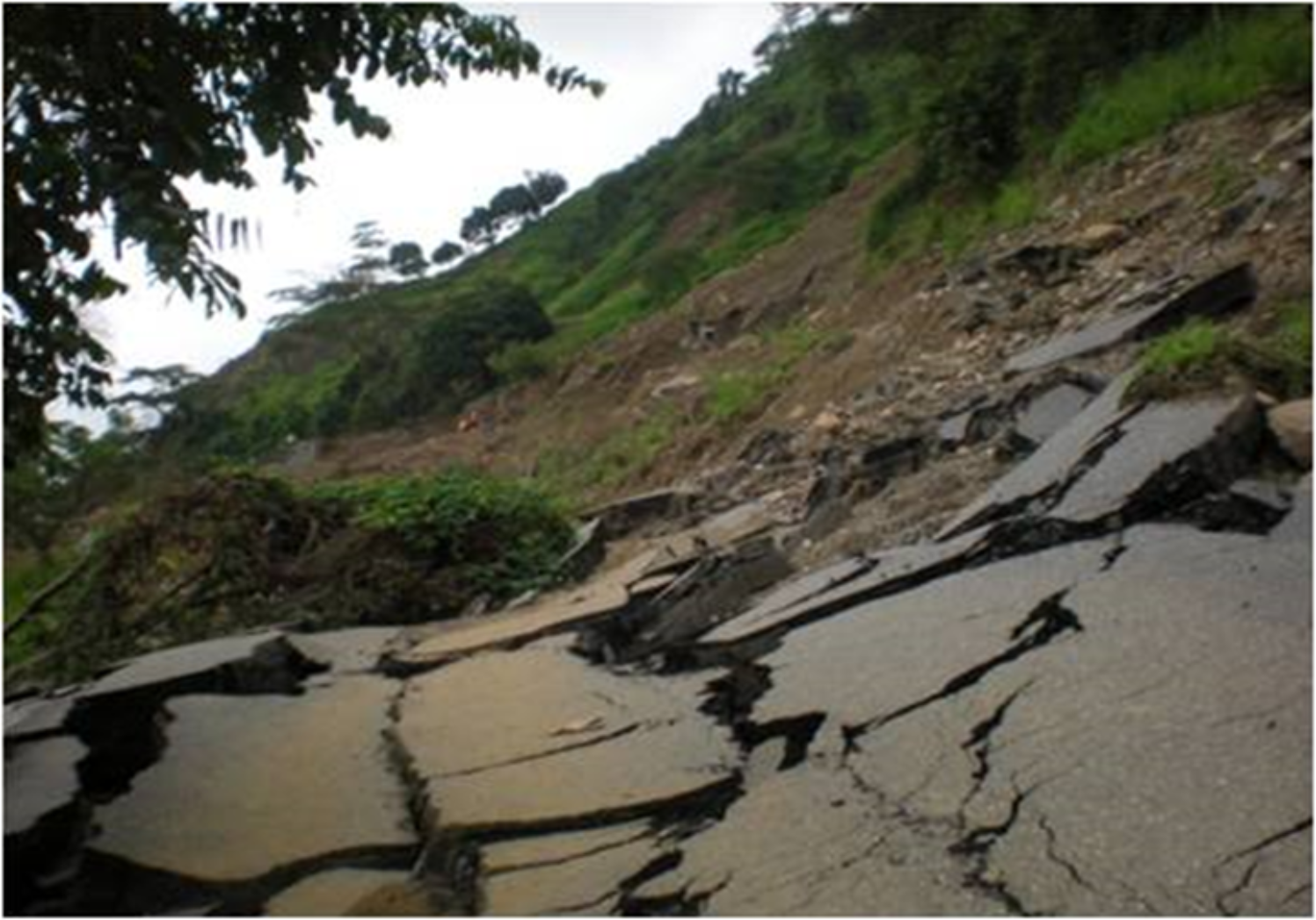}
       \vspace*{0.1in}  
       
    \includegraphics[height=2.5in]{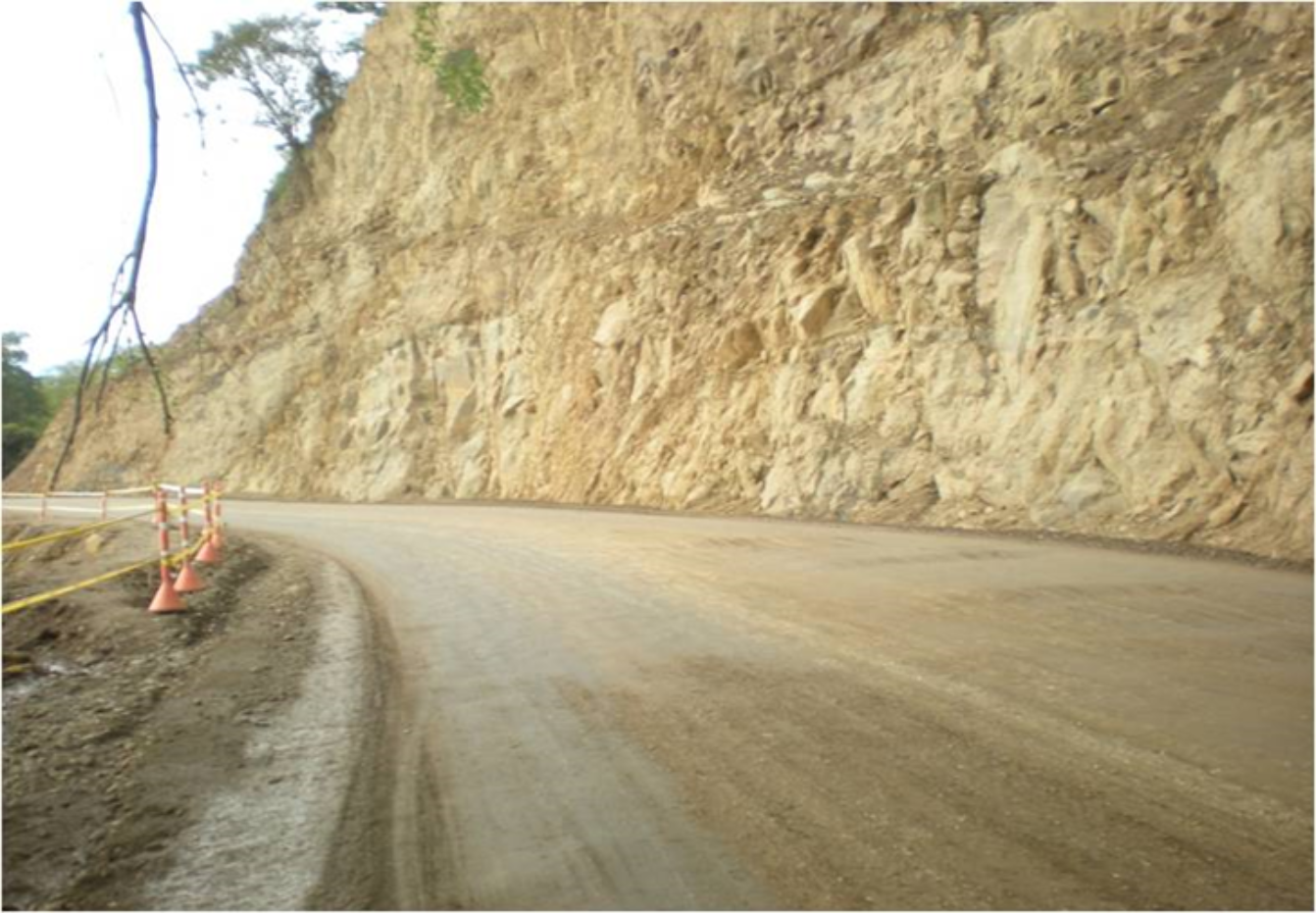}
    \vspace*{0.1in}  
    
    \label{fig:example_entree}
             \scalebox{0.90}{
\begin{minipage}{0.8\textwidth}
\advance\leftskip 0cm
	{\footnotesize {
\textit{Notes}: The figure illustrates an example of a road disruption. In the top panel, there's a photograph of the road shortly after the landslide, and in the bottom panel, there's a photograph of the same road after it was repaired. } \par }
\end{minipage}
 } 
    \end{figure}

\begin{table}[htbp]
\centering
\caption{Distribution of exposed and non-exposed 
relationships by relationship age at the shock}
\label{tab:cohort_balance}
\vspace*{0.1in}
\begin{tabular}{lccc}
\toprule
\toprule
Relationship age & Non-exposed & Exposed & Total \\
(months at shock) & ($E_{ij,0}=0$) & ($E_{ij,0}=1$) & \\
\midrule
\addlinespace[3pt]
0-6 & 297 & 344 &  641 \\
\addlinespace[3pt]
6-12 & 1,333 & 1,357 & 2,688 \\
\addlinespace[3pt]
12-18 & 1,721 & 4,571 & 6,292 \\
\addlinespace[3pt]
18-24 &  1,184 &  1,601 &  2,785 \\
\addlinespace[3pt]
24-30 &  946 & 803 &  1,749 \\
\addlinespace[3pt]
30-36 & 1,230 &  1,006  &  2,236 \\
\addlinespace[3pt]
36-42 &  896 &  568  &   1,464 \\
\addlinespace[3pt]
42-48 &  5,669   &  2,989 &  8,658 \\
\addlinespace[3pt]
\midrule
Total & 13,274 &  13,239 &26,513 \\
\bottomrule
\end{tabular}
\vspace*{0.1in}
\scalebox{0.90}{
\begin{minipage}{1.0\textwidth}
\advance\leftskip 0.5cm
{\footnotesize{\textit{\textbf{Notes:}} The table 
reports the number of relationship-month observations 
by relationship age at the time of the shock (October 
2010) and exposure status. Relationship age is defined 
as the number of months between the first 
observed transaction and October 2010. }\par}
\end{minipage}
}
\end{table}

\begin{table}[htbp]
\centering
\caption{Relationships by product category}
\vspace*{0.1in}
\scalebox{0.85}{
\begin{tabular}{llccc}
\toprule
\toprule
HS Code & Description & Non-exposed & Exposed & Total \\
& & ($E_{ij,0}=0$) & ($E_{ij,0}=1$) & \\
\midrule
\addlinespace[3pt]
\multicolumn{5}{l}{\textit{Live plants and bulbs}} \\
\addlinespace[3pt]
060110 & Bulbs, tubers, rhizomes (dormant) & 0 & 0 & 0 \\
\addlinespace[2pt]
060210 & Unrooted cuttings and slips & 2 & 5 &7 \\
\addlinespace[2pt]
060220 & Trees and shrubs, edible fruit & 0 & 0 & 0 \\
\addlinespace[2pt]
060290 & Other live plants, cuttings (n.e.s.) & 0 & 0 & 0 \\
\addlinespace[5pt]
\multicolumn{5}{l}{\textit{Fresh cut flowers}} \\
\addlinespace[3pt]
060311 & Roses &  1,347  & 344 &   1,691 \\
\addlinespace[2pt]
060312 & Carnations &  748  & 208  & 956 \\
\addlinespace[2pt]
060313 & Orchids & 6 & 4 & 10 \\
\addlinespace[2pt]
060314 & Chrysanthemums & 185 &  828 & 1,013 \\
\addlinespace[2pt]
060319 & Other fresh cut flowers (incl.\ hydrangeas) & 1,039 & 1,658 &  2,697 \\
\addlinespace[5pt]
\multicolumn{5}{l}{\textit{Other cut flowers and foliage}} \\
\addlinespace[3pt]
060390 & Cut flowers, dried or otherwise prepared & 11 & 3 & 14 \\
\addlinespace[2pt]
060491 & Foliage, branches and grasses, fresh & 56  & 292  & 348 \\
\addlinespace[2pt]
060499 & Foliage, branches and grasses, dried & 2 & 1 & 3 \\
\addlinespace[3pt]
\bottomrule
\end{tabular}
}
\vspace*{0.1in}
\scalebox{0.90}{
\begin{minipage}{1.0\textwidth}
\advance\leftskip 0.5cm
{\footnotesize{\textit{\textbf{Notes:}} The table 
reports the average number of relationships per 
month by HS product code (2007 classification) 
and exposure status, computed over the pre-shock period. Non-exposed and exposed 
refer to $E_{ij,0}=0$ and $E_{ij,0}=1$ respectively. 
The three dominant products by relationship count 
are roses (060311), carnations (060312), other fresh cut flowers 
(060319, primarily hydrangeas), and chrysanthemums 
(060314).
}\par}
\end{minipage}
}
\label{tab:rel_products}
\end{table}

\begin{figure}[H]
    \centering
  
 \includegraphics[height=1.95in]{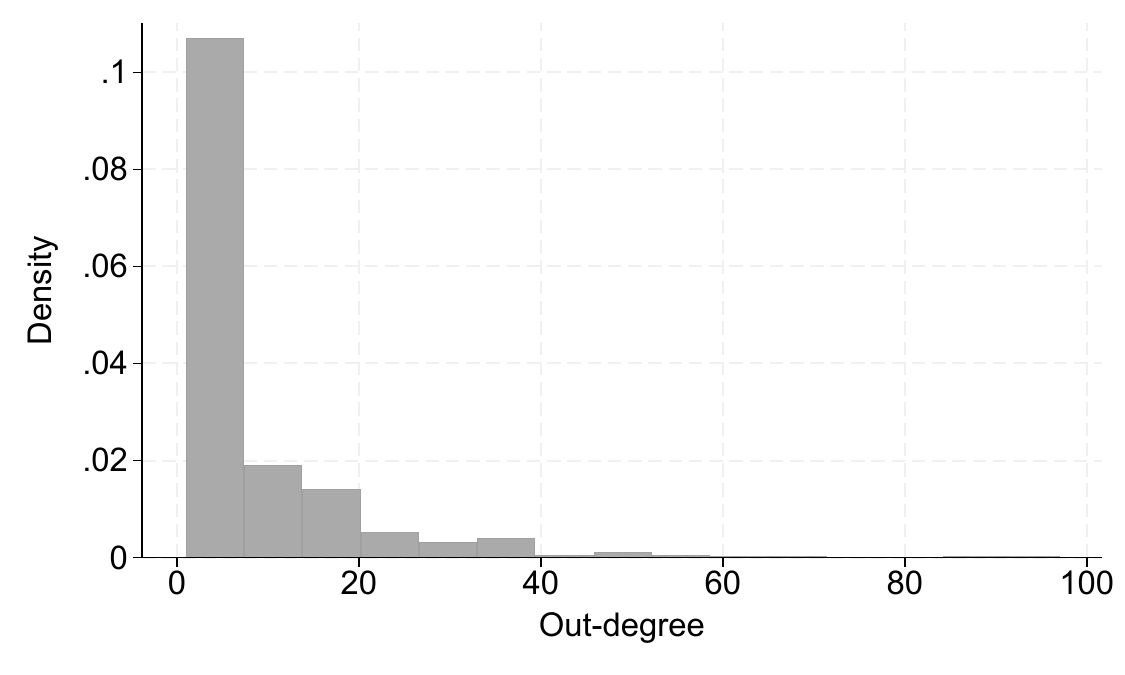} 
 \includegraphics[height=1.95in]{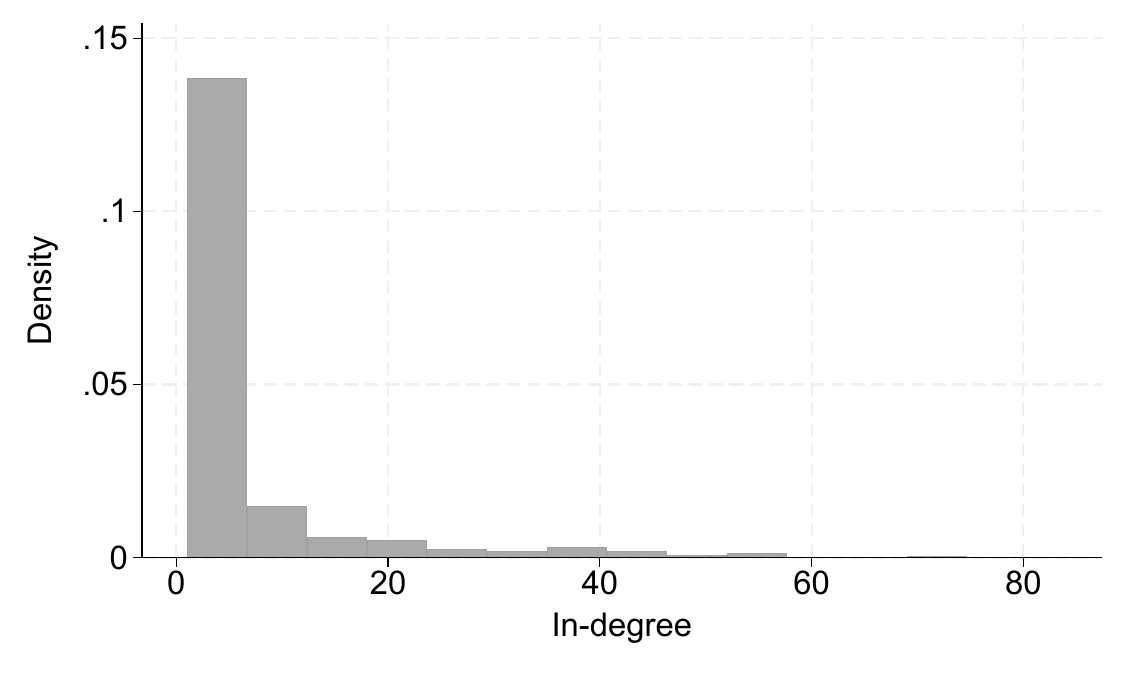}  
     \vspace*{-0.2in}
     \caption{Distribution of firms' connections in the pre-shock period}
          \vspace*{0.1in}

         \scalebox{0.90}{
\begin{minipage}{1.1\textwidth}
\advance\leftskip 0.5cm
	{\footnotesize {
\textit{\textbf{Notes:}}  The figure shows the distribution of the number of active relationships per firm (\textit{degree}) in the pre-shock period, for the exposure sample. The left panel shows exporter out-degree (number of active 
U.S.\ importers per Colombian exporter) and the right panel shows importer in-degree (number of active Colombian exporters per U.S.\ importer).} \par }
\end{minipage}
} 
    \label{fig:distr_ninj}
\end{figure}

\begin{figure}[ht]
    \centering

 \vspace*{0.2in}   
   \includegraphics[height=1.95in]{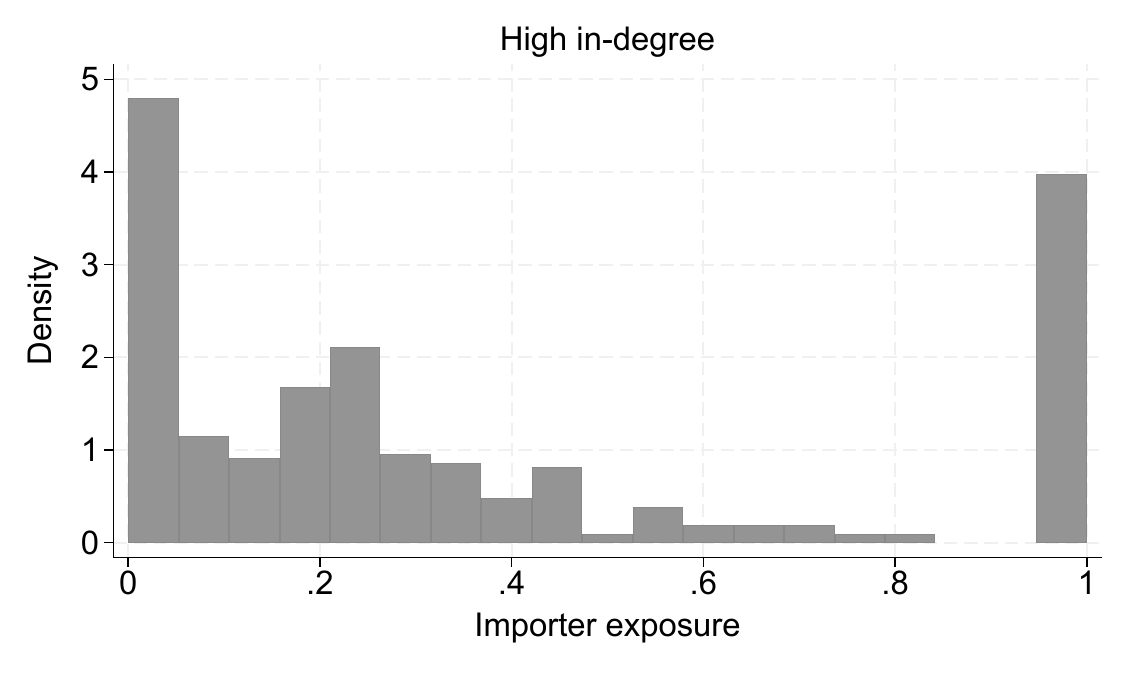}
   \includegraphics[height=1.95in]{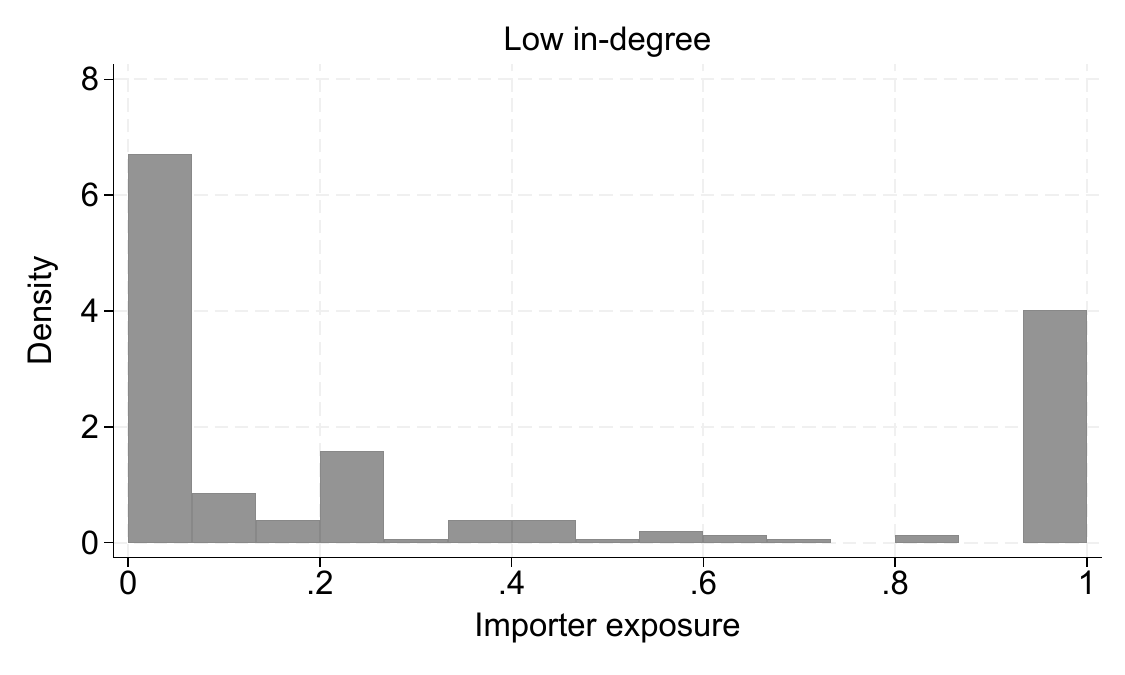}   
         \vspace*{-0.2in}
        \caption{Distribution of exposure measures at the firm-level by degree of importer}
     \vspace*{0.1in}

     \scalebox{0.90}{
\begin{minipage}{1.1\textwidth}
\advance\leftskip 0.5cm
	{\footnotesize {\textit{\textbf{Notes:}} The figure shows the distribution of importer exposure ($IE_j$) for subset of firms.  The left panel shows the distributions for importers with high in-degree (above median) in the pre-shock period and the right panel shows the distributions for importers with low in-degree (below the median).   
} \par }
\end{minipage}
} 
        \label{fig:distr_ie_indegree}
\end{figure}

%%Robustness checks within relationships 

\begin{figure}[H]
 \begin{center}
\includegraphics[height=1.95in]{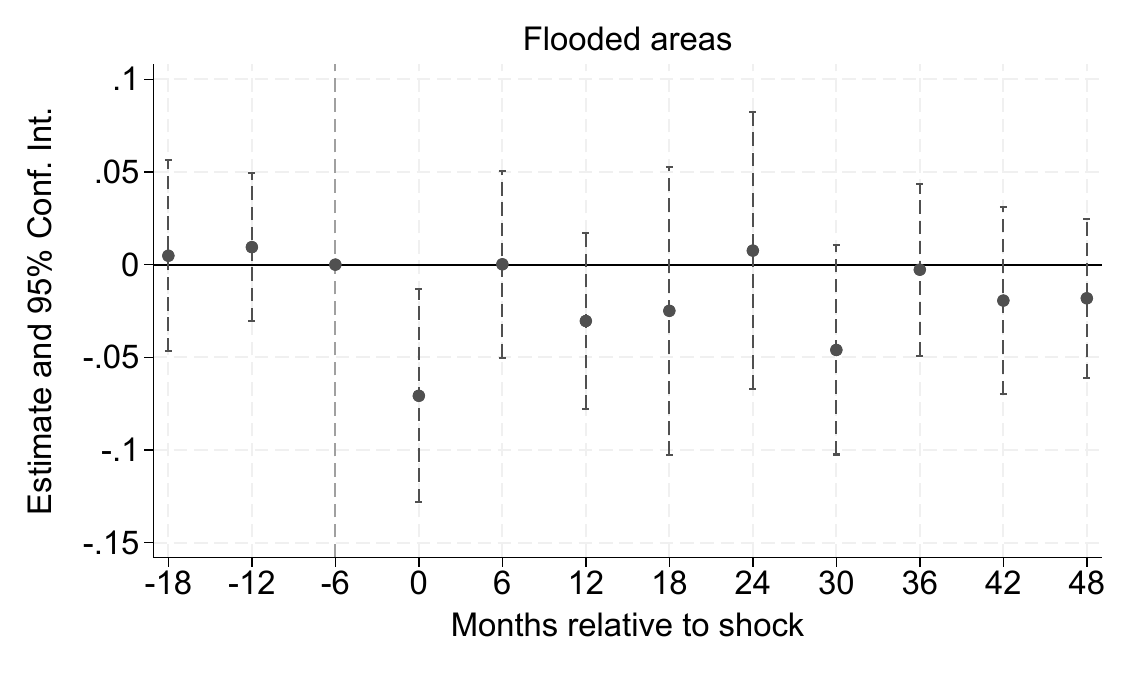}
\includegraphics[height=1.95in]{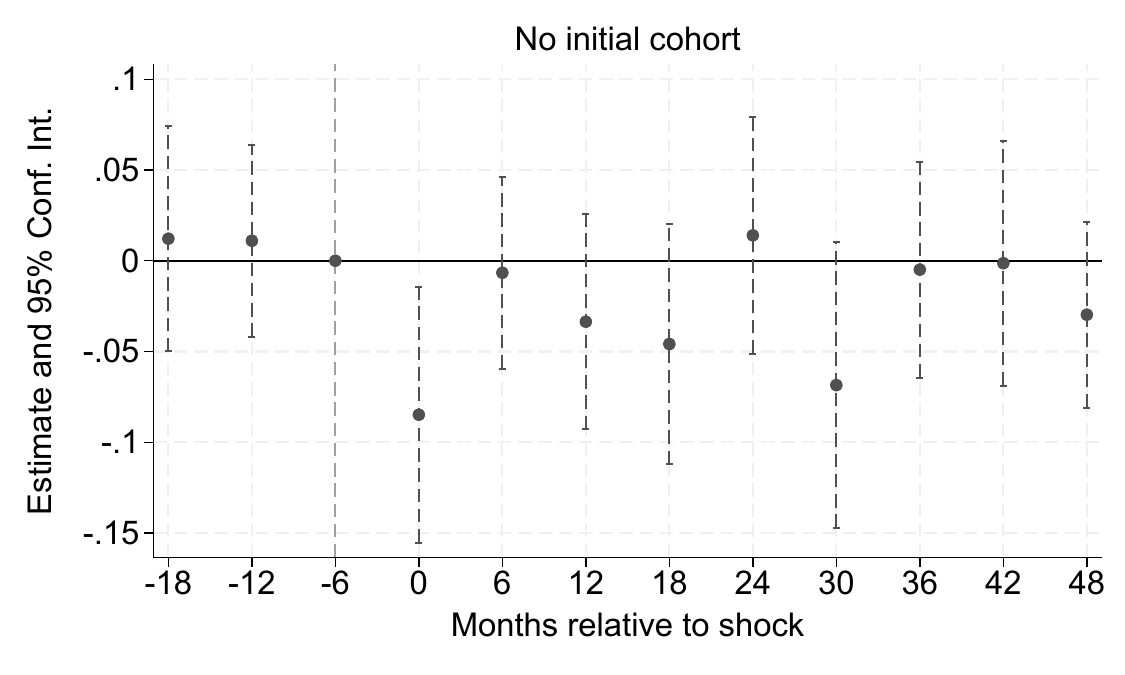}
\end{center}
\vspace*{-0.2in}
   \caption{Robustness: Road disruptions effect on the probability of a relationship ending}
\vspace*{0.1in}
\scalebox{0.90}{
\begin{minipage}{1.1\textwidth}
\advance\leftskip 0.5cm
	{\footnotesize{
\textit{\textbf{Notes:}} All figures plot the $\beta_l$ coefficients from estimating equation (\ref{Y1})  within-importer specification with $\alpha_{jt}$ and  their respective 95\% confidence intervals.  The left panel includes flooded areas, the  right panel excludes initial age group (42-48 months). All regressions include relationship,  and cohort-by-time fixed effects. Confidence intervals are clustered at the exporter level.} \par}
\end{minipage}
}
\end{figure}

\begin{figure}[H]
\begin{center}
\textbf{\footnotesize{\textsc{All relationships}}}
\hspace{1.8in}
\textbf{\footnotesize{\textsc{Within importer}}}\\
\hspace*{-0.35in}
\includegraphics[height=1.9in]{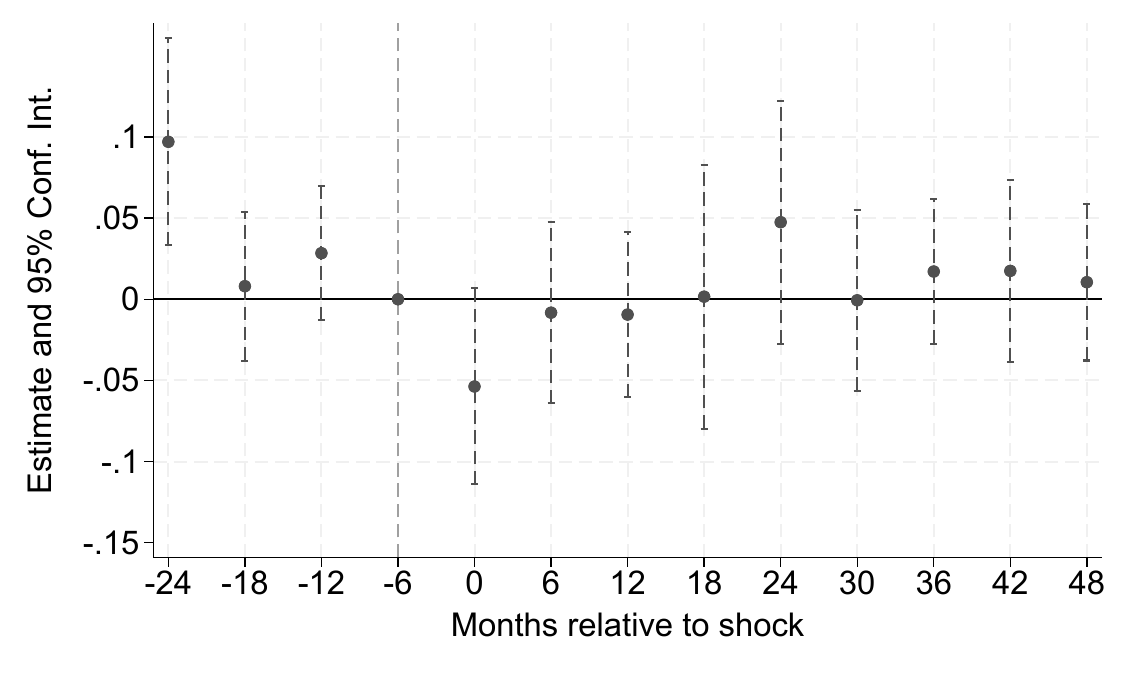}
\includegraphics[height=1.9in]{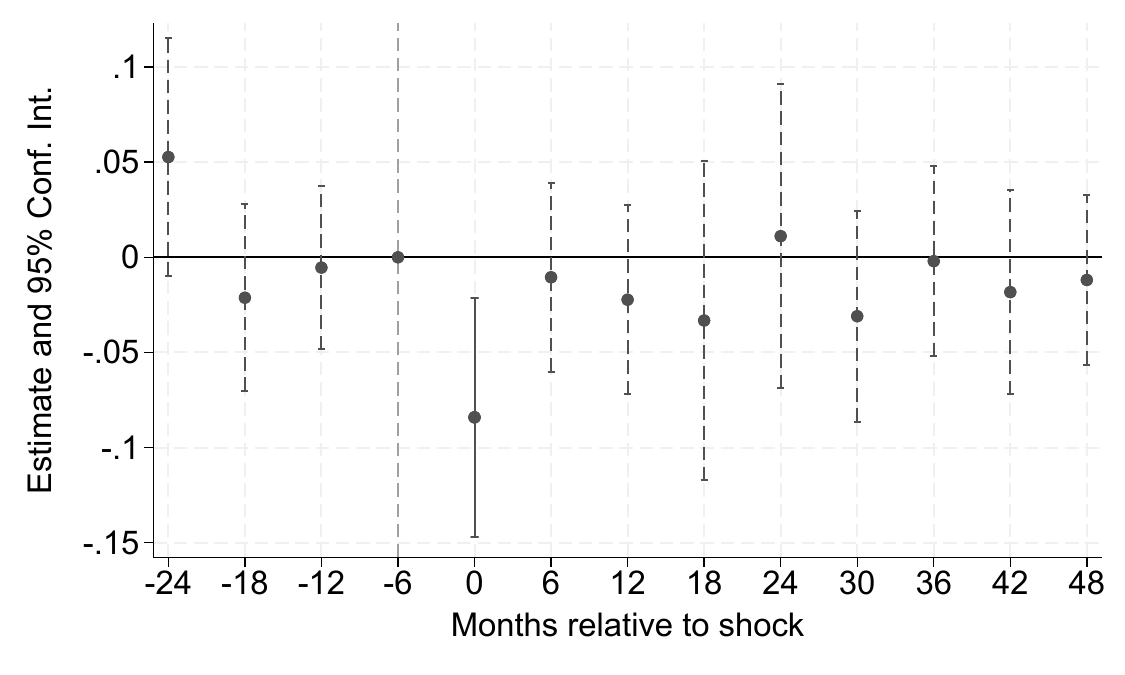}
\end{center}
\vspace*{-0.2in}
    \caption{Robustness:  Road disruptions effect on the probability of a relationship ending (Extended pre-shock)}
    \vspace*{0.1in}
\scalebox{0.90}{
\begin{minipage}{1.1\textwidth}
\advance\leftskip 0.5cm
	{\footnotesize{
\textit{\textbf{Notes:}} The figure plots the $\beta_l$ coefficients 
from equation~\eqref{Y3}, where the outcome equals one if a 
transaction is observed in month $m$ and zero otherwise. The 
estimation sample is the panel of relationships active at some 
point in the sample, excluding relationships first observed after 
September 2010. All specifications include relationship and 
cohort-by-month fixed effects. The right panel additionally includes 
importer-by-month fixed effects $\alpha_{j,m}$, restricting 
identification to within-importer-month variation.  Coefficients are reported with 95\% confidence intervals 
based on standard errors clustered at the exporter level.} \par}
\end{minipage}
}
\end{figure}

\begin{figure}[H]
\begin{center}
\hspace{0.25in}
\textbf{\footnotesize{\textsc{Kilograms}}}
\hspace{2.5in}
\textbf{\footnotesize{\textsc{Trade value}}}\\
\hspace*{-0.35in}
\includegraphics[height=1.95in]{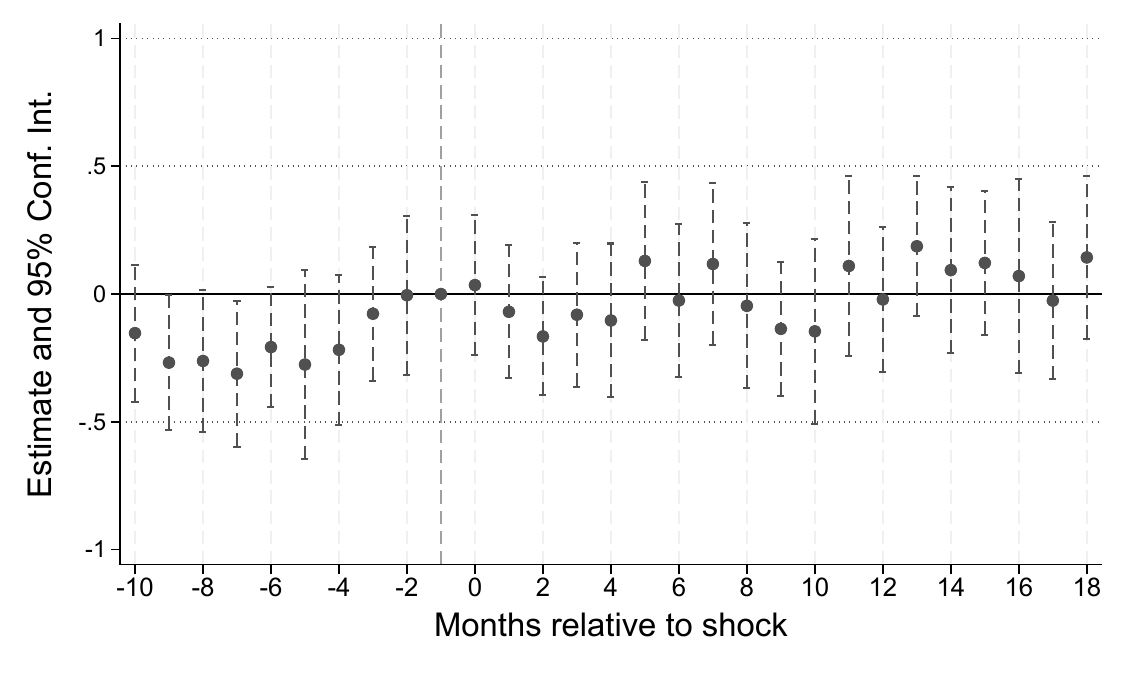}
\includegraphics[height=1.95in]{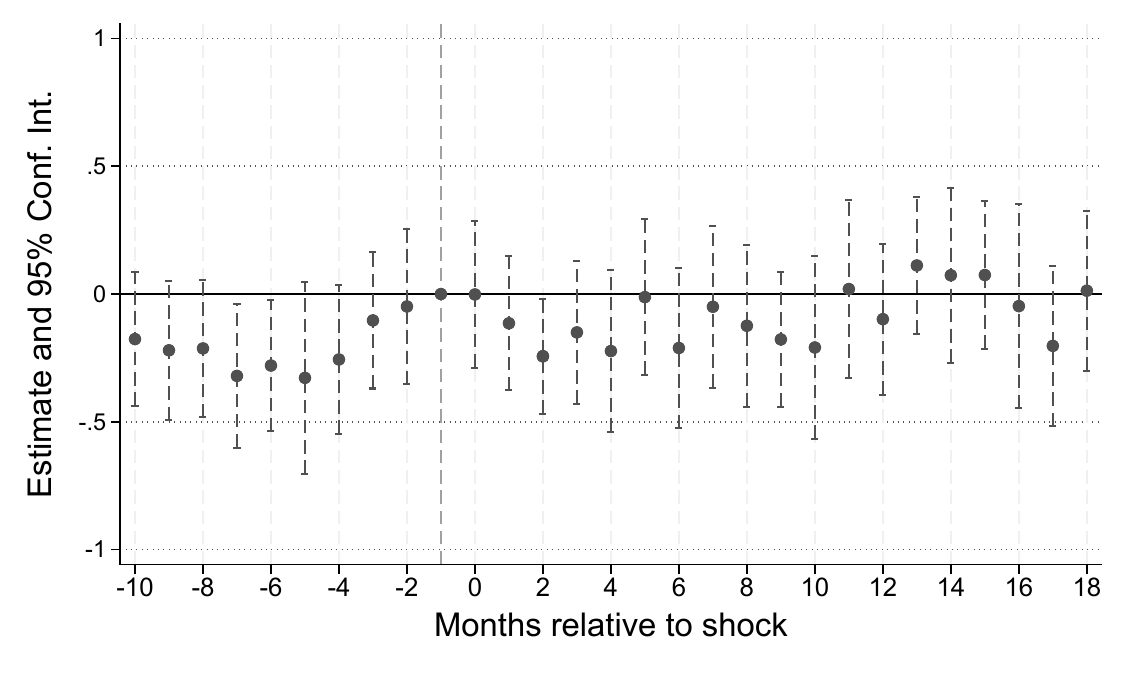}
\end{center}
\vspace*{-0.2cm}
\caption{Robustness: Within importer-time fixed effects}
\vspace*{0.1cm}
\scalebox{0.90}{
\begin{minipage}{1.1\textwidth}
\advance\leftskip 0.5cm
{\footnotesize{
\textit{\textbf{Notes:}} Both panels plot the $\beta_l$ coefficients 
from equation~\eqref{Y4} with 95\% confidence intervals, where the 
dependent variable is the inverse hyperbolic sine of kilograms 
(left panel) and trade value (right panel). The sample is restricted 
to cut-flower HS codes (same HS6 codes as in Panel B of 
Table~\ref{tab:stat}) and to relationship--product cells observed 
in at least two months. Regressions include relationship--product, importer-by-month
and cohort-by-month fixed effects. Standard errors 
are clustered at the exporter level.  }\par}
\end{minipage}
}
\end{figure}

\begin{figure}[H]
\begin{center}
\hspace{0.25in}
\textbf{\footnotesize{\textsc{Unit values}}}
\hspace{2.5in}
\textbf{\footnotesize{\textsc{Price per kg}}}\\
\hspace*{-0.35in}
\includegraphics[height=1.95in]{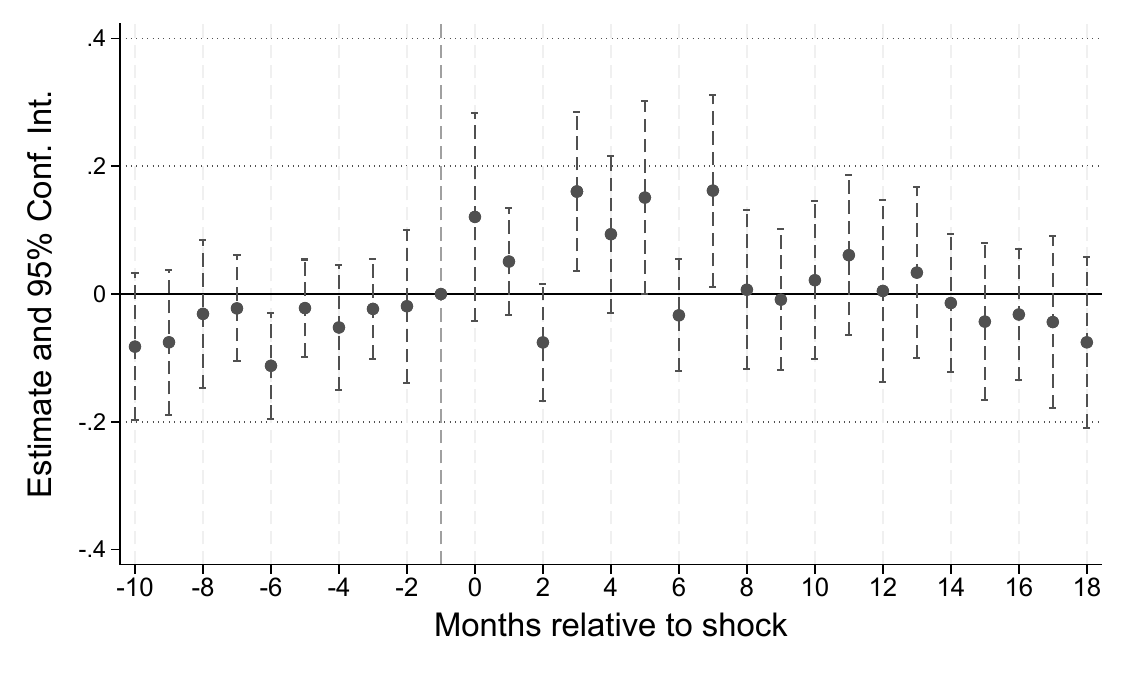}
\includegraphics[height=1.95in]{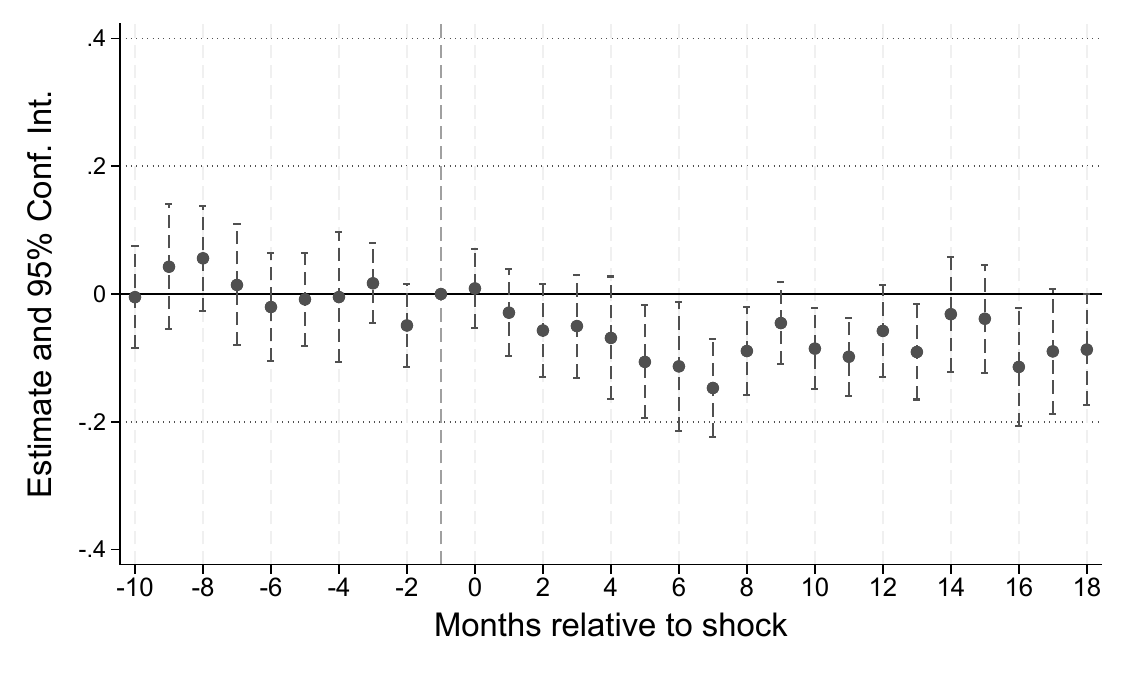}
\end{center}
\vspace*{-0.2cm}
\caption{Robustness: Within importer-time fixed effects}
\vspace*{0.1cm}
\scalebox{0.90}{
\begin{minipage}{1.1\textwidth}
\advance\leftskip 0.5cm
{\footnotesize{
\textit{\textbf{Notes:}}  Both panels plot the $\beta_l$ coefficients 
from equation~\eqref{Y4} with 95\% confidence intervals, where the 
dependent variable is the inverse hyperbolic sine of unit value (left 
panel) and price per kilograms (right panel). The sample is restricted 
to cut-flower HS codes (same HS6 codes as in Panel B of 
Table~\ref{tab:stat}) and to relationship--product cells observed 
in at least two months. Regressions include relationship, importer-by-month, cohort-by-month, and product-by-month
Standard errors are clustered at the exporter level. }\par}
\end{minipage}
}
\end{figure}

\begin{figure}[H]
\begin{center}
\hspace{0.25in}
\textbf{\footnotesize{\textsc{Unit values}}}
\hspace{2.5in}
\textbf{\footnotesize{\textsc{Price per kg}}}\\
\hspace*{-0.35in}
\includegraphics[height=1.95in]{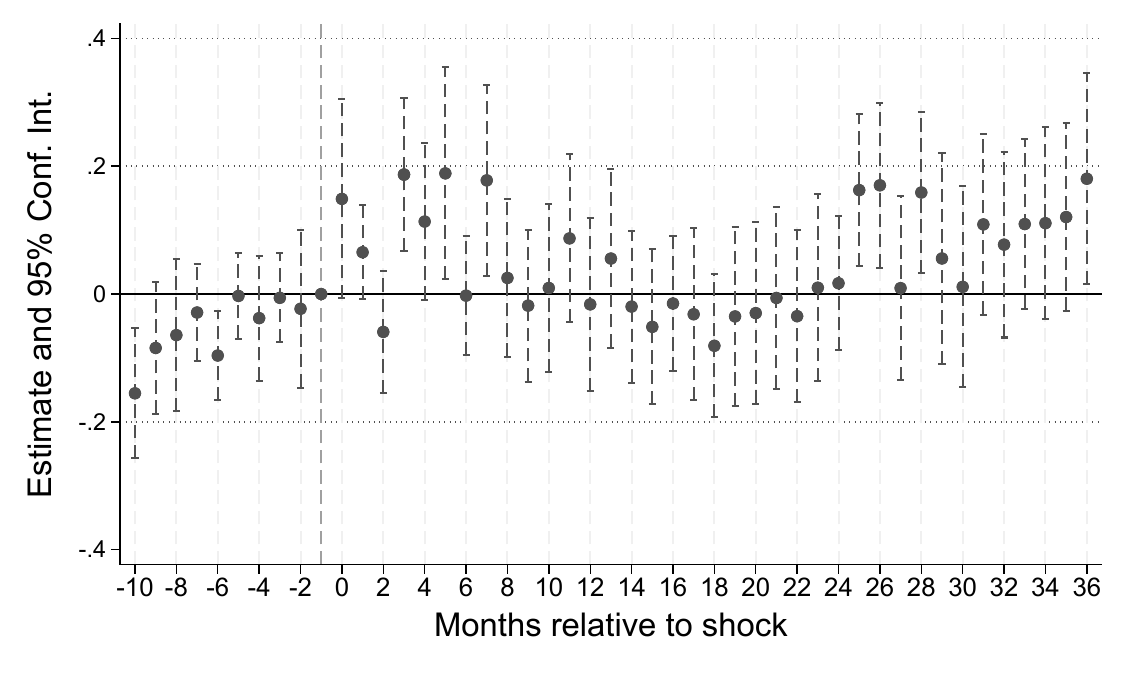}
\includegraphics[height=1.95in]{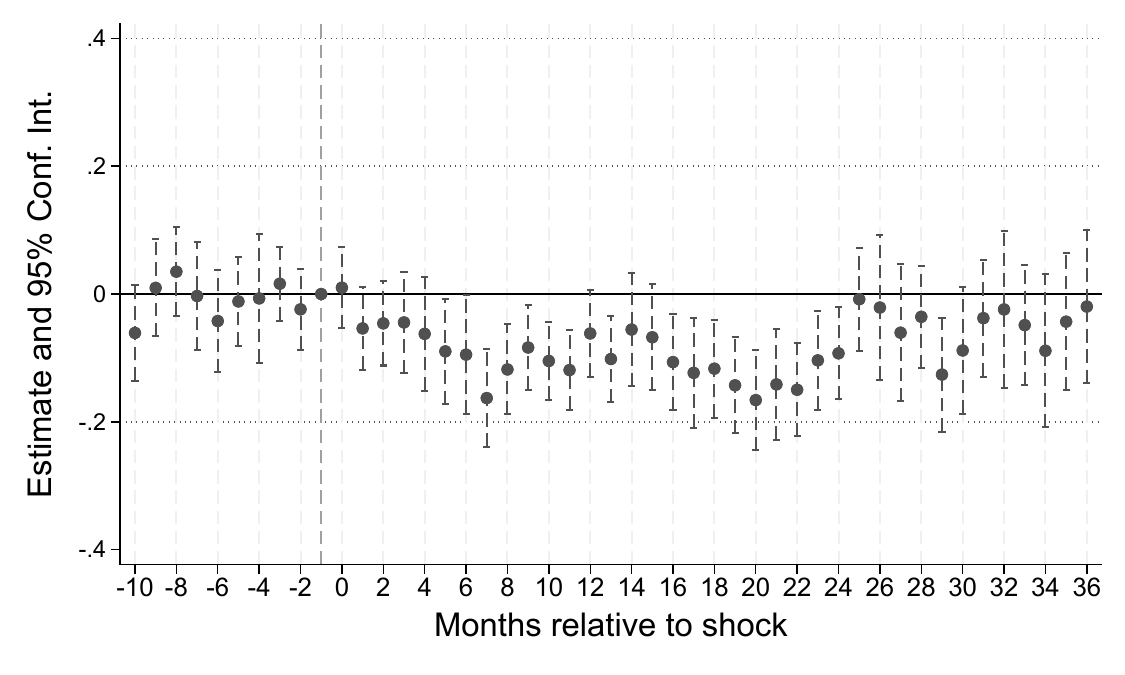}
\end{center}
\vspace*{-0.2cm}
\caption{Robustness: Long post-shock window}
\vspace*{0.1cm}
\scalebox{0.90}{
\begin{minipage}{1.1\textwidth}
\advance\leftskip 0.5cm
{\footnotesize{
\textit{\textbf{Notes:} } Both panels plot the $\beta_l$ coefficients 
from equation~\eqref{Y4} with 95\% confidence intervals, where the 
dependent variable is the inverse hyperbolic sine of unit values (left 
panel) and price per kilogram (right panel). The sample is restricted 
to cut-flower HS codes (same HS6 codes as in Panel B of 
Table~\ref{tab:stat}) and to relationship--product cells observed 
in at least two months. Regressions include relationship, 
cohort-by-month and product-by-month  fixed effects. 
Standard errors are clustered at the exporter level. }\par}
\end{minipage}
}
\end{figure}

\begin{figure}[h]
\begin{center}
\hspace{0.25in}
\textbf{\footnotesize{\textsc{All importers}}}
\hspace{2.5in}
\textbf{\footnotesize{\textsc{By in-degree}}}\\
\hspace*{-0.35in}
\includegraphics[height=1.95in]{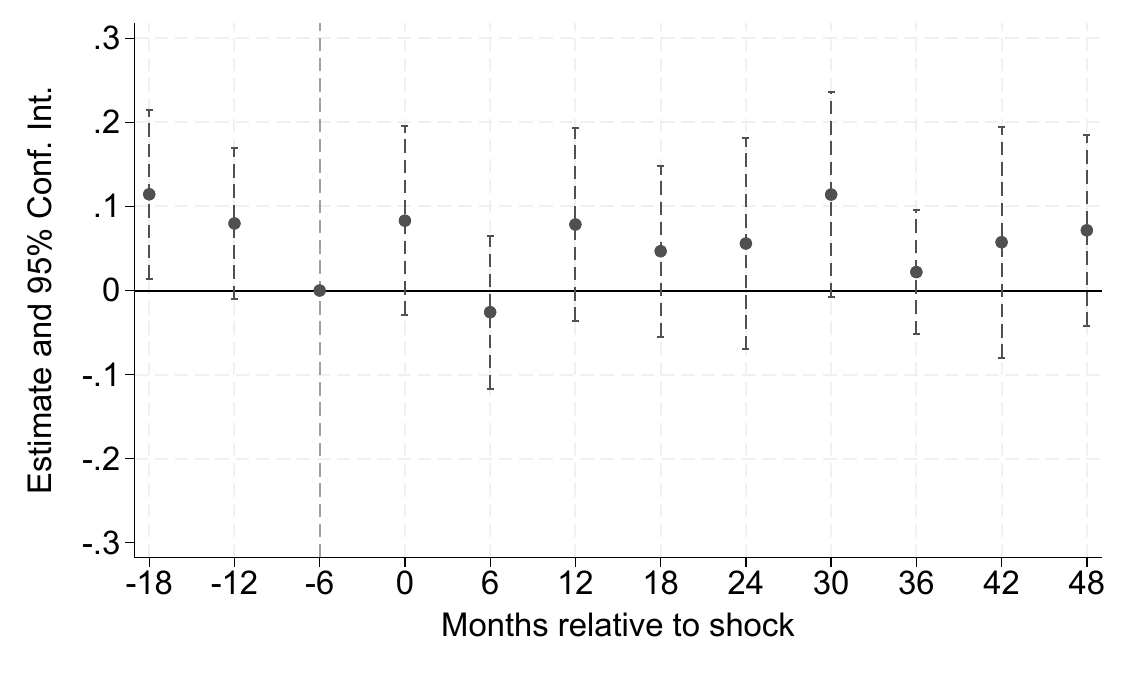}
\includegraphics[height=1.95in]{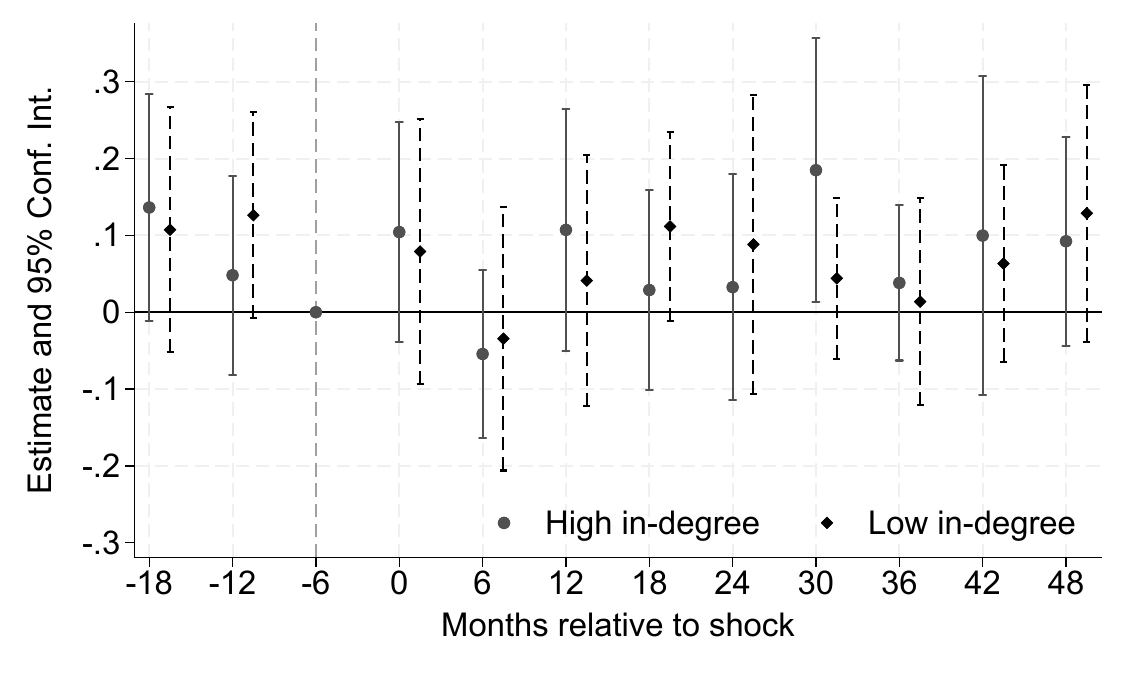}
\end{center}
\vspace*{-0.2cm}
\caption{Robustness: Effect of importer-level exposure ($IE_j$ with NA) on the probability 
of a relationship ending}
\vspace*{0.1cm}
\scalebox{0.90}{
\begin{minipage}{1.1\textwidth}
\advance\leftskip 0.5cm
{\footnotesize{
\textit{\textbf{Notes:}} The figure plots the $\beta_l$ coefficients 
from equation~\eqref{Y2}, where the outcome is the relationship-ending 
hazard (equal to one if a relationship active in $t-1$ ends in $t$). 
The in-degree split defined in 
Section~\ref{sec:empirical}. All regressions include relationship 
and cohort-by-time fixed effects. Coefficients are reported with 
95\% confidence intervals based on standard errors clustered at the 
exporter level. }\par}
\end{minipage}
}
\end{figure}

\begin{figure}[H]
\begin{center}
\hspace{0.25in}
\textbf{\footnotesize{\textsc{All buyers}}}
\hspace{2.5in}
\textbf{\footnotesize{\textsc{By in-degree}}}\\
\hspace*{-0.35in}
\includegraphics[height=1.95in]{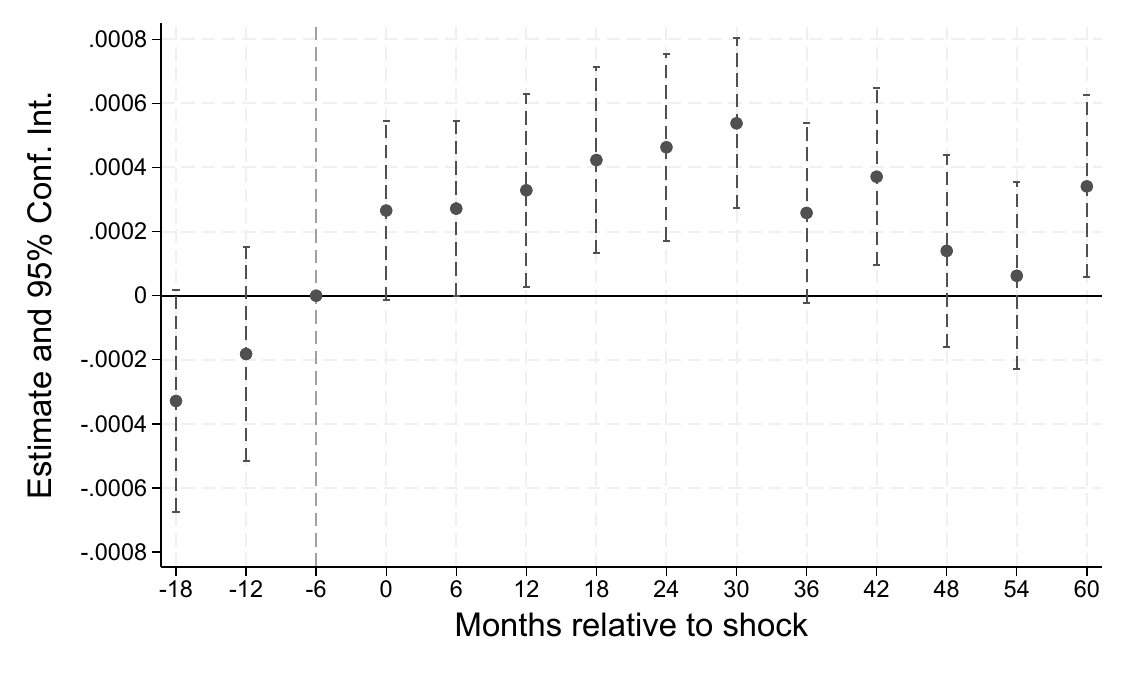}
\includegraphics[height=1.95in]{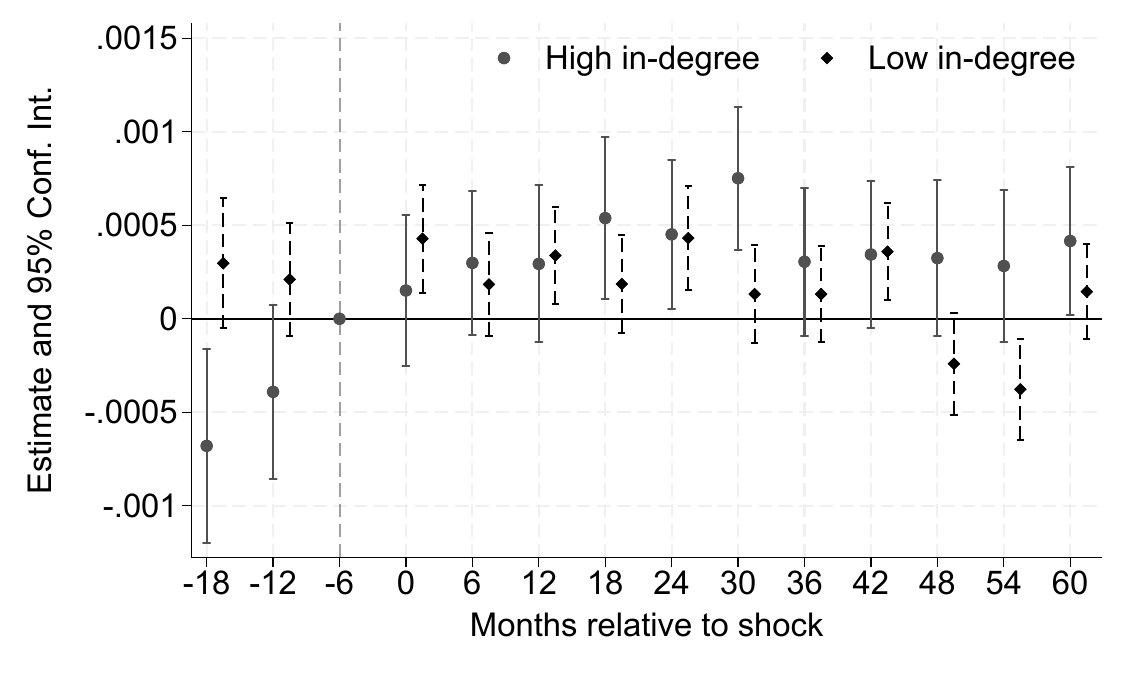}
\end{center}
\vspace*{-0.2cm}
\caption{Effect of road disruptions on new 
relationship formation: binary importer-level exposure}
\label{fig:new_matches_rob}
\vspace*{0.1cm}
\scalebox{0.90}{
\begin{minipage}{1.1\textwidth}
\advance\leftskip 0.5cm
{\footnotesize{
\textit{\textbf{Notes:}} The figure plots the 
$\beta_l$ coefficients from equation~(\ref{Y5n}) for the new relationship formation indicator ($n_{ijt}$), for all importers (left panel) and by in-degree (right panel)  with a 95\% confidence 
intervals. Importer-level 
exposure ($IE_j$) is replaced by a binary indicator equal to one if $IE_j > 0.50$ and zero otherwise.  All regressions, include importer, exporter and time fixed effects. Standard errors 
are clustered at the exporter level.}\par}
\end{minipage}
}
\end{figure}

\begin{figure}[H]
\begin{center}
\hspace{0.25in}
\textbf{\footnotesize{\textsc{All buyers}}}
\hspace{2.5in}
\textbf{\footnotesize{\textsc{By in-degree}}}\\
\hspace*{-0.35in}
\includegraphics[height=1.95in]{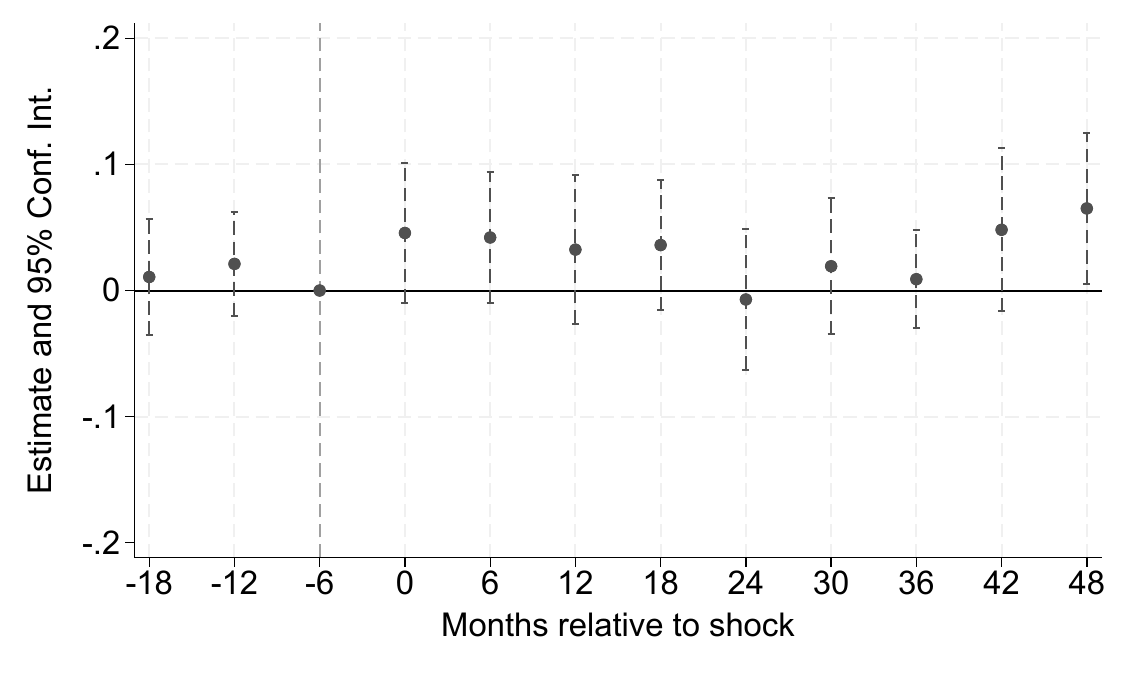}
\includegraphics[height=1.95in]{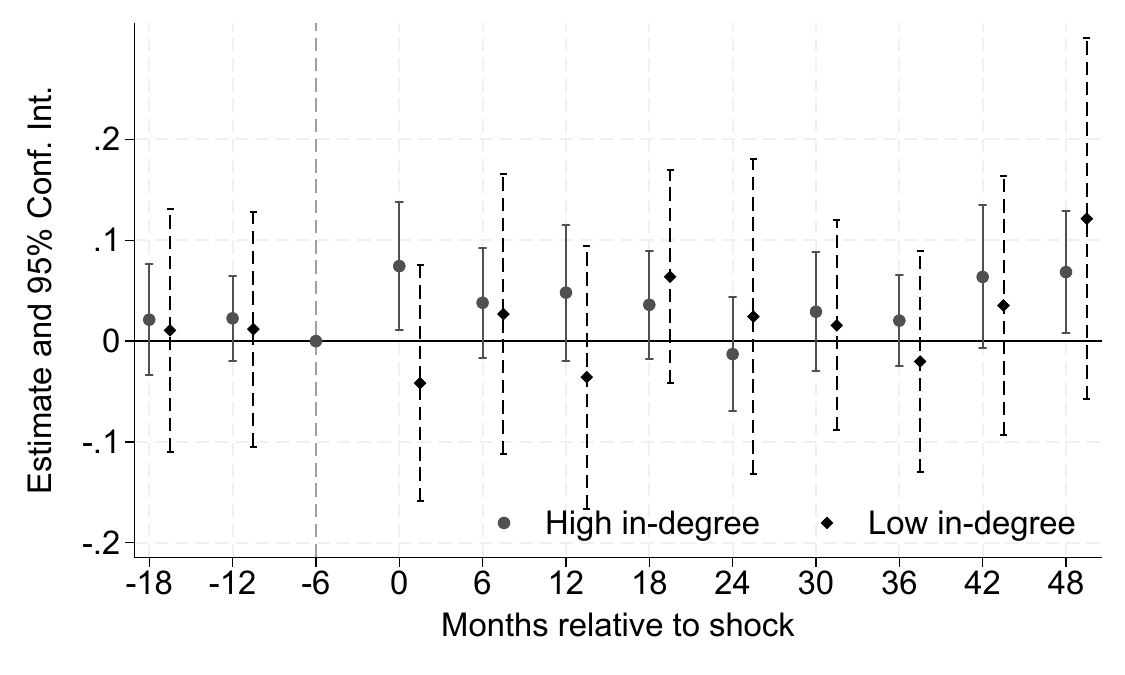}
\end{center}
\vspace*{-0.2cm}
\caption{Probability of a relationship ending:
binary importer-level exposure}

\vspace*{0.1cm}
\scalebox{0.90}{
\begin{minipage}{1.1\textwidth}
\advance\leftskip 0.5cm
{\footnotesize{
\textit{\textbf{Notes:}} The figure plots the 
$\beta_l$ coefficients from equation~(\ref{Y2}) 
with 95\% confidence intervals, where importer-level 
exposure ($IE_j$) is replaced by a binary indicator 
equal to one if $IE_j > 0.50$ and zero otherwise. 
The left panel uses all buyer--seller pairs; the 
right panel estimates effects separately for each 
subsample defined by importer in-degree. All 
regressions include relationship, cohort-time  and period fixed effects. Standard errors 
are clustered at the exporter level.}\par}
\end{minipage}
}
\label{fig:firm_Exposure_rob}
\end{figure}

\begin{figure}[h]
\begin{center}
\hspace{0.25in}
\textbf{\footnotesize{\textsc{All importers}}}
\hspace{2.5in}
\textbf{\footnotesize{\textsc{By in-degree}}}\\
\hspace*{-0.35in}
\includegraphics[height=1.95in]{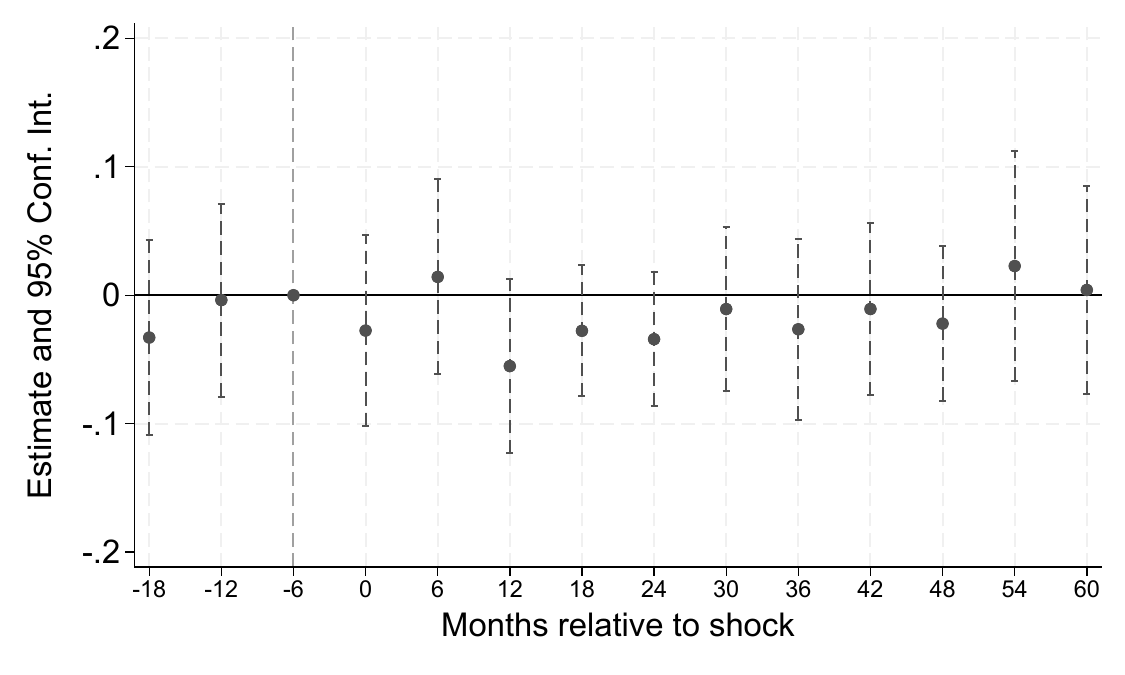}
\includegraphics[height=1.95in]{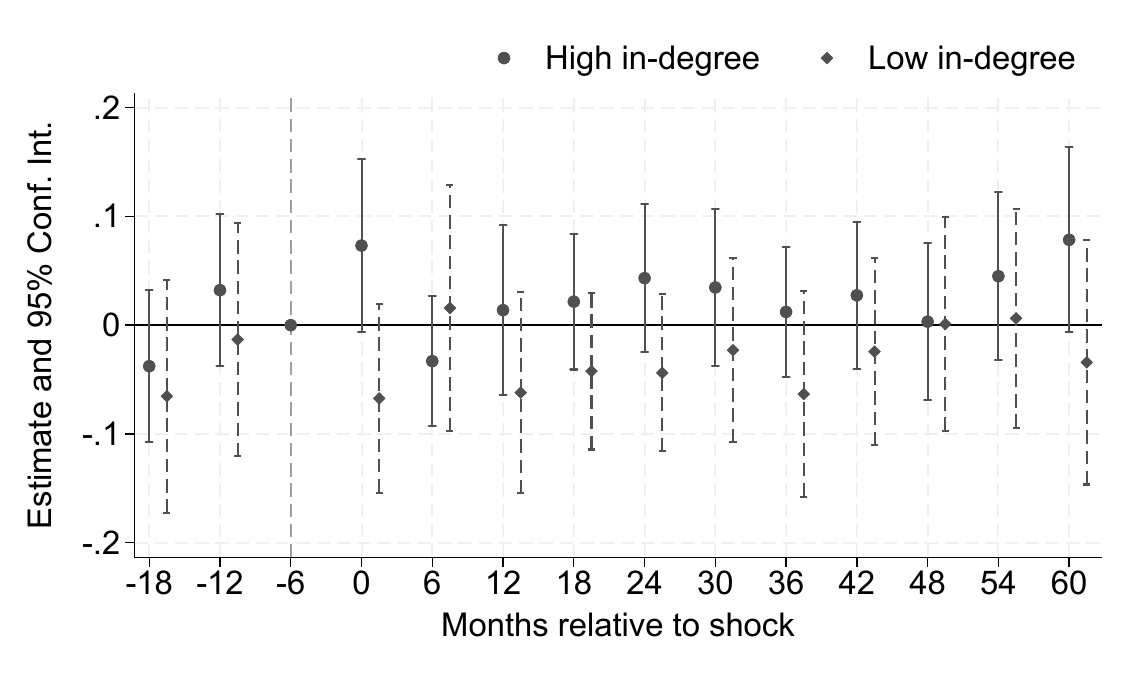} 
\vspace*{-0.2cm}
\end{center}
\caption{Probability of importer exit}
\vspace*{0.1cm}
     \scalebox{0.90}{
\begin{minipage}{1.1\textwidth}
\advance\leftskip 0.5cm
	{\footnotesize {
\textit{\textbf{Notes:}} The figure plots the $\beta_l$ coefficients from estimating equation (\ref{Y6}) with 95\% confidence intervals.  The left-side panel shows all importers. The right-panel shows by in-degree.  Importer-level 
exposure ($IE_j$) is replaced by a binary indicator equal to one if $IE_j > 0.50$ and zero otherwise. Regressions include importer and time fixed effects. N=  4,688 all, high = 4,168 low= 2,155. } \par }
\end{minipage}
}  
\end{figure}

\clearpage
\includepdf[pages=-]{Flooded_area_savannah.pdf}

\includepdf[pages=-]{Flooded_area_antioquia.pdf}

\end{document}